\documentclass[review]{elsarticle}

\usepackage{lineno,hyperref}
\usepackage{amsmath}
\usepackage{graphicx}
\usepackage{bm}
\usepackage{caption}
\usepackage{subcaption}
\usepackage{float}
\usepackage{tikz, graphicx}
\usepackage{color}
\usepackage{tikz}
\usetikzlibrary{shapes}
\usepackage{mathtools}
\usepackage{csquotes} 
\usepackage{amsmath}
\usepackage{amssymb}
\usepackage{graphicx}
\usepackage{bm}
\usepackage{caption}
\usepackage{subcaption}
\usepackage{float}
\usepackage{mathrsfs}
\usepackage{color}
\usepackage{tikz}
\usetikzlibrary{shapes}
\usepackage{mathtools}
\usepackage{xcolor}
\usepackage{amssymb}
\usepackage{bm}
\usepackage{nameref}
\usepackage[toc,page]{appendix}
\usepackage[english]{babel}
\usepackage[T1]{fontenc}
\usepackage{lipsum}
\usepackage[normalem]{ulem}
\newcommand{\sz}[1]{{\color{black}#1}}

\newcommand{\mr}[1]{{\color{black}#1}}

\journal{Advances in Water Resources}

\begin{document}

\begin{frontmatter}

\title{The role of wetting on the flow of two immiscible fluids in porous media}

\author{Armin Shahmardi\fnref{KTH}}
\author{Salar Zamani Salimi\fnref{NTNU}}

\fntext[KTH] {FLOW, Department of Engineering Mchanics, Royal Institute of Technology (KTH), S-100 44 Stockholm, Sweden}

\author{Outi  Tammisola\fnref{KTH}}

\author{Luca Brandt\fnref{KTH,NTNU,Torino}}
\fntext[NTNU] {Department of Energy and Process Engineering, Norwegian University of Science and Technology (NTNU), Trondheim, Norway}
\fntext[Torino] {Department of Environmental, Land and Infrastructure Engineering, Politecnico di Torino, 10129 Turin, Italy}

\author{Marco Edoardo Rosti\fnref{Okinawa}}
\fntext[Okinawa]{ Complex Fluids and Flows Unit, Okinawa Institute of Science and Technology Graduate University, 1919-1 Tancha, Onna-son, Okinawa 904-0495, Japan}
\address{$^1$ FLOW, Department of Engineering Mchanics, Royal Institute of Technology (KTH), S-100 44 Stockholm, Sweden}

\address{$^2$ Department of Energy and Process Engineering, Norwegian University of Science and Technology (NTNU), Trondheim, Norway}
\address{$^3$ Department of Environmental, Land and Infrastructure Engineering, Politecnico di Torino, 10129 Turin, Italy}
\address{$^4$ Complex Fluids and Flows Unit, Okinawa Institute of Science and Technology Graduate University, 1919-1 Tancha, Onna-son, Okinawa 904-0495, Japan}

\author[arminsh@mech.kth.se]{arminsh@mech.kth.se}

\begin{abstract}
We study the role of the capillary number, $Ca$ and of the surface wettability on the dynamics of the interface between an invading and a defending phase in a porous medium by means of numerical simulations. We employ a hybrid phase field-immersed boundary approach to successfully model the contact line dynamics over the solid objects. \sz{Using a phase-field method which naturally incorporates dynamic wetting we eliminate the need for empirical contact line models to address contact line singularity.}
We map the \sz{two} dominant modes governing the motion of the interface, namely, capillary fingering, and stable penetration, in the ($Ca$-$\theta$) plane, with $\theta$ the static contact angle prescribed at the solid pores. Capillary fingering dominates at lower values of $Ca$ and pores hydrophobic to the invading phase, while a stable penetration is observed on hydrophillic surfaces. \sz{We present new measurements and analyses, including curvature probability density functions (pdf) and average curvature}. We also show that the pressure needed for the invading phase to advance at constant flow rate decreases with the capillary number, and increases with the contact angle at the capillary numbers considered. The latter is due to a significant  increase of the length of the interface in the case of capillary fingering. Finally, we show that it is possible to identify the different interfacial modes by measuring the penetration length and velocity during the medium filling.

\end{abstract}

\begin{keyword}
\texttt{Porous Media}\sep \texttt{Phase Field Model}, \texttt{Immersed Boundary Method} \sep \texttt{Contact-line dynamic}
\MSC[2010] 00-01\sep  99-00
\end{keyword}

\end{frontmatter}


\section{Introduction}\label{sec:introduction}
Fluid flow in porous media is of great importance in nature as well as in many industries such as secondary oil recovery \cite{10.2118/21621-PA}, metal infiltration \cite{LEGER201557}, subsurface fluid contaminant transport \cite{PhysRevE.70.016303}, paper and textile industry \cite{LU20071248}, transport in biological tissues \cite{KHALED20034989}. Numerous theoretical studies \cite{PhysRevA.46.7680}, numerical simulations \cite{https://doi.org/10.1029/2003JB002504}, and experiments \cite{PhysRevA.45.R8313} have been performed to understand the complex dynamics and the physics underlying the displacement of different fluids, from Newtonian to non-Newtonian \cite{DEVITA201810,Rosti2020}, immiscible and miscible \cite{PhysRevFluids.2.103902}, in different flow regimes, viscous dominated \cite{Homsy} or capillary dominated \cite{doi:10.1146/annurev-fluid-010518-040342} and for different porous media arrangements. 

During the last decade, the flow of two immiscible fluids within a porous medium has attracted even more attention due to a growing range of applications in industry such as $CO_2$ sequestration and subsurface water flow \cite{PERAZZO2018305}. Despite numerous experiments and numerical simulations on the flow of immiscible fluids through porous media, many details of the displacement dynamics are not fully clear, both as concerns the behavior at the pore scale and at the macroscopic scale. 
This difficulty arises because of the numerous parameters affecting the dynamics of the fluid displacement such as geometry of the medium, the porosity and permeability, the viscosity and density ratios between the two fluids, heterogeneities, surface wettability, etc.  \sz{McClure et al.~\cite{McClure_2016} investigate the dynamics of fluid-fluid interfaces and common curves in two-fluid-phase flow within porous media using a combined level-set and lattice-Boltzmann method. The study demonstrates accurate simulation of interface and common curve velocities, even in the presence of spurious currents. Fakhari and Bolster \cite{FAKHARI2017620} introduce an efficient lattice Boltzmann method for simulating immiscible multiphase flows with large density and viscosity ratios. The proposed model, which employs a diffuse-interface phase-field approach, accurately specifies three-phase contact angles on curved boundaries. Zhao et al.~\cite{Zhao_Chris} provides an objective comparison of various state-of-the-art pore-scale models for multiphase flow in porous media, using a dataset from recent microfluidic experiments as a benchmark. The findings highlight that no single model excels across all conditions, with significant challenges identified in accurately simulating thin films and corner flow, crucial for predicting macroscopic flow patterns in natural and industrial processes.}

Our current understanding on the role of these different parameters, as emerged from several previous studies, is shortly reviewed here, focusing on the works more closely related to the present one. In particular,  the pioneering work by \mr{Lenormand \cite{LENORMAND1989}} employs
experiments and numerical simulations and introduces three main categories of displacement for a non-wetting invading phase within a two dimensional porous medium, namely, viscous fingering, capillary fingering, and stable penetration. \sz{Viscous fingers are identified as thin elongated regions, whereas capillary
fingers are characterised by a network of thick interconnected flow patterns, the
stable penetration being characterised by flat fronts}, see also \cite{AFKHAMI20095370}.
\sz{ \mr{Holtzman and Segre \cite{HOLTZMAN}} found that increased wettability stabilizes the immiscible displacement of viscous fluids in disordered porous media by promoting cooperative pore filling, although this effect diminishes with higher flow rates due to viscous instabilities. The impact of disorder and wettability on quasi-static immiscible displacement in porous media was investigated by \mr{Hu et al.~\cite{Hu_Lan_Wei_Chen_2019}}, and a phase diagram was proposed to show how increased disorder affects displacement patterns depending on the contact angle. Later, \mr{Lan et al.~\cite{LAN-TIAN}} use a theoretical model to reveal that critical contact angles for the transition from compact displacement to capillary fingering decrease with capillary number, improving understanding of fluid invasion patterns in porous media.} Finally,  in a more recent study, \mr{Primkulov et al.~\cite{primkulov_pahlavan_fu_zhao_macminn_juanes_2021}}  provide a complete map of the dominant flow regimes by meticulously employing a moving capacitor-pore network model.
 

If the interfacial tension overcomes the viscous stresses, the invading phase penetrates the defending fluid by forming capillary fingers or through a stable displacement. \cite{PhysRevLett.60.2042, PhysRevB.41.11508} suggested that the local instabilities at the fluid front (the interface of the two immiscible fluids) and the wettability of the pore surface dictate whether the invading phase advances inside the defending phase by forming capillary fingers or through stable penetration. In general, these interfacial instabilities can be divided into three main categories, namely burst, touch, and overlap \cite{doi:10.1146/annurev-fluid-010518-040342}. The burst mode \cite[introduced by][and hence also called Haines jump]{haines_1930}  is seen when the invading phase jumps to an adjacent pore due to a decrease of the interface curvature and as a consequence of a local drop in capillary pressure. The touch mode is very similar to the burst mode, except for the interface accelerating towards the onward pore (instead of bursting onto that) so to fulfil the wettability requirements. Finally,  at least two neighbour menisci sharing a pore are required in the overlap mode, as the interface displaces by the coalescence of neighboring menisci. In the literature, burst and the touch modes are usually referred to as non-cooperative instability modes, whereas the overlap mode is denoted as the cooperative mode.

The numerical simulations and experiments in \cite{PhysRevFluids.1.074202} indicate that the crossover between the interface stable penetration and the capillary fingers is determined by the wettability of the pore solid surface. 
In the case of an invading phase flowing through a hydrophobic porous medium (drainage), the interface undergoes instabilities in the form of capillary fingers. The displacement of the meniscus between the two pores  is independent on the advancement of the other neighboring menisci due to the non-cooperative nature of these interfacial instabilities.
\sz{However, in the case of capillary fingering, the menisci can retract as neighboring pores experience Haines jumps.}
Conversely, when the invading phase penetrates a hydrophilic medium (imbibition), the displacements of neighboring menisci sharing a pore are entangled through cooperative interfacial instabilities. Nevertheless, in the absence of strong surface tension forces, viscous fingers form for high injection rates or when the viscosity of the invading phase is less than that of the defending phase \cite{https://doi.org/10.1029/2020WR028149}. Viscous fingering is beneficial in the many industrial applications requiring fluid mixing  such as oil recovery and microfluidic lab-on-a-chip devices \cite{https://doi.org/10.1002/2014WR015811}.

The goal of this study is twofold. Firstly, although previous studies have addressed the features of the dominant displacement modes of two immiscible fluids in a porous medium (viscous fingering, capillary fingering and stable penetration of the interface), discussing the different types of instability  of capillary driven flows (burst, touch, and overlap), an 
in-depth analysis relating the different displacement modes to quantitative measurements is missing. In this study, we therefore aim to present the signatures of the different displacement modes in terms of  quantitative measurements such as the pressure drop, length of the interface, interface curvature, etc. Secondly, we note that
 most of the previous studies have separately  investigated either the quasi steady state of the system (once the volume of the invading phase in the porous medium is saturated) or the transient properties (when the volume of the invading phase in the porous medium is increasing). However, in-depth knowledge of the two scenarios is crucial in several applications. Therefore, we have conducted  comprehensive analyses on both these scenarios,  i.e., the quasi steady and the transient dynamics. 
 In particular, we focus here on the effects of different capillary numbers and wettability conditions of the porous media at fixed viscosity ratio, density ratio, and pore arrangement.  

The manuscript is organised as follows. The mathematical model and the governing equations  are presented in section \ref{Sec:PFM}, whereas a brief explanation of the  numerical algorithm adopted is provided in section \ref{sec:Numercis}, before introducing the simulation setup and parameters in section \ref{Sec:Setup}. 
In section \ref{Sec:Results}, we discuss the results of the simulations in four subsections dedicated to flow visualisations, an overall map of the flow regimes (depending on the capillary number and surface wettability), results of the quasi steady analyses, and finally analysis  of the transient dynamics. Finally, in section \ref{sec:conclusion} we summarise the most important outcomes of this research and suggest some possible future extensions.

\par 
\sz{Our work offers several novel contributions, including a comprehensive study that encompasses various analyses of transient and quasi-steady states across a range of capillary numbers and wettability conditions. We present new measurements and analyses, such as curvature probability density functions (pdf) and average curvature. Notably, we introduce a numerical simulation using a diffuse interface method in a porous medium to eliminate the contact line singularity, thereby addressing and removing concerns associated with this issue.}

\section{Mathematical model and numerical solution}

\subsection{Governing equations} \label{Sec:PFM}
We solve the incompressible Navier-Stokes equations to model the fluid flow. The interface between the two immiscible fluids (the invading and the defending phases) is 
solved with  a free-energy based Phase-Field model, i.e., the so-called Cahn-Hilliard equation \cite[see][among others]{Carlson2012}. We employ a recently developed hybrid phase-field immersed boundary method to impose the contact angle and slip-velocity boundary conditions at the solid-pore boundary \cite{SHAHMARDI2021110468}.

The two immiscible fluid phases are distinguished by an indicator function, the order parameter $C$, which is $1$ in the invading phase and $-1$ in the defending phase, and changes rapidly but smoothly within a finite interface thickness. Hence, the interface of the two phases can be considered as the zero level of the order parameter, i.e.,~$C =0$.

We solve the Cahn-Hilliard equation to obtain the evolution of the order parameter and the dynamics of the interface\cite{john1961spinodal,cahn1958free}:
\begin{linenomath}\begin{equation} \label{eq:Cahn-Hilliard}
\begin{aligned}
\begin{gathered}
\frac{\partial C}{\partial t} + u_i \frac{\partial C}{\partial x_i} = \frac{3}{2 \sqrt 2} \frac{\partial }{\partial x_i}  \left(   M  \frac{\partial \phi}{\partial x_i}  \right) ,\\
\end{gathered}
\end{aligned}
\end{equation}\end{linenomath}
where $M$ is the mobility coefficient (which is assumed to be a constant in this study), $u_i$ is the fluid velocity vector, and $\phi$ is the chemical potential of the system defined as
\begin{linenomath}\begin{equation} \label{Chemical-Potential}
 \begin{aligned}
\begin{gathered}
 \phi =  \frac{\sigma}{\epsilon} {\psi}'(C) -\sigma \epsilon  \frac{\partial }{\partial x_i}  \left( \frac{\partial C}{\partial x_i} \right) .
 \end{gathered}
\end{aligned}
\end{equation}\end{linenomath}
In equation (\ref{Chemical-Potential}), $\sigma$ is the surface tension coefficient, $\epsilon$ is the thickness of the interface, $\psi={(C^2-1)}^2/4$ a double-well function with two minima, each corresponding to one of the stable phases \cite{jacqmin_2000,Carlson2012,SHAHMARDI2021110468}.  
The conservation of momentum (Navier-Stokes) and continuity equations read:
\begin{linenomath}\begin{equation} \label{eq:Navier-Stokes}
\begin{gathered}
\frac{\partial (\rho u_i)}{\partial t} + \frac{\partial}{\partial x_j}(\rho u_i u_j) = -\frac{\partial P}{\partial x_i} + \frac{\partial}{\partial x_j} \left( \mu ( \frac{\partial u_i}{\partial x_j}+ \frac{\partial u_j}{\partial x_i})\right) + f_{s_i} + f_{I_i} ,\\
\frac{\partial u_i}{\partial x_i} = 0 ,\\
\end{gathered}
\end{equation}\end{linenomath}
where $\rho$ and $\mu$ are the density and dynamic viscosity of the fluid, which in general vary from $\rho_1$ and $\mu_1$ in one phase to $\rho_2$ and $\mu_2$ in the other phase. The local values are obtained as a weighted average of the values in the two phases:
\begin{linenomath}\begin{equation} \label{eq:properties interpolation}
\begin{gathered}
\rho = \left[(C+1)\rho_2-(C-1)\rho_1\right]/2,\\
\mu = \left[(C+1)\mu_2-(C-1)\mu_1\right]/2.
\end{gathered}
\end{equation}\end{linenomath}
In equation (\ref{eq:Navier-Stokes}), $P$ is the pressure,  $f_{s_i}$ represents the surface tension force at the interface,  and $f_{I_i}$ is the immersed boundary force (needed to model the presence of the solid boundaries). The surface tension force at the interface $f_{s_i}$ is computed as the product of the chemical potential and the gradient of the order parameter, i.e., $\phi \frac{\partial C}{\partial x_i} $ \cite{jacqmin_2000}.

At the triple point, we prescribe as boundary conditions 
 a value for the contact angle, a slip velocity parallel to the solid body, and the impermeability condition (no mass penetration inside the pores), which read
\begin{linenomath}\begin{equation} \label{WallBCS}  
\begin{gathered}
  \mu_f \epsilon \left( \frac{\partial C}{\partial t}+ u_i \frac{\partial C}{\partial x_i} \right)= \sigma \epsilon \frac{3}{2 \sqrt 2}\frac{\partial C}{\partial x_i}n_i + \sigma cos(\theta) g'(C),\\
  \frac{\mu}{l_{s}}u_{j_{s}}t_j= \mu \frac{\partial (u_j  t_j)}{\partial (x_in_i)}- \left[ \frac{3}{2 \sqrt 2}\frac{\partial C}{\partial x_i}n_i + \sigma cos(\theta) g'(C)\right]\frac{\partial C}{\partial x_j}t_j ,\\
  \frac{\partial \phi}{\partial x_i}n_i =0,
   \end{gathered}
\end{equation}\end{linenomath}
where $\mu_f$ is the contact line friction coefficient\cite{Carlson2012}, $\theta$ is the equilibrium contact angle, and $n_i$ represents the vector normal to the solid surface. The first boundary condition in equation (\ref{WallBCS}) models the dynamics of the contact line motion, where the contact line friction coefficient is inversely proportional to the time needed by the contact line to relax to its prescribed contact angle \cite{Xu2018}. Here, $g(C)=({2}+{3}{C}-{C}^3)/4 $ is a smoothing function varying between zero and 1 from one stable phase to the other. The second equation models the slip velocity at the solid surface where $u_{s}$, $l_{s}$, and $t_j $ are the slip velocity, slip length, and the unit vector tangent to the surface.  Finally, the third boundary condition in equation (\ref{WallBCS}) guarantees impermeability at the wall. To correctly impose the above-mentioned boundary conditions, we employ the hybrid phase field-immersed boundary method introduced in our previous work; for the details of the method, we refer the interested readers to \cite{SHAHMARDI2021110468}.

\subsection{Numerical method} \label{sec:Numercis}

We solve the previous system of equations on a Cartesian mesh with a staggered arrangement, where the velocity components are defined at the cell faces and the pressure, the chemical potential and the order parameter are defined at the cell centres. The second-order finite-volume scheme is used for spatial discretisation and the different terms are advanced in time explicitly using the second order Adams-Bashforth scheme. Finally, the fractional-step method for incompressible two-fluid systems is implemented as proposed by \cite{DODD2014416}. The baseline solver has been extensively validated in previous works \cite[see among others][]{rosti_brandt_2017, shahmardi_zade_ardekani_poole_lundell_rosti_brandt_2019,devita,PhysRevFluids.5.041301, SHAHMARDI2021110468}.

\begin{figure}
\begin{center}
\includegraphics[width=0.9\textwidth]{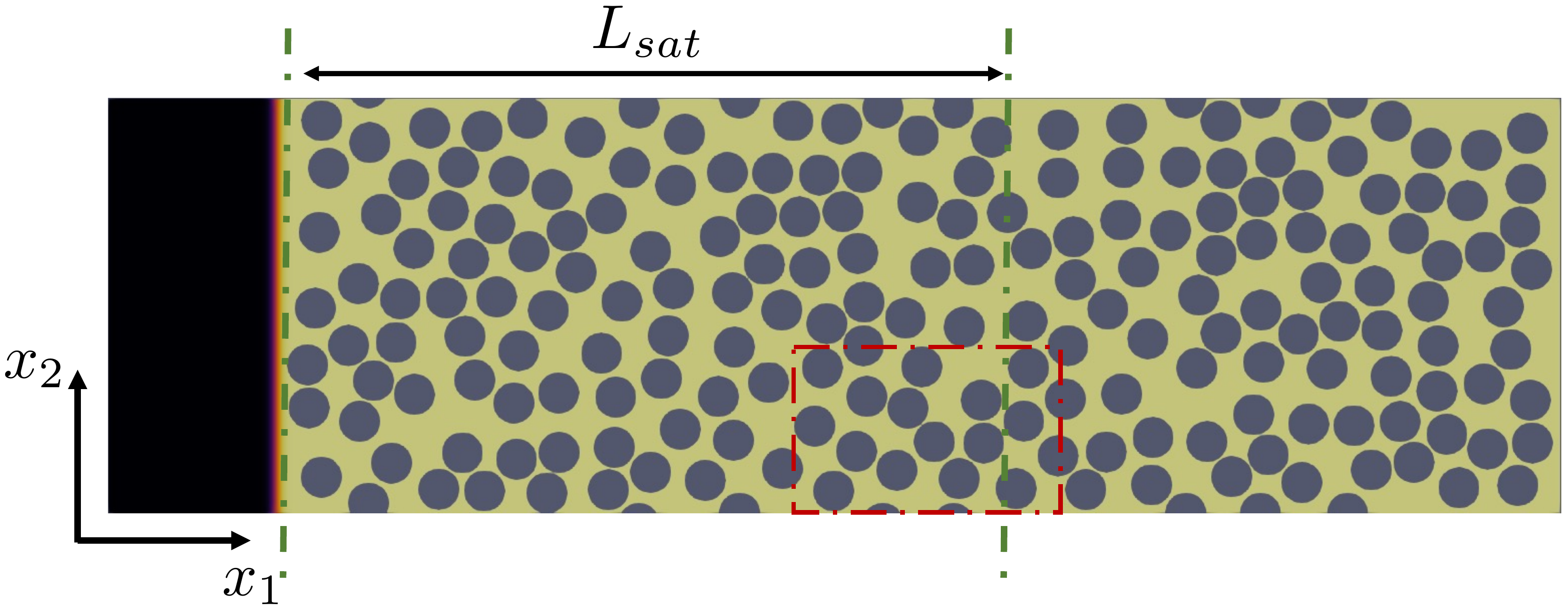}
\caption{Schematic of the simulation setup and coordinate system: the continuous injection of an invading fluid (black) over a random distribution of solid cylinders (grey) with the defending phase (yellow) initially at rest. The green dashed lines represent the subdomain, with length $L_{sat}$, considered for the quasi-steady (saturated)  analysis, whereas the red dashed box indicates the part of the domain where we present the occurrence of successive jumps.}
\label{fig:Setup}
\end{center}
\end{figure}

\section{Problem description} \label{Sec:Setup}
The porous medium considered for this study, shown in figure \ref{fig:Setup}, consists of a random arrangement of circular solid objects (displayed in grey), all of the same size. 
A defending phase (represented in yellow) is opposing the invading phase (illustrated in black). The computational domain is $[0,70R] \times [0,20R]$ where $R$ is the radius of each solid cylinder. The governing equations are discretised on a numerical mesh consisting of $1024 \times 3584$ grid points. 
We start the simulations with a layer of the invading phase (initially at rest) at the inlet of the simulation domain. Considering the thickness of the inlet layer depleted of cylinders ($\approx 8R$), the radius of each cylinder, and the total number of cylinders $(192)$, the porosity of the designed  medium is about $ 50\%$. Inlet-outlet boundary conditions are imposed in the streamwise direction ($x_1$), while periodicity is forced in the cross-stream direction ($x_2$). To drive the invading phase towards the porous medium, a constant mass flux 
of the invading phase is injected from the left boundary (i.e.\ constant velocity in direction $x_1$ at the inlet). 
 We specify the fluid properties and flow regime by four relevant non-dimensional numbers, namely, the Reynolds number $Re$, the capillary number $Ca$,  the P\'eclet  number $Pe$, and the Cahn number $Cn$, defined as follows:
\begin{linenomath}\begin{equation} \label{NonDimensionalParam}  
\begin{gathered}
  Re = \frac{\rho_{ref} U_{ref} R}{\mu_{ref}}, \quad Ca = \frac{\mu_{ref} U_{ref}}{\sigma}, \quad  Pe=\frac{2\sqrt{2}U_{ref}R\epsilon}{3M\sigma}, \quad Cn = \frac{\epsilon}{R}.
   \end{gathered}
\end{equation}\end{linenomath}
Note that, in this study we consider same density and same viscosity for the invading and the defending phases. Thus, in equation \ref{NonDimensionalParam}, $\rho_{ref}$ and $\mu_{ref}$ are the density and the dynamic viscosity of both phases. $U_{ref}$ is the constant inflow velocity,  and the radius of the solid cylinders ($R$) is assumed as the reference length. 
The Cahn number is set to $Cn=0.1$ and, to minimise the effects of flow inertia, we chose $Re<1$. The focus of the work is on the role of surface tension and wettability on the flow in the porous medium. Therefore, while we assumed a constant mobility coefficient for all the cases by setting $Pe=10^8$, 
simulations are  performed for \sz{three} different values of capillary number ($10^{-2}$, $10^{-3}$, $10^{-4}$) and for five values of contact angle ($\theta$ = $45^\circ$, $67.5^\circ$, $90^\circ$, $112.5^\circ$, $135^\circ$) defined as the angle between the solid substrate and the contact line inside the invading phase. \sz{Note that, here and in the following we use the terms hydrophobic and hydrophilic for simplicity, although our analysis is applicable to any generic Newtonian fluid.}
Note that all the performed analyses are based on the selected parameters and any other choice of parameters (for instance mobility coefficient) can affect the results.

A static contact angle boundary condition ($\mu_f=0$) and a small slip velocity at the solid surface ($l_s=0.1R$) are imposed for all the cases. It is worth to mention that to the best of our knowledge, 
no measures are taken to address the well-known contact line singularity
in the previous numerical simulations of the problem at hand. 
In general, one of the advantages of diffuse interface methods, such as the Phase Field Method, is to remove the singularities by the diffusion of the interface. However,  to avoid excessive diffusion in the narrow channels, we chose a relatively small mobility number. Thus, to assure that any possible singularity is successfully removed, we resort to another common approach and let the contact line slip on the wall with an imposed very small slip velocity.

\section{Results and discussions} \label{Sec:Results}
We will present our results in four different sections. First, in section \ref{Sec:Visualization}, we visualise the penetration of the invading fluid phase within the defending one for selected values of the contact angle ($\theta= 45^\circ, 90^\circ,$ and $135^\circ$) and the different capillary numbers under investigations at four different non-dimensional time instances. This would provide a first qualitative analysis of the interface advancement. Next, see section \ref{sec:map}, we will provide a map of the different flow regimes identified, function of capillary number and contact angle. In section \ref{Sec:Quasi-steady}, we will present a first analysis of the flow, focusing on the portion of the porous medium delimited by the green dashed lines in figure \ref{fig:Setup} (with length $L_{sat}= 24R$): here, the flow features eventually saturate and hence, the fluid flow is in a quasi-steady state. To conclude, we will examine the transient dynamics of the system, section \ref{Sec:Transient}, and study the evolution of different observables over the full length of the domain. 

\subsection{Fluid flow visualization} \label{Sec:Visualization}
\begin{figure} 
\begin{center}
\includegraphics[width=0.15\textwidth]{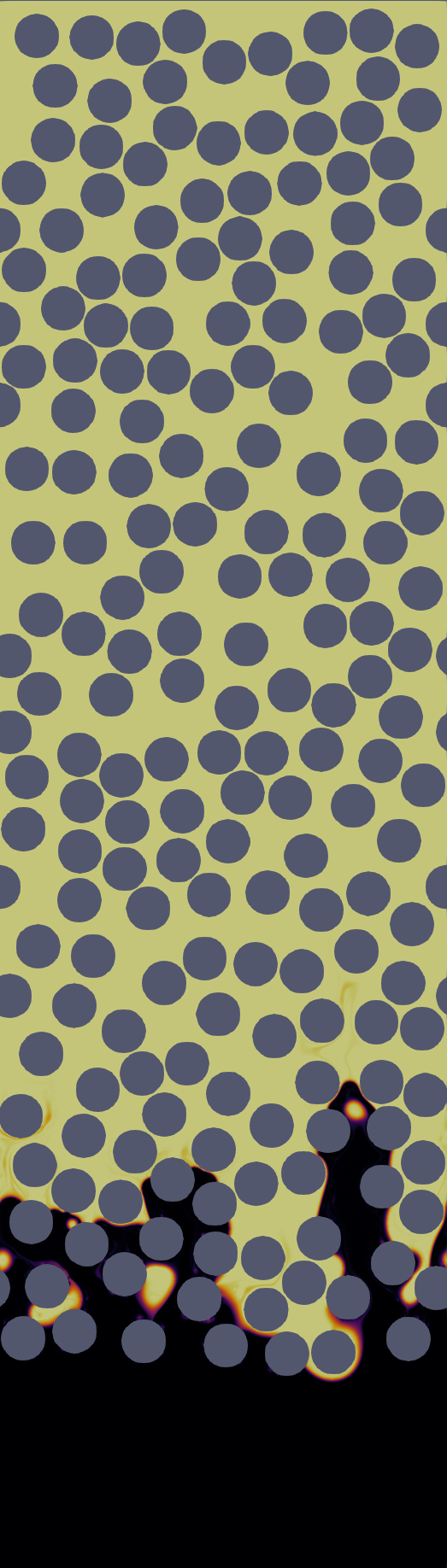}
\includegraphics[width=0.15\textwidth]{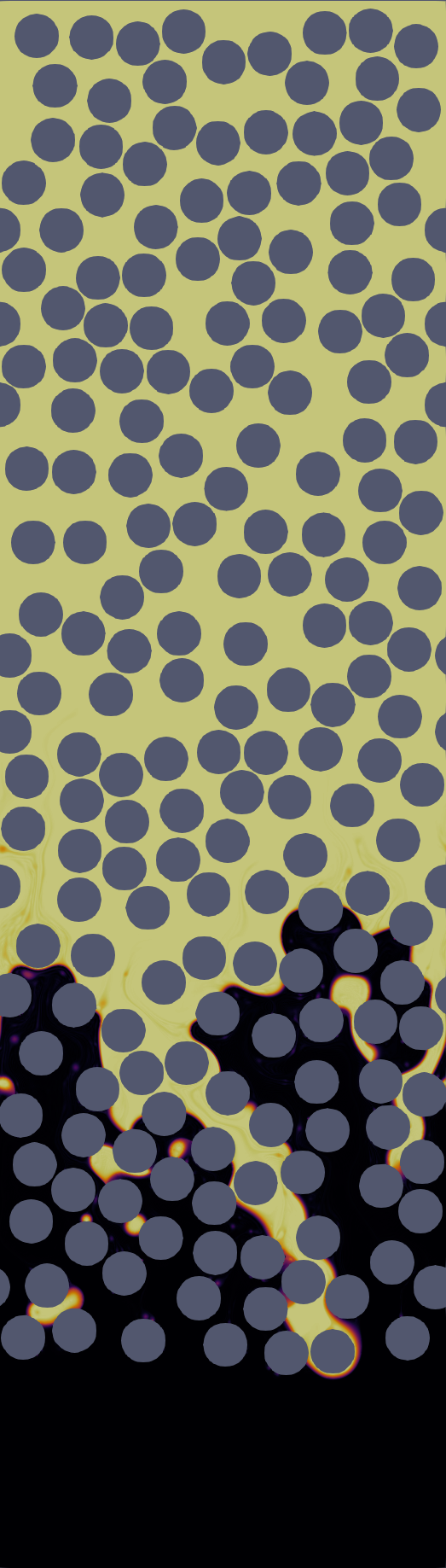}
\includegraphics[width=0.15\textwidth]{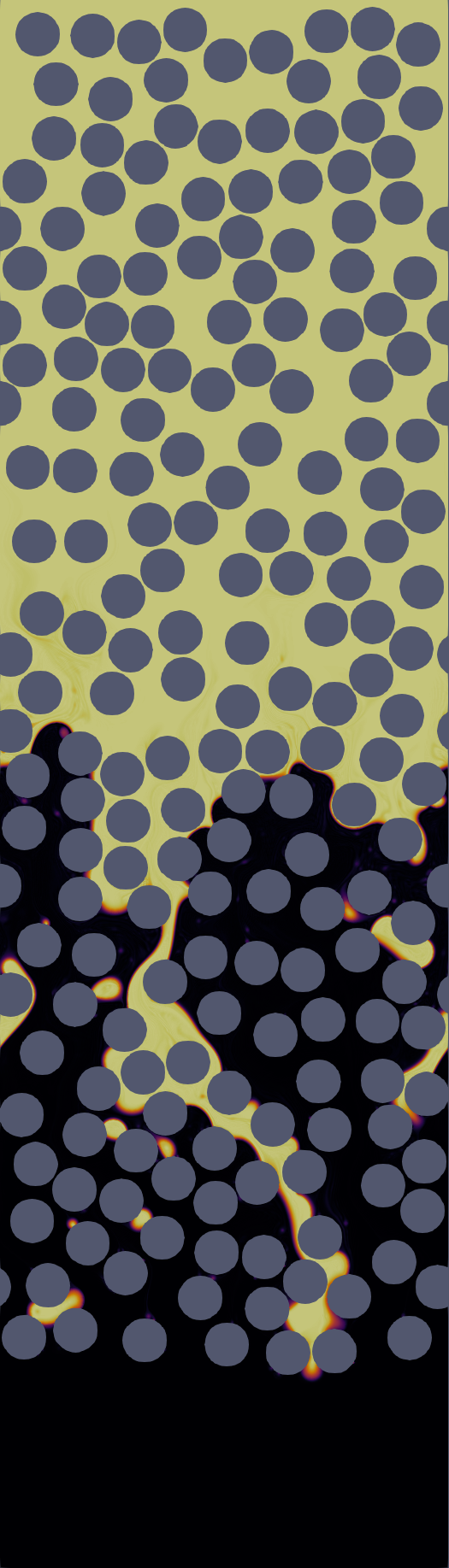}
\includegraphics[width=0.15\textwidth]{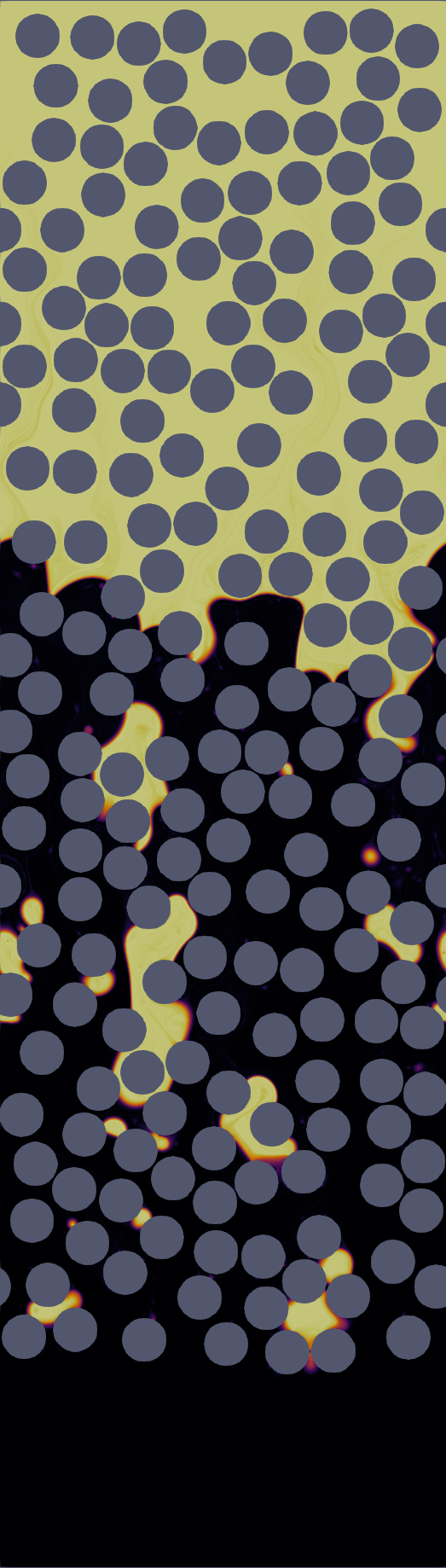}
\put(-230,207){\rotatebox{0}{\normalsize $t^*=4.19$}}
\put(-170,207){\rotatebox{0}{\normalsize $t^*=8.34$}}
\put(-112,207){\rotatebox{0}{\normalsize $t^*=12.57$}}
\put(-48,207){\rotatebox{0}{\normalsize $t^*=16.75$}}
\put(-255,85){\rotatebox{90}{\Large $\theta=45^\circ$}}\\

\includegraphics[width=0.15\textwidth]{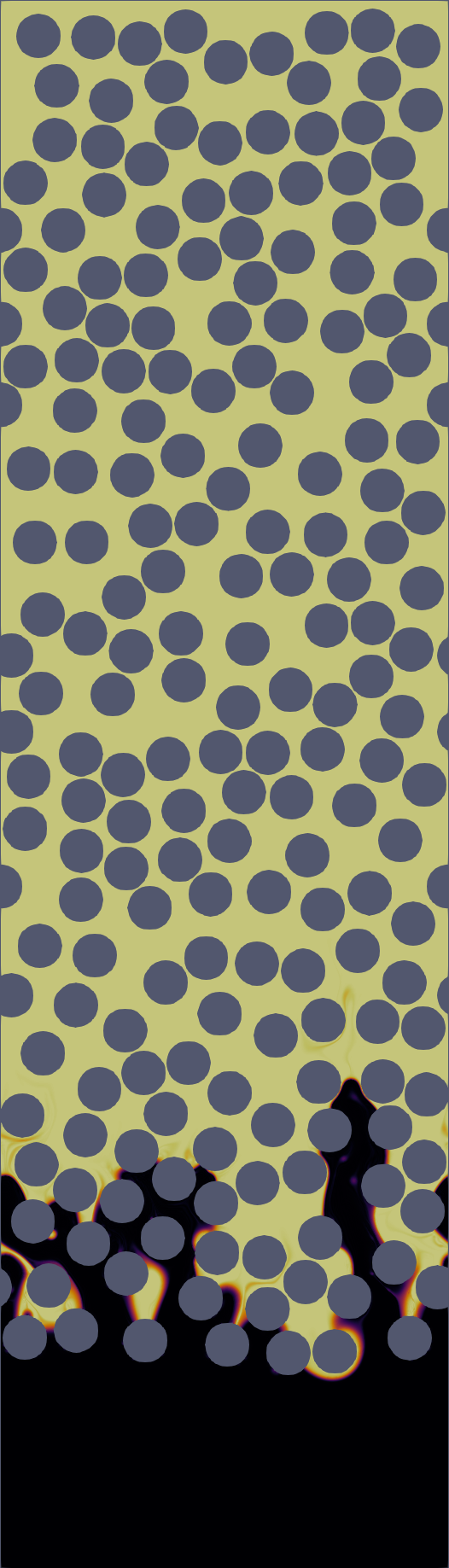}
\includegraphics[width=0.15\textwidth]{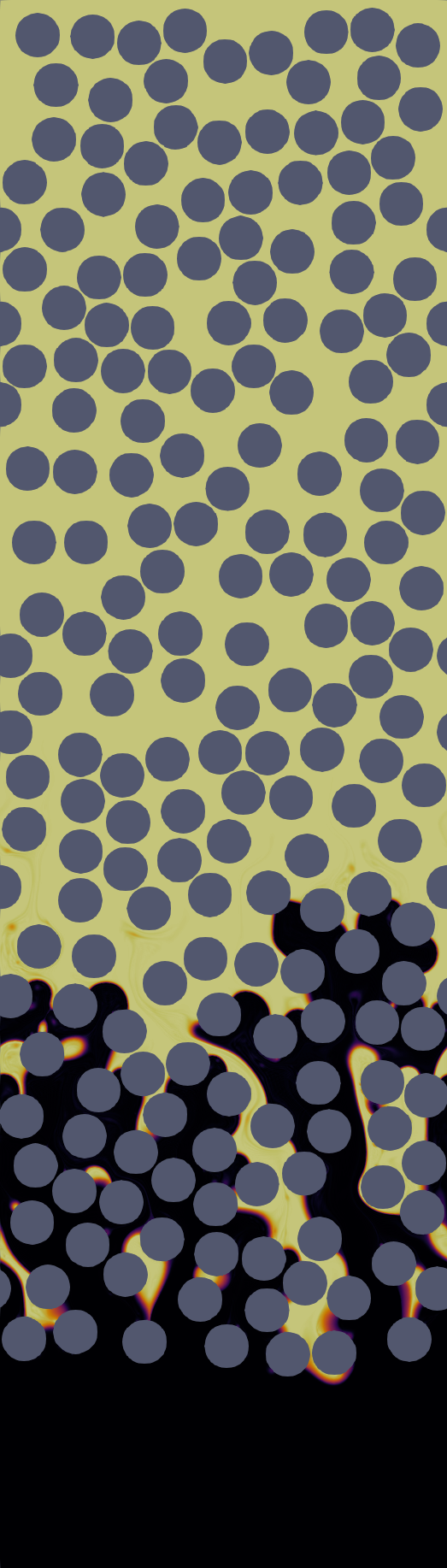}
\includegraphics[width=0.15\textwidth]{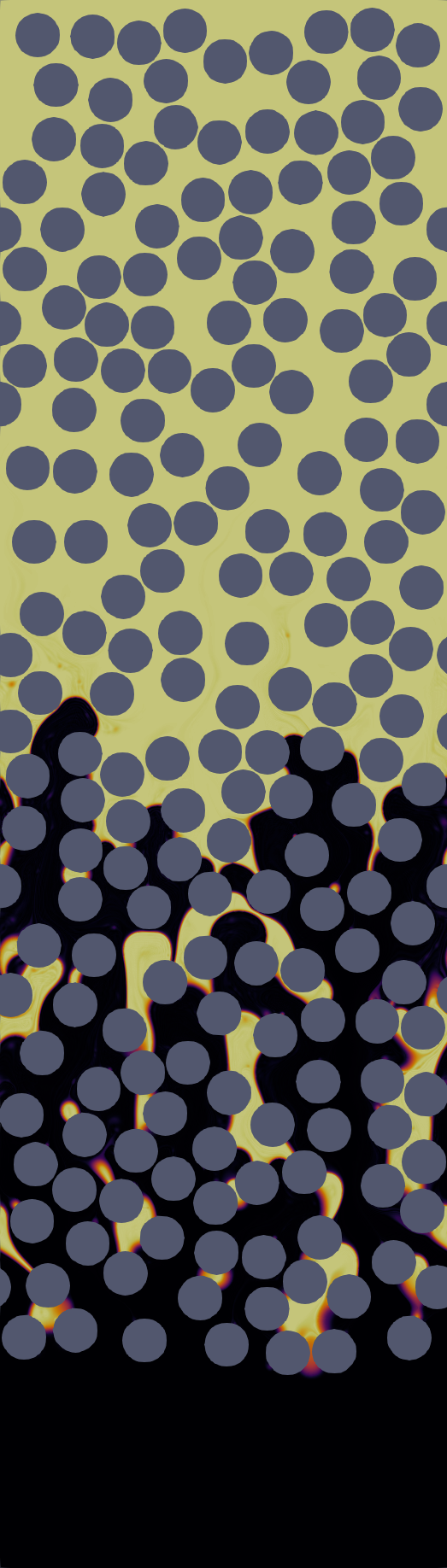}
\includegraphics[width=0.15\textwidth]{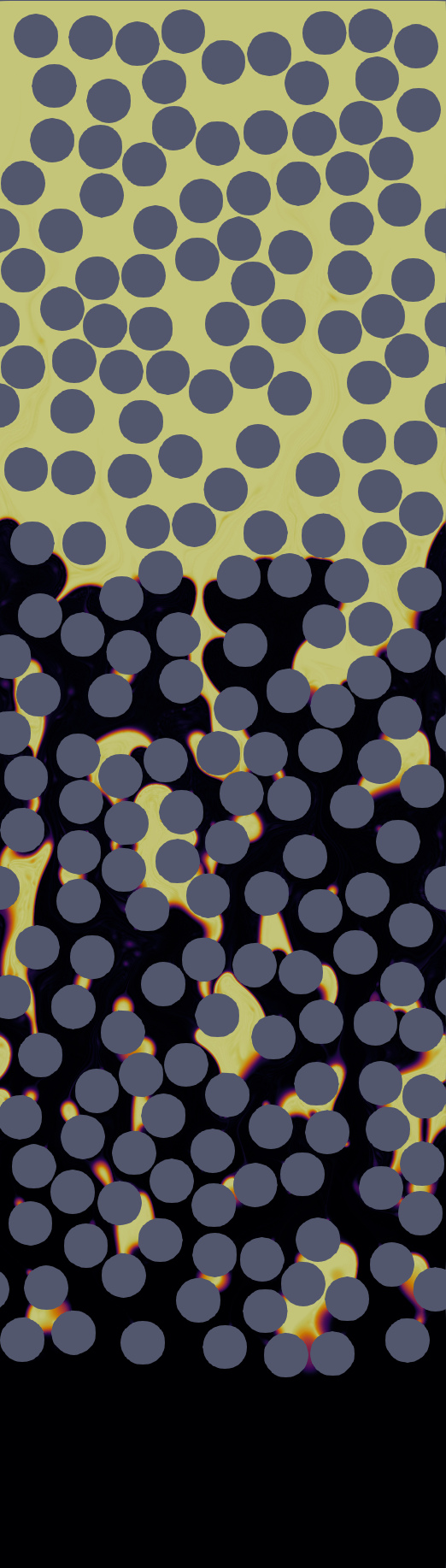}
\put(-255,85){\rotatebox{90}{\Large $\theta=90^\circ$}}\\

\includegraphics[width=0.15\textwidth]{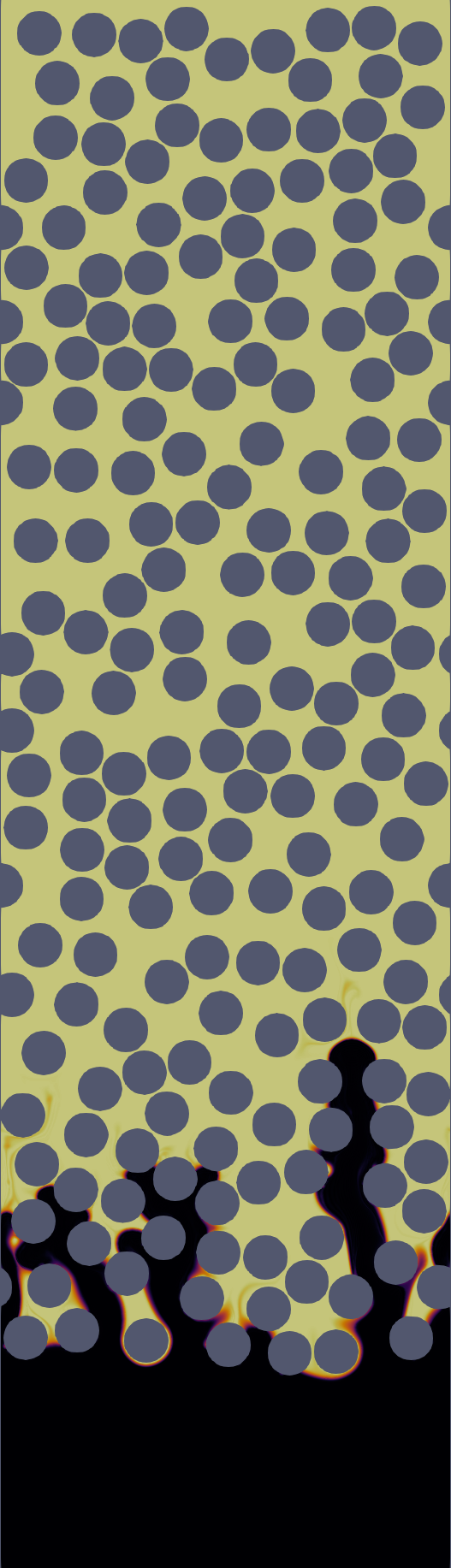}
\includegraphics[width=0.15\textwidth]{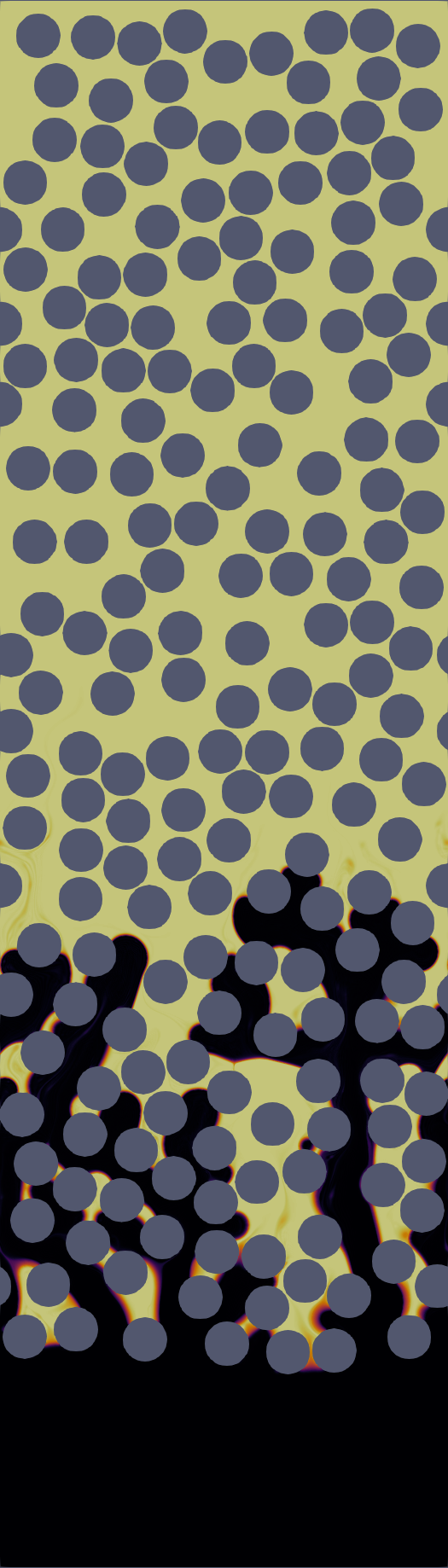}
\includegraphics[width=0.15\textwidth]{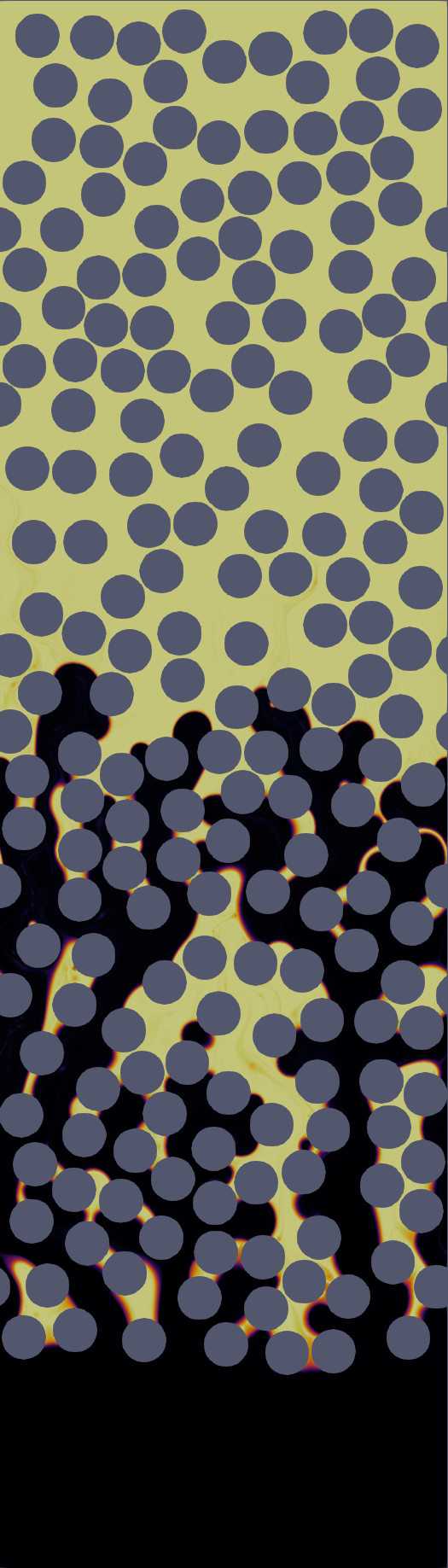}
\includegraphics[width=0.15\textwidth]{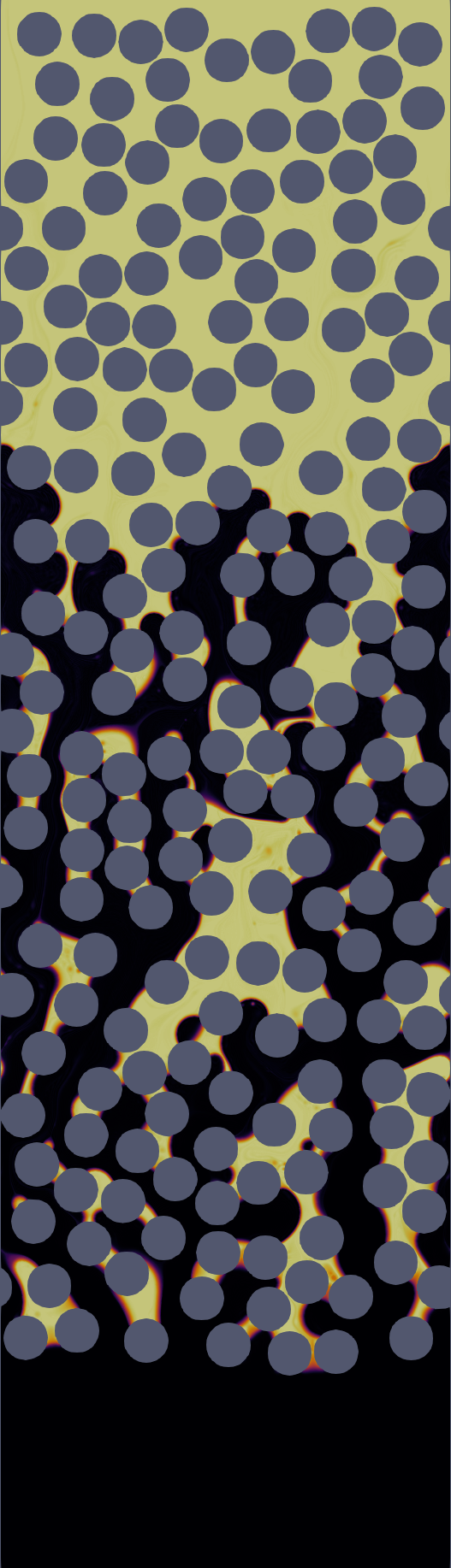}
\put(-255,85){\rotatebox{90}{\Large $\theta=135^\circ$}}\\
\caption{Snapshots of the two-fluid flow in the porous medium for contact angle $\theta=45^\circ$ (top panel), $\theta=90^\circ$ (middle panel) and $\theta=135^\circ$ (bottom panel) and capillary number $Ca=10^{-2}$.}
\label{fig:VisCa0.01}
\end{center}
\end{figure}

\begin{figure} 
\begin{center}
\includegraphics[width=0.15\textwidth]{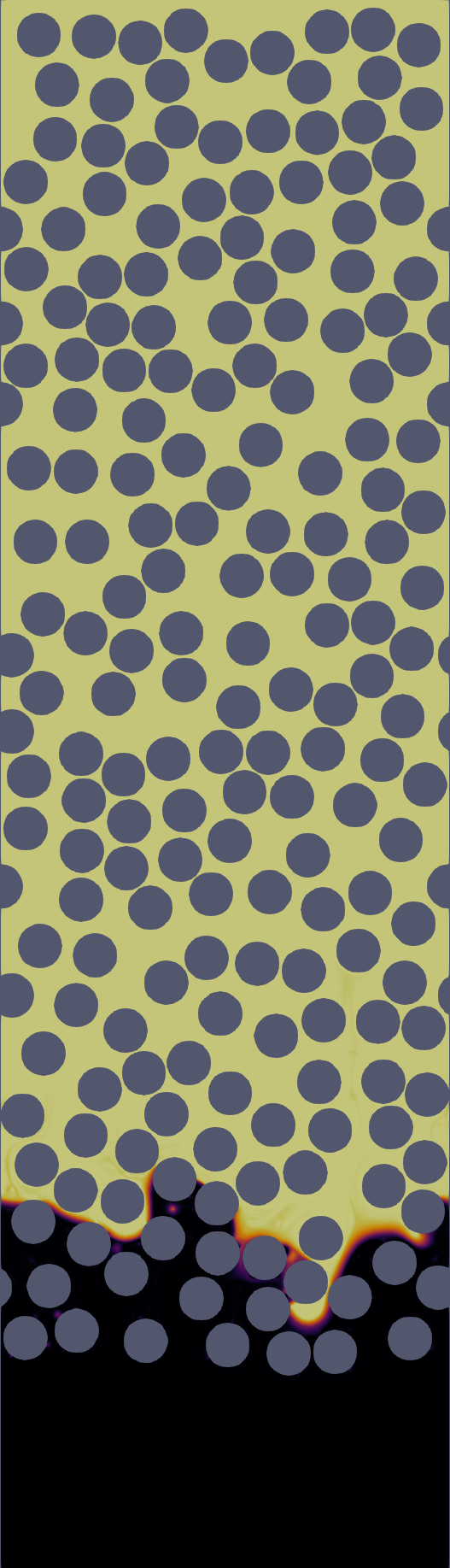}
\includegraphics[width=0.15\textwidth]{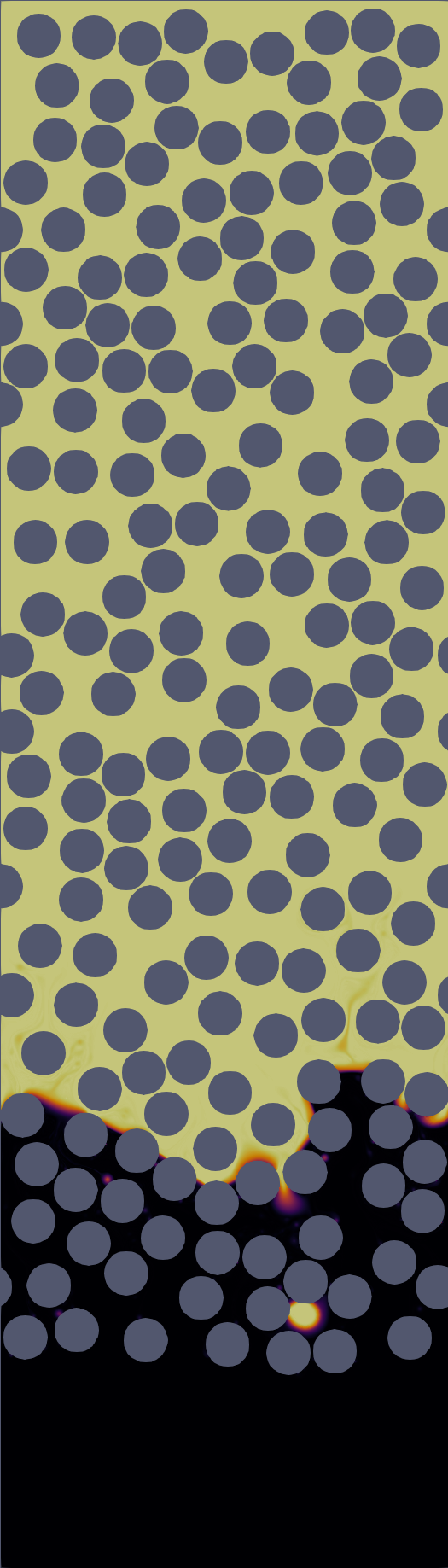}
\includegraphics[width=0.15\textwidth]{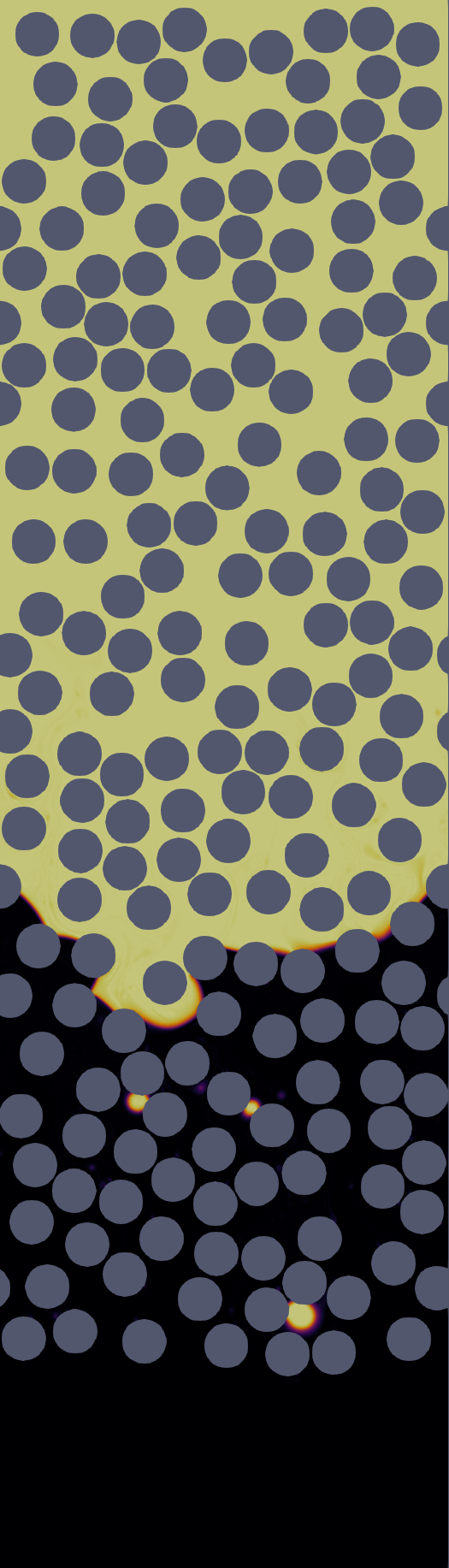}
\includegraphics[width=0.15\textwidth]{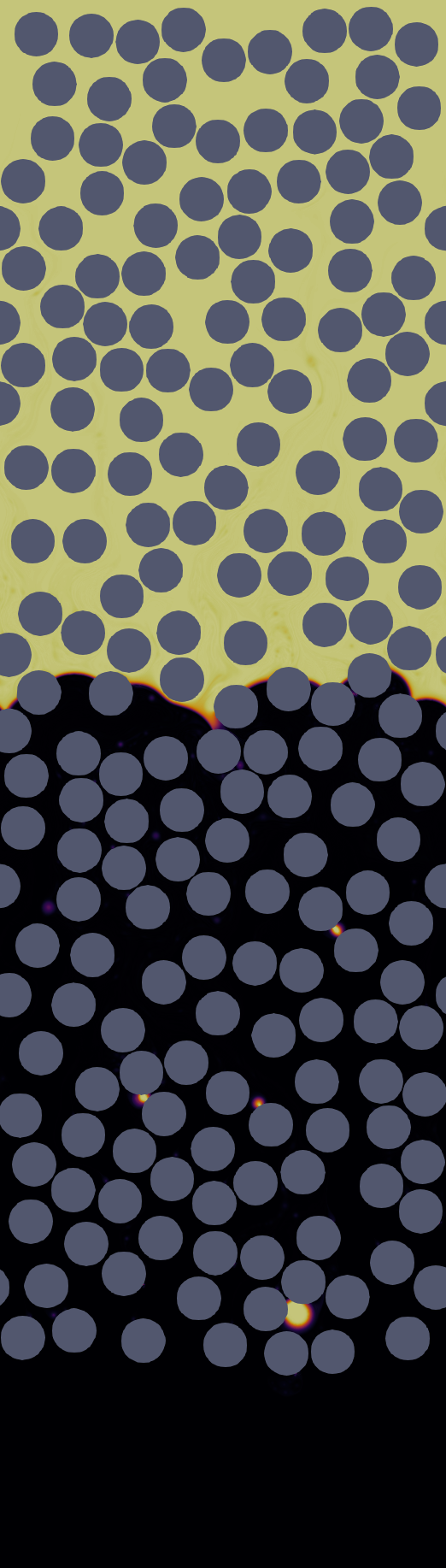}
\put(-230,207){\rotatebox{0}{\normalsize $t^*=4.19$}}
\put(-170,207){\rotatebox{0}{\normalsize $t^*=8.34$}}
\put(-112,207){\rotatebox{0}{\normalsize $t^*=12.57$}}
\put(-48,207){\rotatebox{0}{\normalsize $t^*=16.75$}}
\put(-255,85){\rotatebox{90}{\Large $\theta=45^\circ$}}\\

\includegraphics[width=0.15\textwidth]{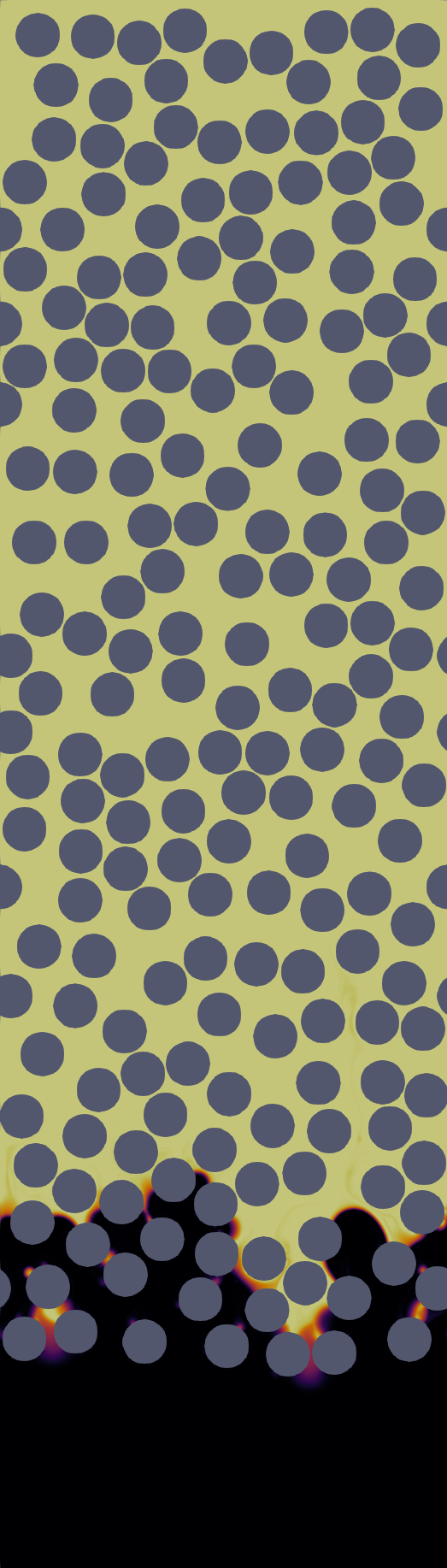}
\includegraphics[width=0.15\textwidth]{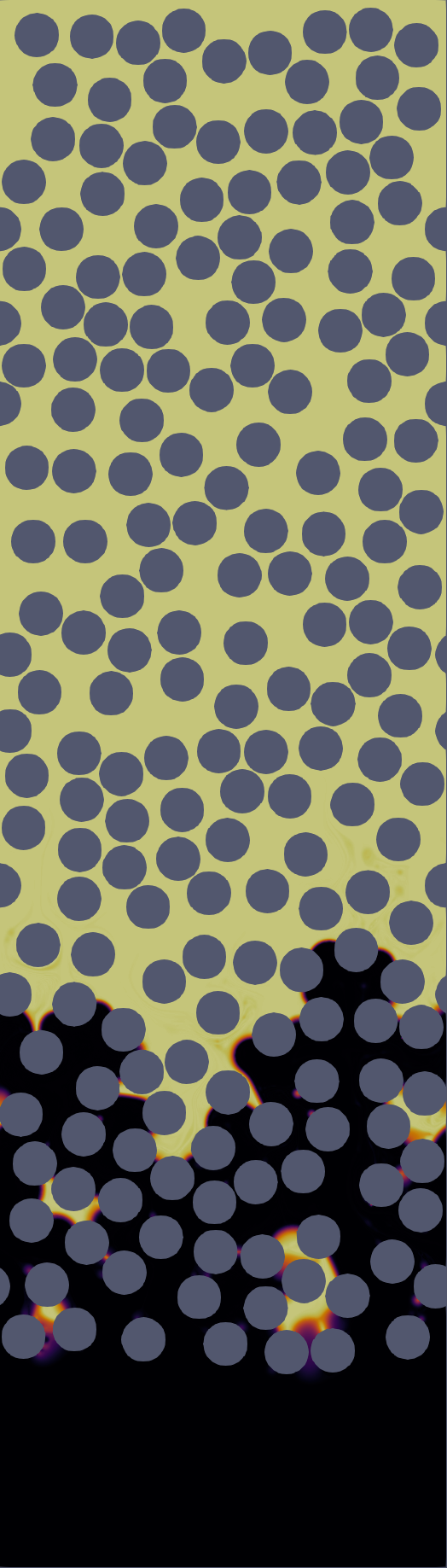}
\includegraphics[width=0.15\textwidth]{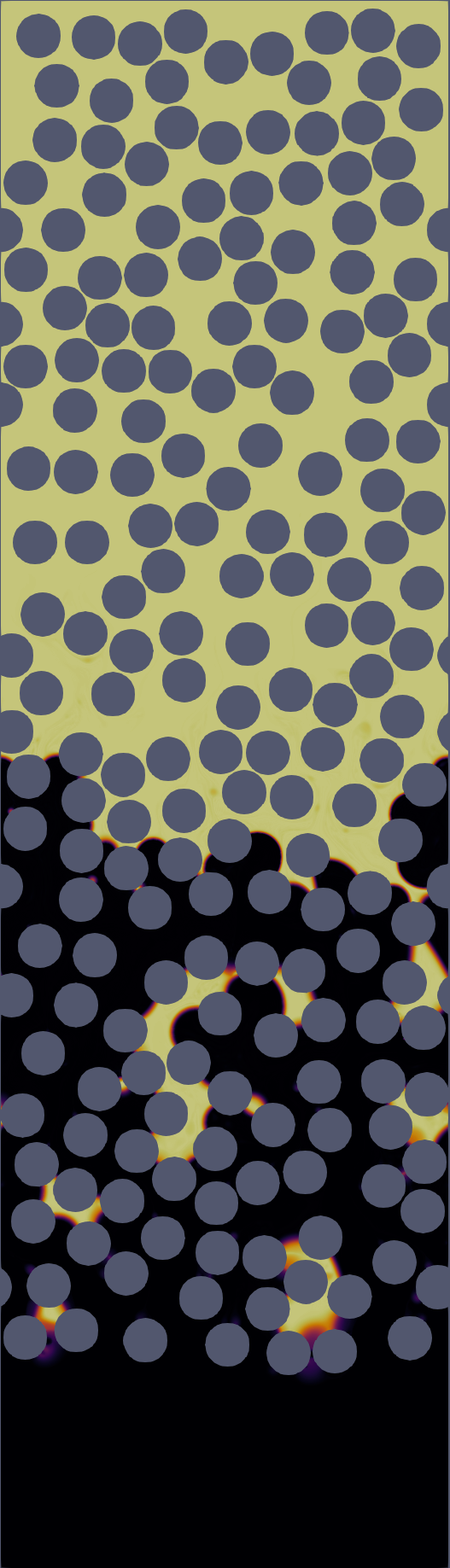}
\includegraphics[width=0.15\textwidth]{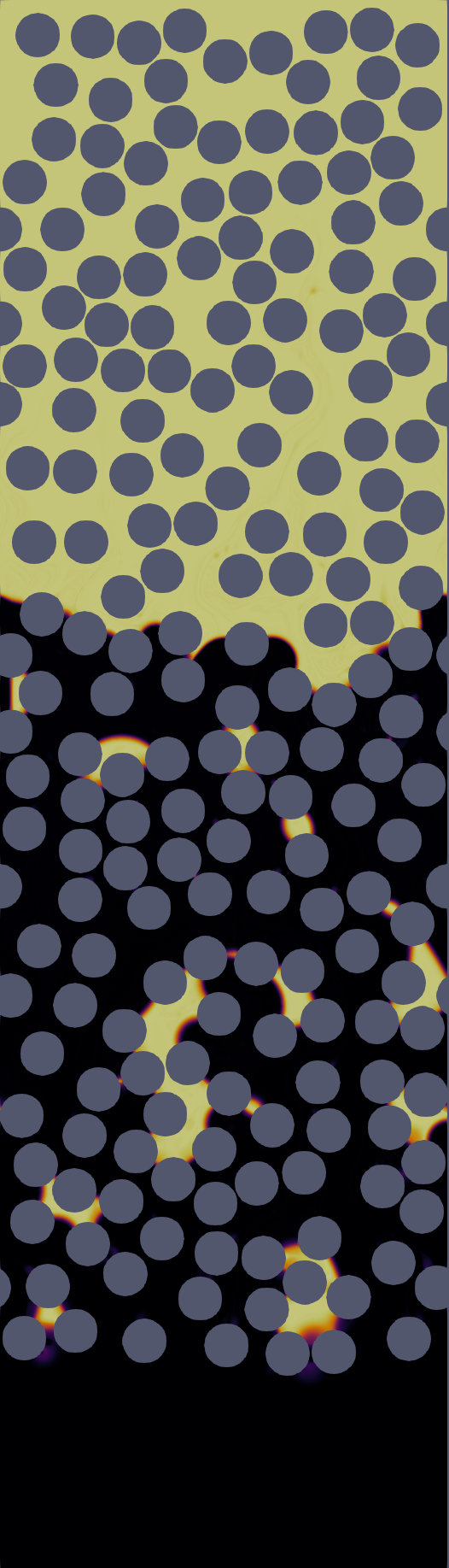}
\put(-255,85){\rotatebox{90}{\Large $\theta=90^\circ$}}\\

\includegraphics[width=0.15\textwidth]{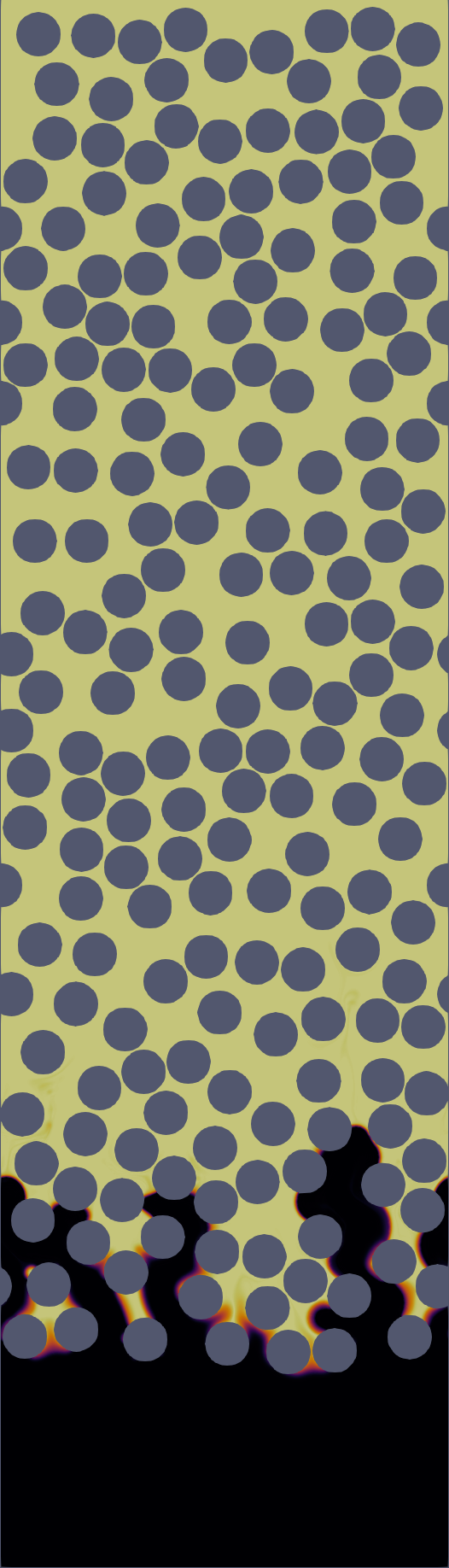}
\includegraphics[width=0.15\textwidth]{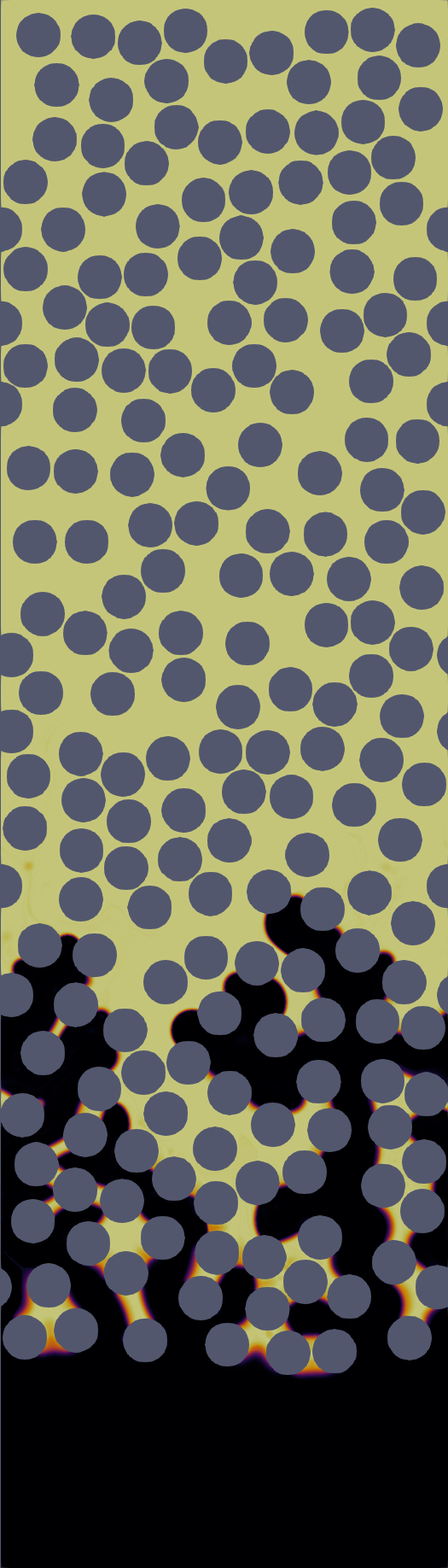}
\includegraphics[width=0.15\textwidth]{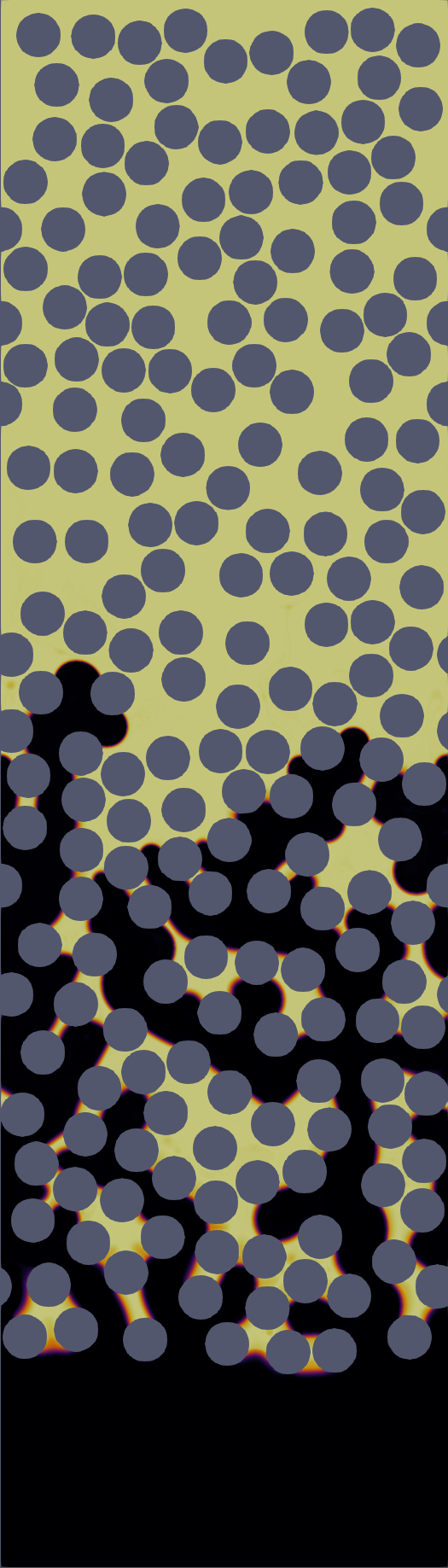}
\includegraphics[width=0.15\textwidth]{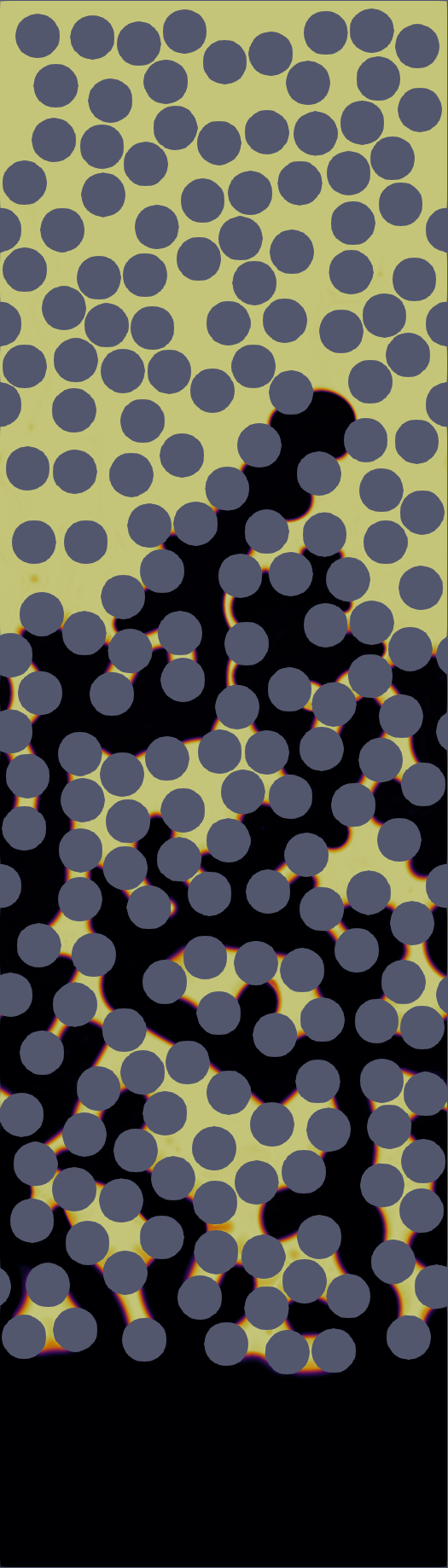}
\put(-255,85){\rotatebox{90}{\Large $\theta=135^\circ$}}\\
\caption{Snapshots of the two-fluid flow in the porous medium for contact angle $\theta=45^\circ$ (top panel), $\theta=90^\circ$ (middle panel) and $\theta=135^\circ$ (bottom panel) and capillary number $Ca=10^{-3}$.}
\label{fig:VisCa0.001}
\end{center}
\end{figure}

\begin{figure} 
\begin{center}
\includegraphics[width=0.15\textwidth]{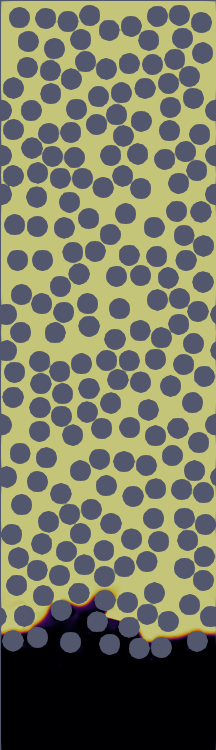}
\includegraphics[width=0.15\textwidth]{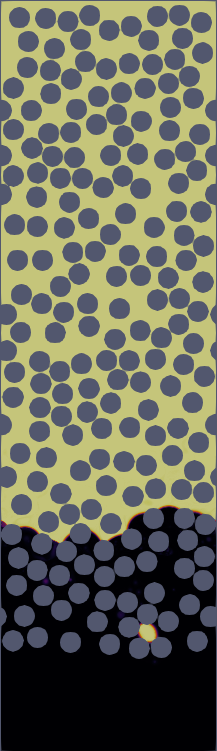}
\includegraphics[width=0.15\textwidth]{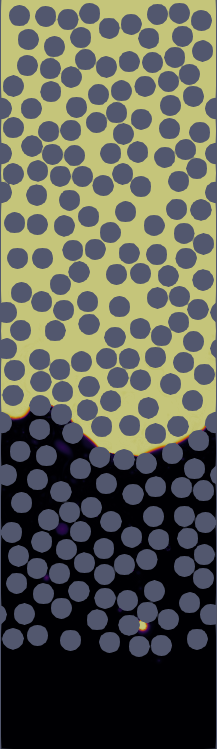}
\includegraphics[width=0.15\textwidth]{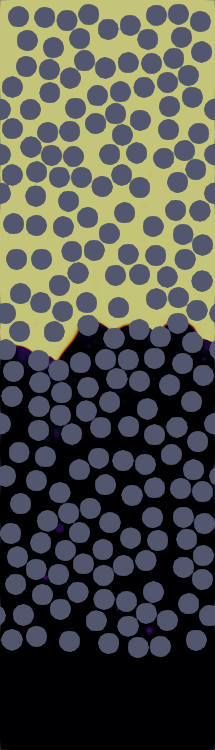}
\put(-230,207){\rotatebox{0}{\normalsize $t^*=4.19$}}
\put(-170,207){\rotatebox{0}{\normalsize $t^*=8.34$}}
\put(-112,207){\rotatebox{0}{\normalsize $t^*=12.57$}}
\put(-48,207){\rotatebox{0}{\normalsize $t^*=16.75$}}
\put(-255,85){\rotatebox{90}{\Large $\theta=45^\circ$}}\

\includegraphics[width=0.15\textwidth]{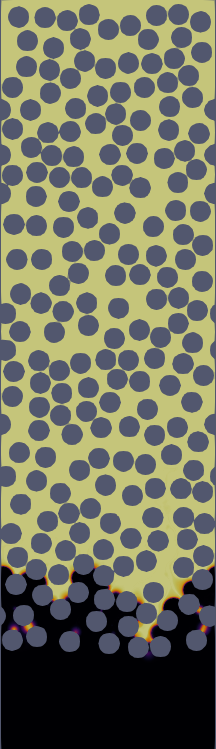}
\includegraphics[width=0.15\textwidth]{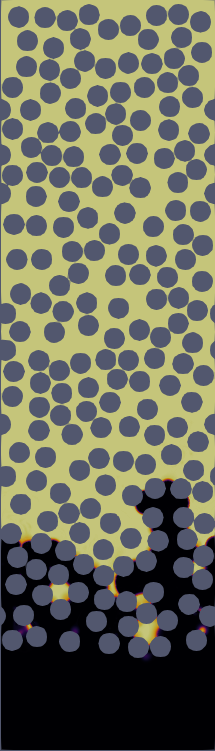}
\includegraphics[width=0.15\textwidth]{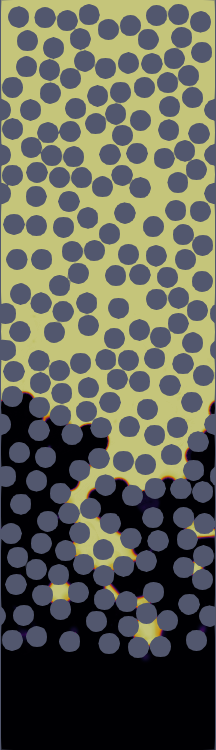}
\includegraphics[width=0.15\textwidth]{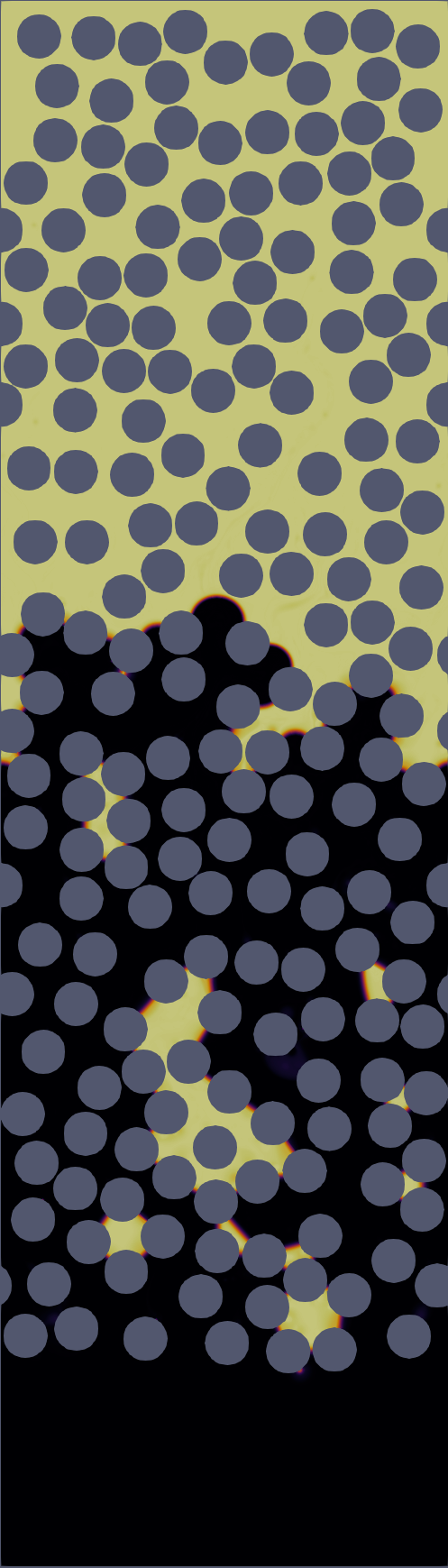}
\put(-255,85){\rotatebox{90}{\Large $\theta=90^\circ$}}

\includegraphics[width=0.15\textwidth]{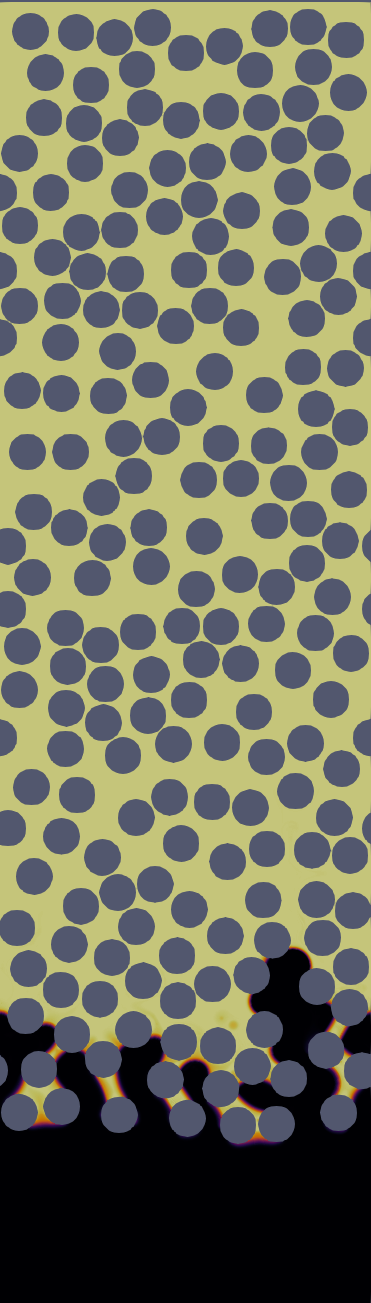}
\includegraphics[width=0.15\textwidth]{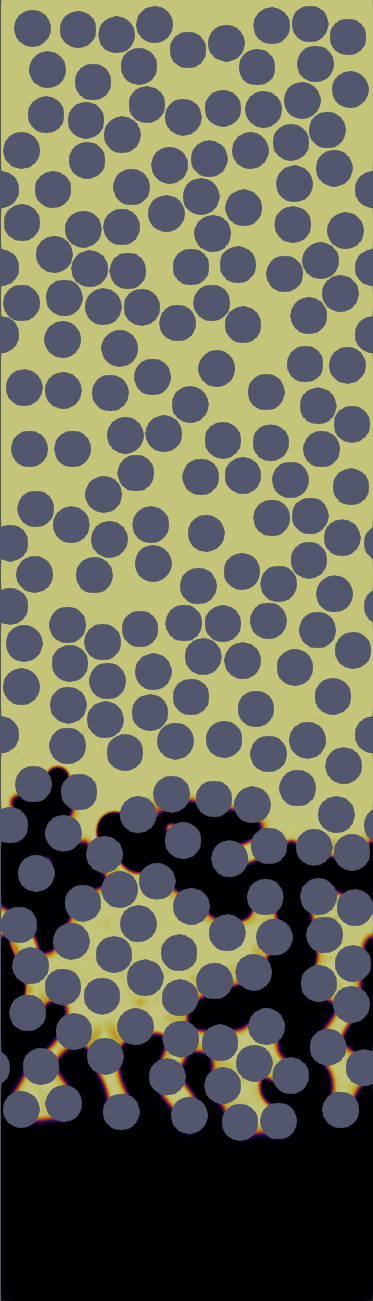}
\includegraphics[width=0.15\textwidth]{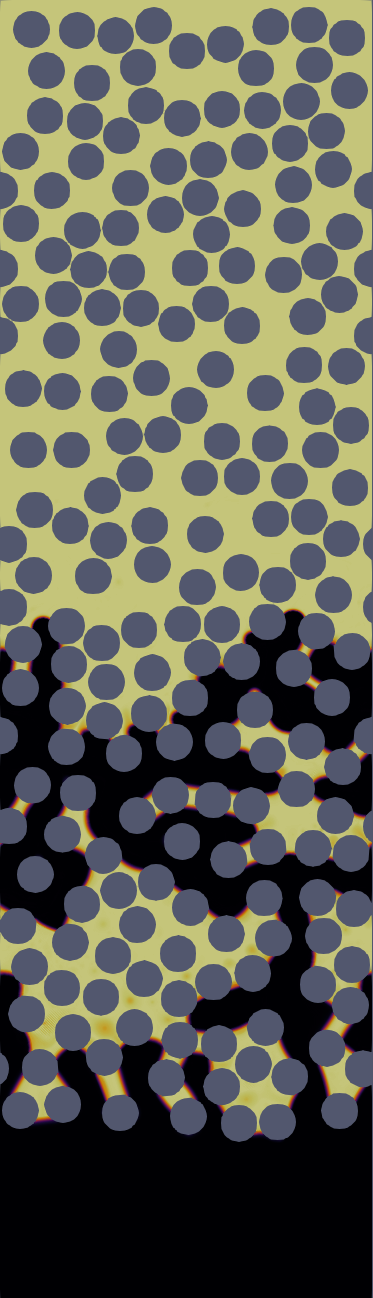}
\includegraphics[width=0.15\textwidth]{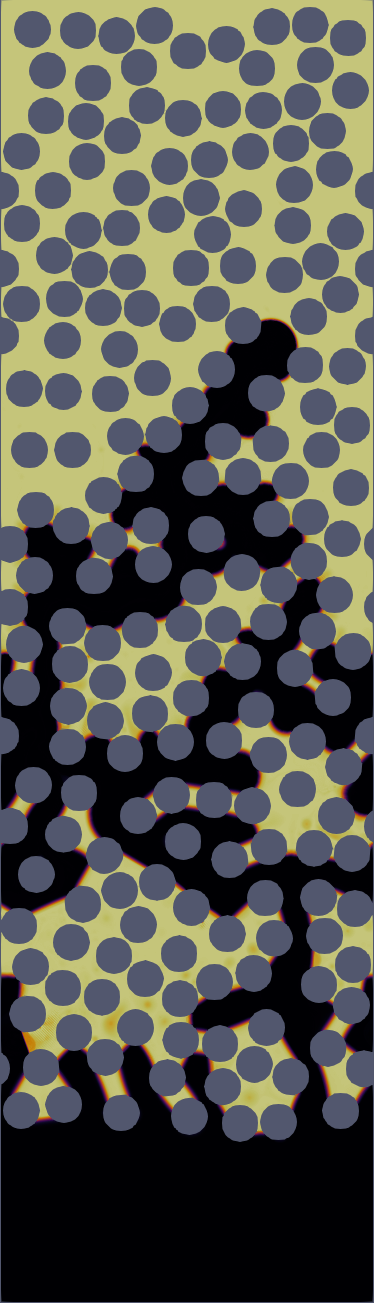}
\put(-255,85){\rotatebox{90}{\Large $\theta=135^\circ$}}\\
\caption{ Snapshots of the two-fluid flow in the porous medium for contact angle $\theta=45^\circ$ (top panel), $\theta=90^\circ$ (middle panel) and $\theta=135^\circ$ (bottom panel) and capillary number $Ca=10^{-4}$.}
\label{fig:VisCa0.0001}
\end{center}
\end{figure}

Figures \ref{fig:VisCa0.01} to \ref{fig:VisCa0.0001} illustrate the fluid flow in the porous medium for three different contact angles from top to bottom ($45^\circ$, $90^\circ$, and $135^\circ$) at four different non-dimensional time instances, $t^*=4.19$, $8.34$, $12.57$, and $16.75$ from left to right, where time is made non-dimensional with the convective scale, $t^*=tU_{ref}/R$.


As shown in figure \ref{fig:VisCa0.01}, the invading fluid penetrates the defending phase by forming thick fingers (known as capillary fingering)
regardless of the surface wettability when $Ca=10^{-2}$. 
In addition, we note 
that portions 
 of the defending phase remain trapped inside the invading fluid while the interface advances, \sz{ known as trapping \cite{doi:10.1146/annurev-fluid-010518-040342}-\cite{ZHOU2010262}-\cite{HU2020120508}-\cite{https://doi.org/10.1002/2015WR017852}}.
 
 Comparing the different rows of figure \ref{fig:VisCa0.01}, we observe that the volume of these pockets decreases as the solid surface becomes more hydrophilic (i.e., as $\theta$  decreases). Finally, there is a noteworthy difference in the shape and thickness of the fingers between cases with different value of the contact angle. For the most hydrophobic case (bottom row), the interface advances by forming several capillary fingers, while  the capillary fingers become thicker and fewer for lower values of the contact angle, the interfacial forces becoming more important. 

Thus, irrespective of the values of the capillary number and the contact angle, the interface of the invading phase advances by forming unstable capillary fingers. From the data in figure \ref{fig:VisCa0.001}, however, pertaining a smaller capillary number ($Ca=10^{-3}$), we conclude 
that the invading phase moves inside a hydrophilic porous medium with a stable motion of the interface (i.e.\, without forming any unstable fingers), whereas the penetration of the invading phase is still mainly governed by unstable capillary fingering when the porous medium is hydrophobic.

Finally, figure \ref{fig:VisCa0.0001} illustrates the simulation results for the smallest capillary number considered in this study ($Ca=10^{-4}$). The results show that even for very small values of the capillary number, when the interfacial tension strongly overcomes the viscous stress, one can still observe unstable fingers for large values of the contact angle. In addition, comparing figures \ref{fig:VisCa0.001}  and \ref{fig:VisCa0.0001}, we note that by decreasing the capillary number from $Ca=10^{-3}$ to $Ca=10^{-4}$, the area of the pockets of trapped defending phase reduces in a hydrophilic medium, whereas it increases in the hydrophobic case.

\begin{figure}
\begin{center}
\includegraphics[width=0.8\textwidth]{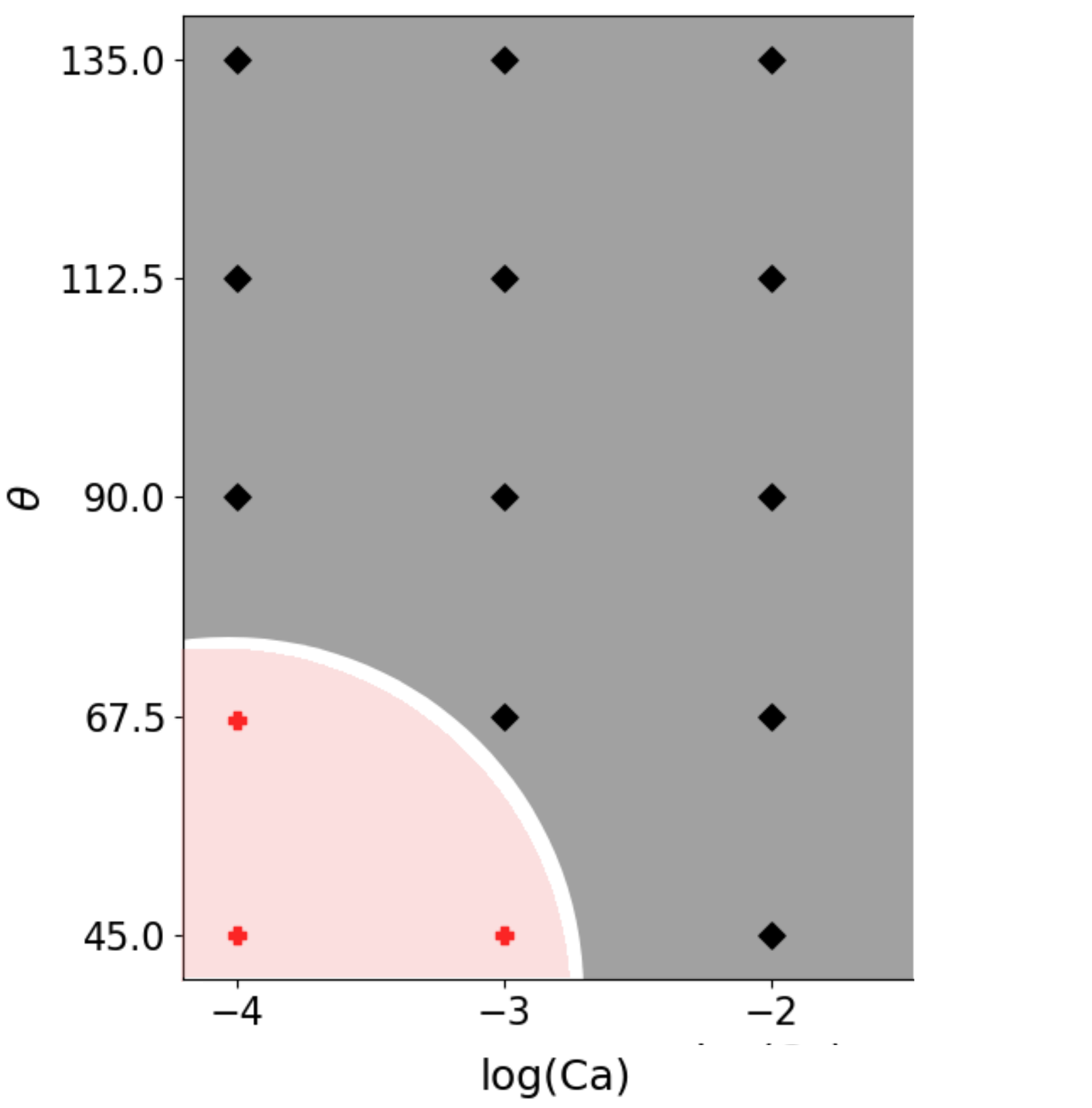}
\put(-85,110){\rotatebox{0}{\large $(1)$}}
\put(-200,55){\rotatebox{0}{\large $(2)$}}
\caption{\sz{Map of the dominant interface motion mechanisms as function of the contact angle $\theta$ and capillary number $Ca$ (note the log scale). The black, and pink background colour represent the estimated regions where capillary fingering and stable penetration of the interface are expected to occur. The available simulation data are represented by the markers, with black diamonds, and red crosses indicating capillary fingering, and the stable penetration of the interface.}}
\label{fig:map}
\end{center}
\end{figure}

\subsection{Map of the dominant interface motion mechanisms }\label{sec:map}
In figure \ref{fig:map}, we summarise the results of the previous section into a map with the dominant modes of the interface motion as a function of the capillary number and the level of surface wettability. We attribute a dominant mode of interface motion to each of the \sz{15} cases under study, using black diamonds for capillary fingering and red crosses for stable penetration of the interface, see figure \ref{fig:map}.
%
 We also use colours to indicate the regions of the map corresponding to the different mechanisms of interface motion (grey, and pink for capillary fingering, and stable penetration, respectively). Since the exact values of capillary number and contact angle where the transition between the dominant modes occurs cannot be determined, we leave a blank margin between the shaded areas. 

As seen in  figure \ref{fig:map}, most of the cases chosen for this study display capillary fingers (the grey region labeled 1 in the figure), whereas only in three cases the invading phase advances with a stable penetration of the interface (labeled \sz{2} in the figure). 
The curved boundary between regions 1 and \sz{2} indicates the importance of the combined effects of capillary number and surface wettability for \sz{ the range of $Ca$ considered in this study}. In particular, when the interfacial forces become more important (small values of $Ca$), the interface moves by forming capillary fingers in a hydrophobic medium whereas the dominant interface motion is a stable penetration when the pore walls are hydrophilic in character. Finally, as clearly observed in the map, there is no direct transition from the \sz{ capillary} fingering mode to the stable penetration.

\begin{figure} 
\begin{center}

\includegraphics[width=0.45\textwidth]{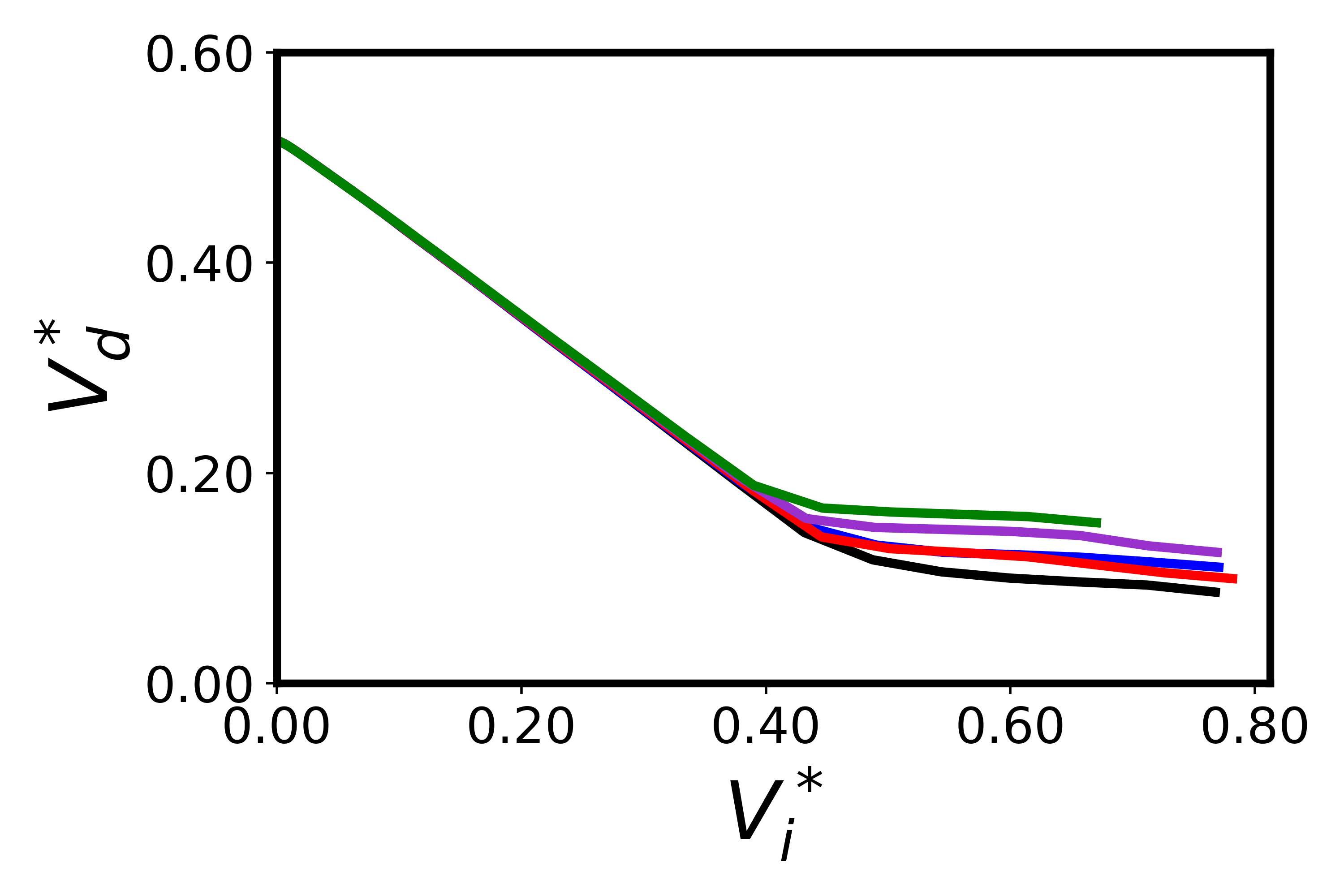}
\put(-130,115){\rotatebox{0}{\large a) $Ca= 10^{-2}$}}
\includegraphics[width=0.45\textwidth]{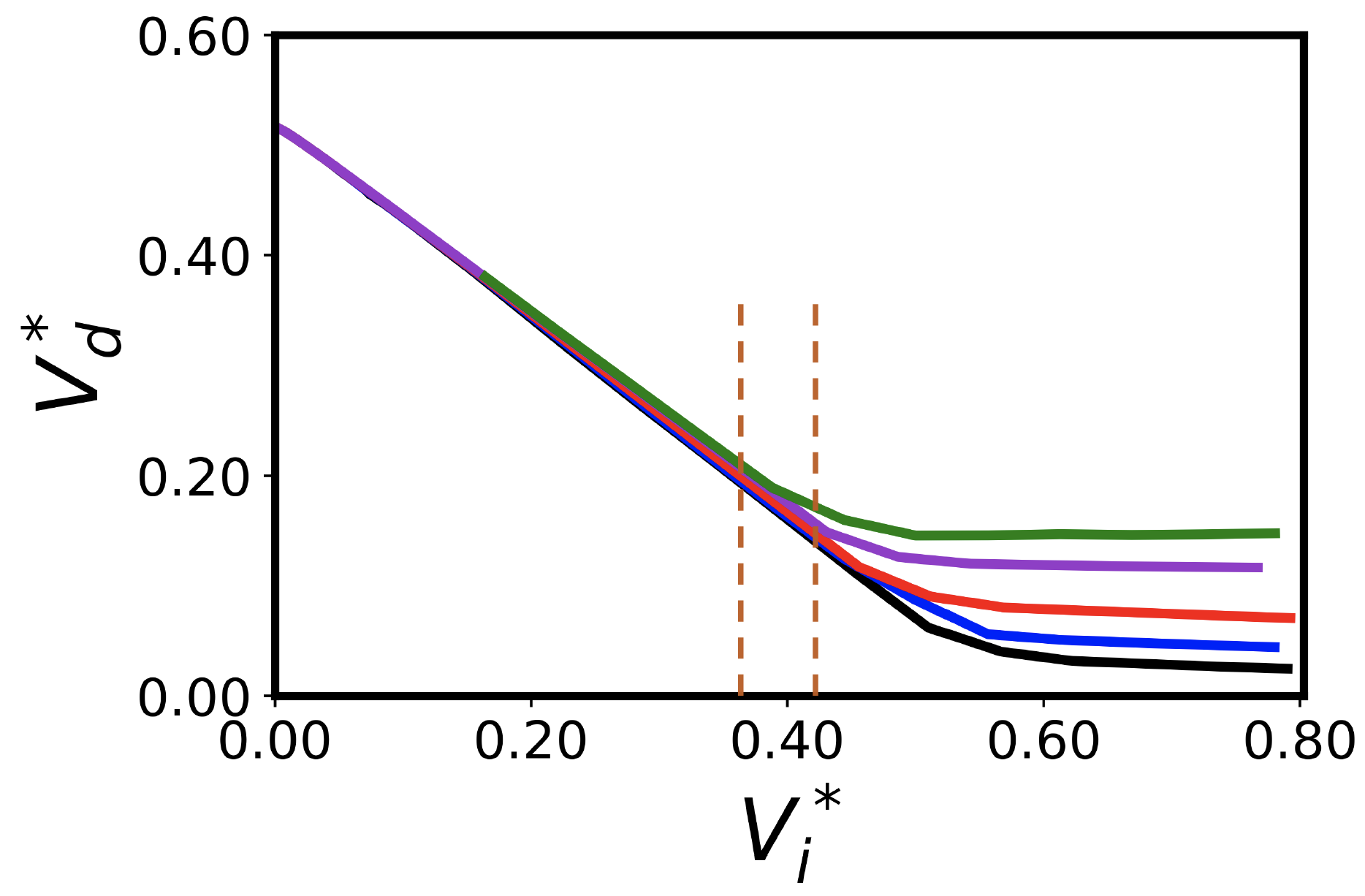}
\put(-130,115){\rotatebox{0}{\large b) $Ca= 10^{-3}$}}

\includegraphics[width=0.45\textwidth]{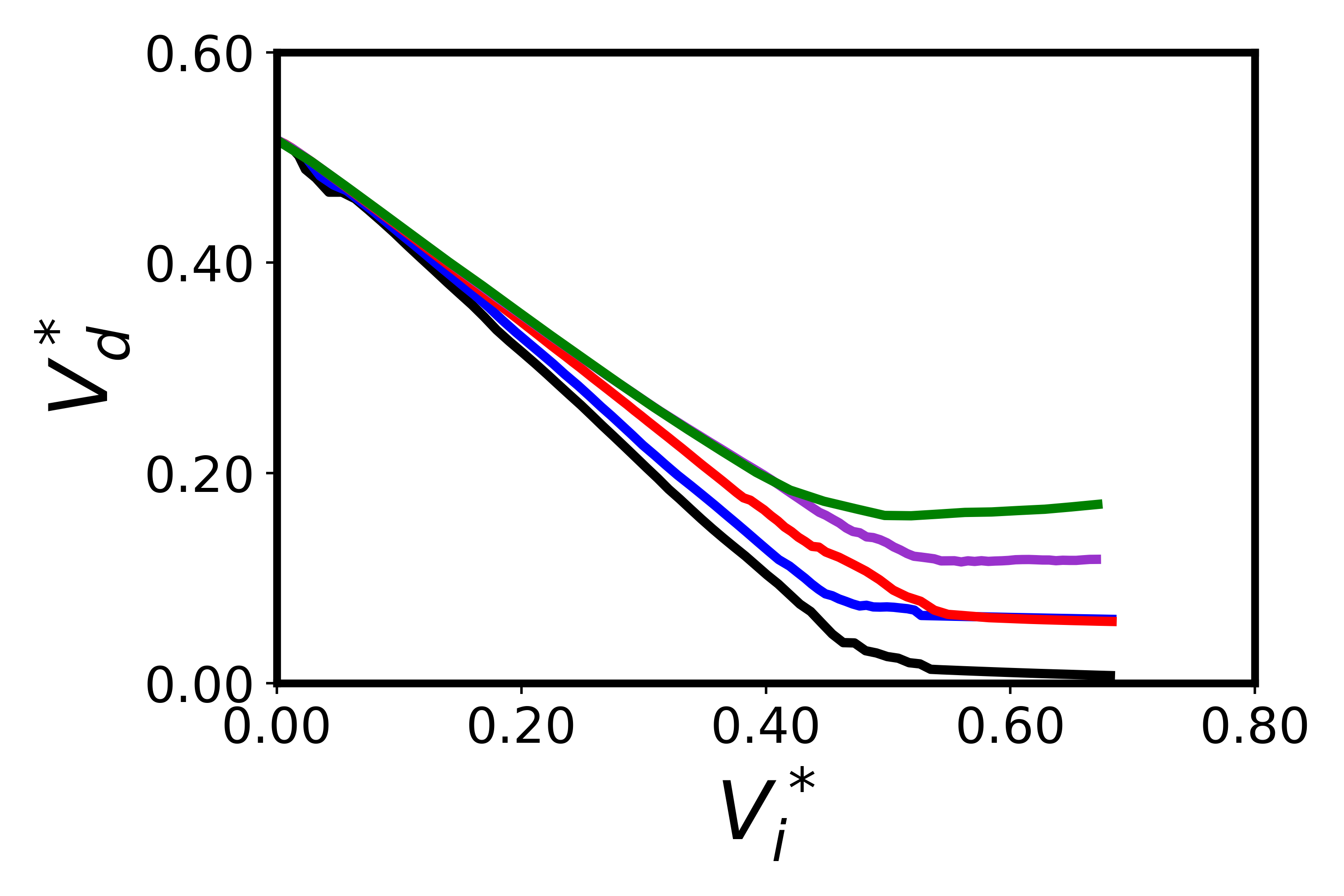}
\put(-130,115){\rotatebox{0}{\large c) $Ca= 10^{-4}$}}

\includegraphics[width=0.9\textwidth]{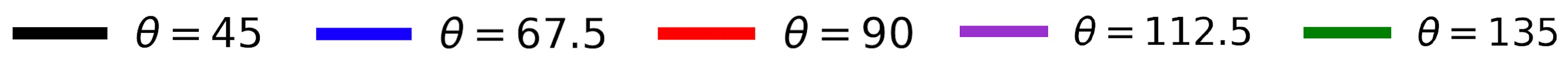}
\caption{\sz{Volume of the defending phase ($V^*_d$) versus the volume of the injected invading phase ($V^*_i$), for different values of the contact angle and  the different capillary numbers under investigation.}}
\label{fig:Saturation}
\end{center}
\end{figure}
\subsection{Quasi-steady analysis} \label{Sec:Quasi-steady}

In this section, we present and discuss the results pertaining the saturated region of the porous medium. The aim is to gain some 
understanding of the variety of the dynamics of the interface, particularly the stable and unstable penetration modes. 
Note that throughout the paper, we will use the superscript $^*$ to refer to non-dimensional values (e.g.~$\alpha^*$ is the non-dimensional value of the arbitrary variable $\alpha$), the over-bar for time averaged values (i.e.\,.~$\overline{\alpha}$ is the time averaged value of $\alpha$) and the subscript $\Gamma$ to indicate the value of quantities over the interface ($\alpha_\Gamma$ is hence the value of $\alpha$ averaged over the whole length of the interface). For the time averaging, we consider the time interval starting just before and ending just after the saturation which differ for the different values of the capillary number and the contact angle. As example, we indicate with two vertical dashed linesin figure \ref{fig:Saturation}b the time interval used for averaging the saturation properties for the case with $Ca=10^{-3}$ and $\theta=135^\circ$. 
For the time averaged quantities, error bars are presented, which are 
 computed based on the standard error of the samples.
Averages over the length of the interface are computed as:
\begin{linenomath}\begin{equation} \label{averaged_over_length}
 \begin{aligned}
\begin{gathered}
\alpha_\Gamma = \frac{ \Sigma_{j=1}^{n_y} \Sigma_{i=1}^{n_x} \alpha_{i,j}  sgn(|\nabla C(i,j)|)}{ \Sigma_{j=1}^{n_y} \Sigma_{i=1}^{n_x} sgn(|\nabla C(i,j)|)}
 \end{gathered}
\end{aligned}
\end{equation}\end{linenomath}
where $n_x$ and $n_y$ indicate the number of grid points in each direction and $sgn$ stands for the sign function. Hence, $ sgn(|\nabla C(i,j)|)$ is zero everywhere but at the interface where the order parameter changes ($|\nabla C(i,j)|>0$). Thus,  the denominator of equation (\ref{averaged_over_length}) counts the number of grid points at the interface, whereas the numerator is the sum of the values of the desired variable over the interfacial points.

Figure \ref{fig:Saturation} illustrates the evolution of the non-dimensional volume of the defending phase ($V^*_d$) versus the non-dimensional volume of the injected invading phase ($V^*_i$) for different values of the contact angle and the capillary number. Note that although our simulations are two-dimensional and the results in figure \ref{fig:Saturation} essentially represent the area occupied  by the different phases, for generality and consistency with the terminology in the literature, we denote them as the volume of each phase. Here, the total area of the domain (including the solid objects) is taken as the reference value for non-dimensionalization. 

As a general observation, we note that, for all the cases, the volume of the defending phase ($V^*_d$) decreases initially almost linearly as the invading fluid is injected into the domain. Close to the saturation of the region under consideration, the reduction of $V^*_d$ decreases until it approaches a plateau. This enables us to define the time when  the system has reached the quasi-steady state. Note again that we only consider the volume delimited by the green dashed lines in figure \ref{fig:Setup}.

 \sz{For the cases with higher capillary numbers (see figures \ref{fig:Saturation}a and b)}, the difference between configurations with different contact angles become evident already before saturation: the volume of the defending phase decreases faster for the hydrophilic cases. Moreover, the values of the saturated volume are found to be lower for the hydrophilic cases than for the hydrophobic solid pores. This is consistent with the observations in \ref{Sec:Visualization} where we identified thicker capillary fingers for $Ca = 10^{-2}$ and stable penetration of the interface for $Ca = 10^{-3}$ and $Ca = 10^{-4}$ 
 in a hydrophilic porous medium (low values of $\theta$); this is also associated with a reduction of the volume of the pockets of the defending phase.

\begin{figure} 
\begin{center}
\includegraphics[width=0.4\textwidth]{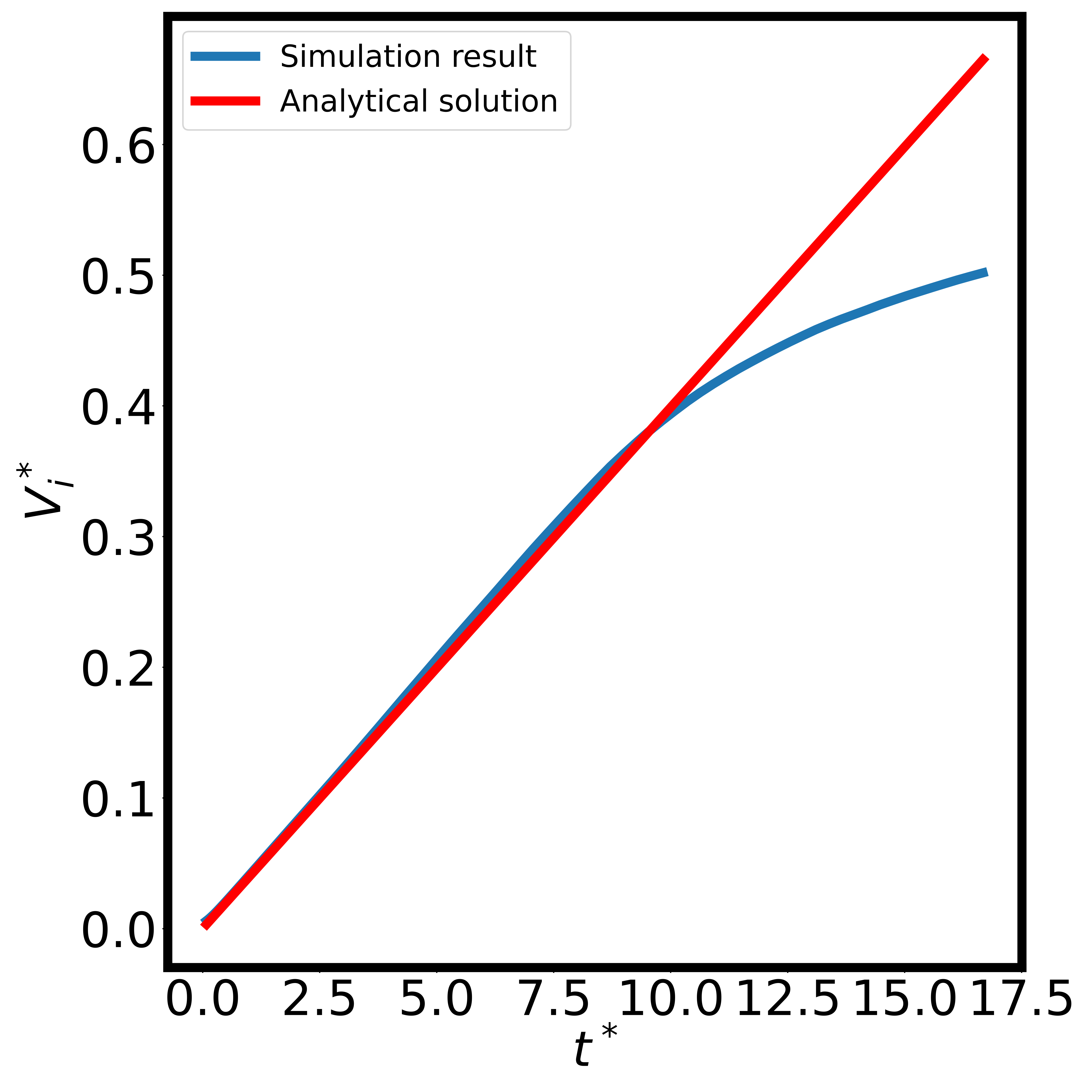}
\caption{Evolution of the volume of the injected invading phase obtained from simulation and analytical solution for the case where $Ca=1$ and $\theta=45^\circ$.}
\label{fig:mass_conservation}
\end{center}
\end{figure}

\sz{In addition, capillary fingering modes} can be recognised by the equal slopes of the $V^*_d$-$V^*_i$ curve for different contact angles, see figures \ref{fig:Saturation}a and b; the reduction in capillary number accelerates the saturation in a hydrophobic configuration while it postpones it in a hydrophilic medium. 
This quantitative analysis confirms the observation in section \ref{Sec:Visualization} of a stable penetration of the invading phase in a hydrophilic medium, which leaves behind only small pockets of the defending phase for small capillary numbers. Therefore, larger volumes of the invading phase should be injected to fill the medium before the saturation state is reached. On the other hand, in the case of small capillary numbers, there are sizeable pockets of invading fluid trapped in the medium for larger contact angles, hence, less injected fluid is required to reach the final saturated state. \sz{In Fig.\ref{fig:Saturation}, the plateau values of the curves represent the remaining defending phase in the initial part of the domain, which is approximately 20\% during fingering and nearly zero in case of
stable penetration. The slope of the curves provides an indirect measure of the time required to expel the defending phase and reach a steady state, due to the constant injection rate. Consequently, saturation is achieved more rapidly in cases with stable penetration.}

\begin{figure} 
\begin{center}
\includegraphics[width=0.47\textwidth]{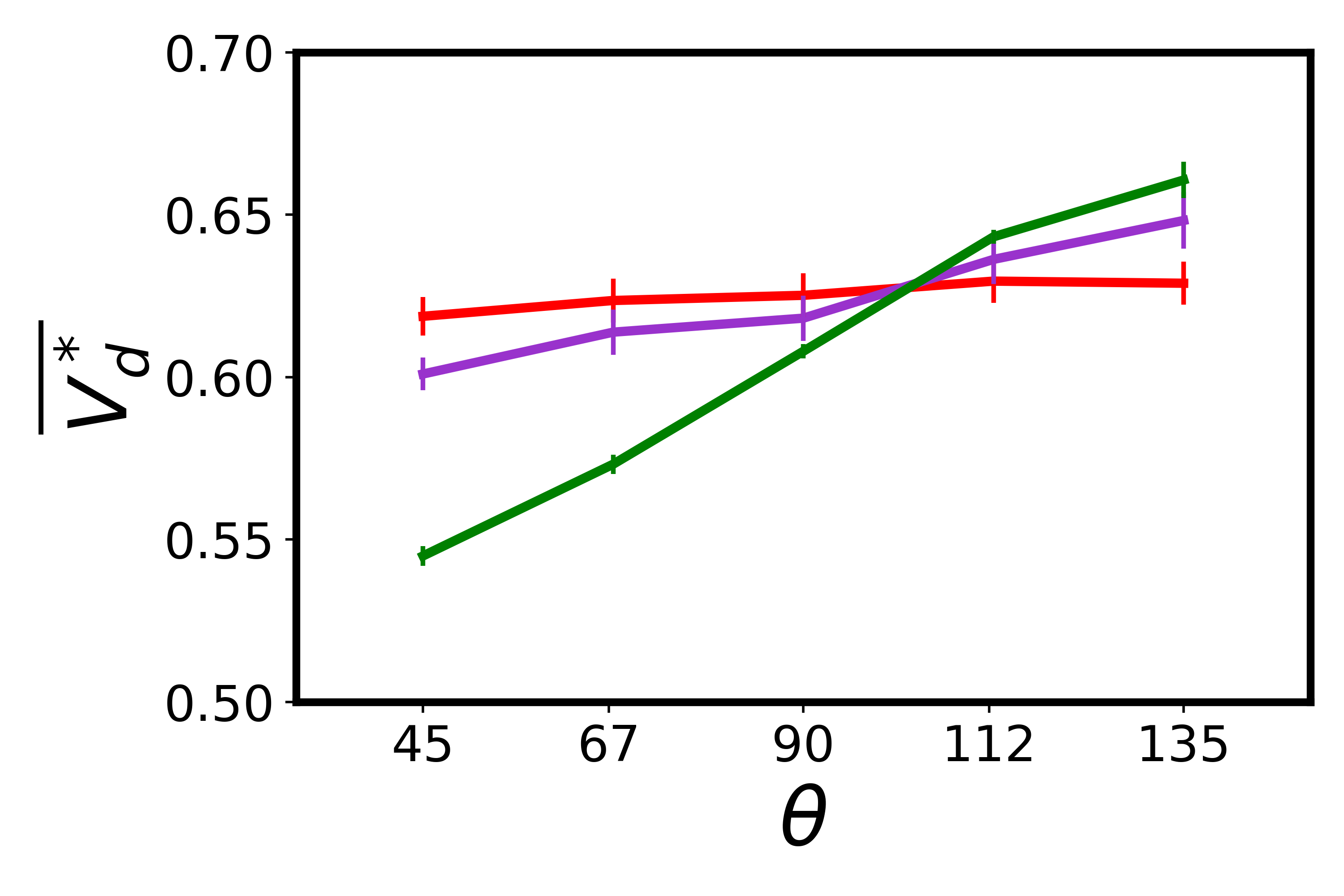}
\put(-135,115){\rotatebox{0}{\large a)}}
\includegraphics[width=0.47\textwidth]{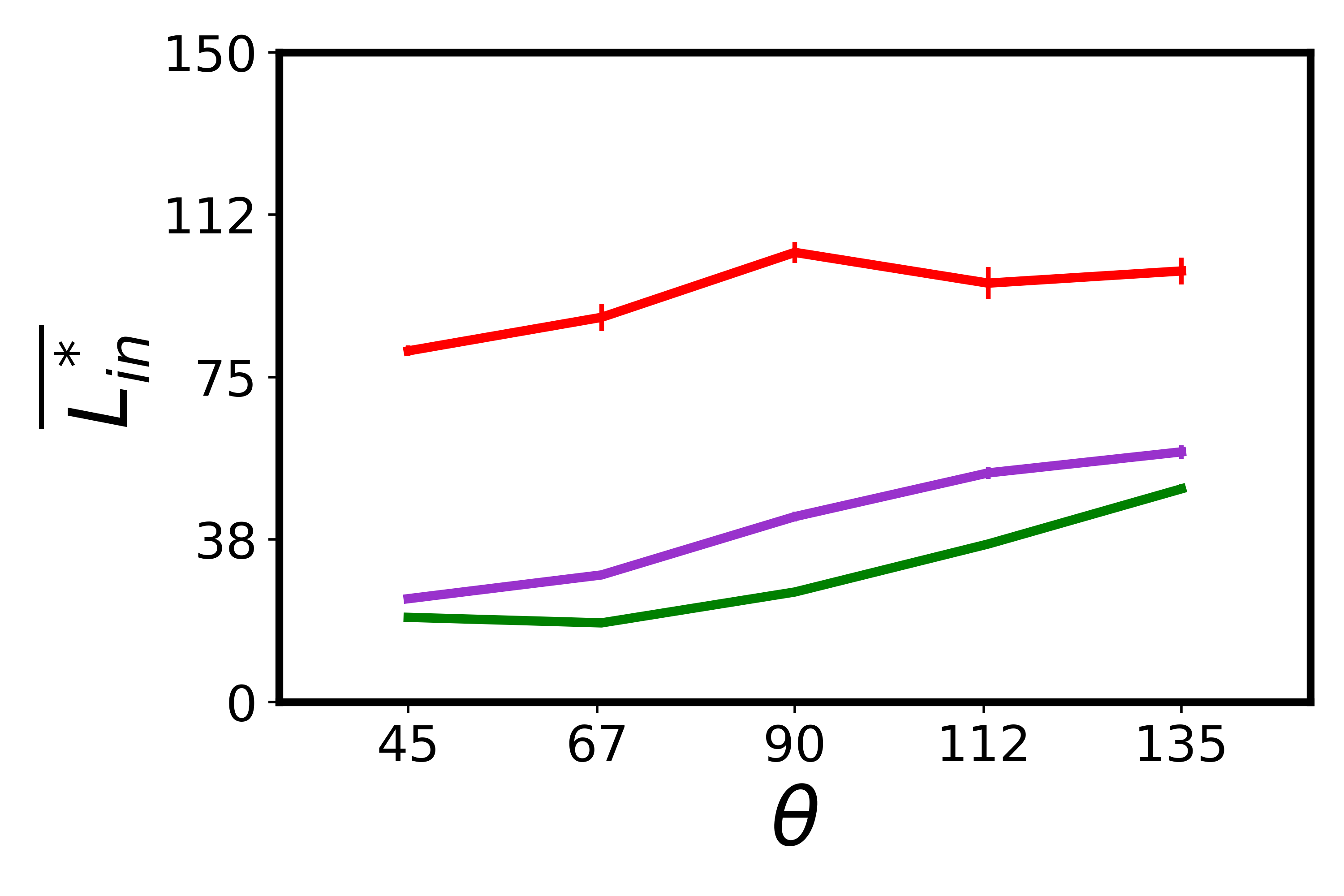}
\put(-135,115){\rotatebox{0}{\large b)}}

\includegraphics[width=0.7\textwidth]{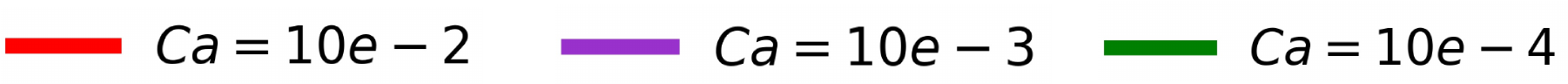}
\caption{\sz{Time averaged volume of the defending phase (panel a) and length of the interface (panel b) at saturation versus  the contact angle and for different values of the capillary number (see legend). The vertical bars on each plot represent the standard error of the samples.}}
\label{fig:Mean_Void_Length}
\end{center}
\end{figure}

Even though the employed numerical framework has been substantially validated in our previous work (\cite{SHAHMARDI2021110468}), we provide here 
more analyses confirming the validity of the numerical algorithm. In particular, we consider the mass conservation and diffusion of the interface for the most diffused case, $Ca=1$ and $\theta=45^\circ$.
Figure \ref{fig:mass_conservation} compares the evolution of the volume of the invading phase with the analytical solution within the saturated domain. As mentioned in section \ref{Sec:Setup}, the invading phase is injected with a constant velocity at the inlet. Thus, the volume of the injecting phase as a function of time is $ V_a=U_{ref} L_y t $, where $L_y$ is the width of the domain. As figure \ref{fig:mass_conservation} suggests, the total volume (thus the mass since the fluid is incompressible) of the injected phase matches the analytical value before the saturation (around $t^*\approx 11$), when some of the  the injected invading phase is exiting the domain.

Figure \ref{fig:Mean_Void_Length}a reports the time averaged value at saturation of the remaining volume of the defending phase ($\overline{V^*_d}$) for different values of the contact angle and capillary number. In general, the volume of the defending phase is directly related to the type of interface motion, namely, capillary fingering and stable penetration. If the interface steadily advances  (stable penetration), the amount of defending phase trapped in the saturated domain is smaller than for
the \sz{capillary} fingering mode. 
For a hydrophilic case (consider e.g.~$\theta=45^\circ$), decreasing the capillary number results in a decrease in $\overline{V^*_d}$. 
In this case, a larger portion of the defending phase remains trapped between the fingers. 
For the smallest capillary numbers, the interfacial forces becomes even more important, which results in a stable motion of the interface, the gradual disappearance of the {\sz{capillary}} fingers  and  the decrease of the trapped portion of the defending phase. For the hydrophobic cases, instead, (consider e.g.~$\theta=135^\circ$), the interface advances forming fingers, irrespective of the capillary number\sz{.} \sz{Decreasing} the capillary number the fingers are \sz{fewer} and thicker. In the case of large capillary numbers, the amount of the defending phase trapped between the many fingers is lower than for flows with smaller capillary numbers. 

Figure \ref{fig:Mean_Void_Length}b shows the time averaged length of the interface for all the cases under investigation. First, the results reveal that increasing the capillary number results in a longer interface, irrespective of the contact angle. This confirms that, as the capillary number increases, the interface motion mechanism shifts from stable penetration (for hydrophilic cases) or the formation of few capillary fingers (for hydrophobic cases) to the formation of many \sz{capillary} fingers, which corresponds to a larger interfacial area. 
Moreover, as shown in figure \ref{fig:Mean_Void_Length}b, the interface length (and correspondingly the dominant interface motion mechanism) changes more with  the capillary number than with the contact angle. Nevertheless, as the system becomes more hydrophobic, the length of the interface slightly increases. This increase is more evident for small capillary numbers due to the transition from the stable penetration (for a hydrophilic medium) to capillary fingering (for a hydrophobic medium).

\begin{figure}
\begin{center}
\includegraphics[width=0.45\textwidth]{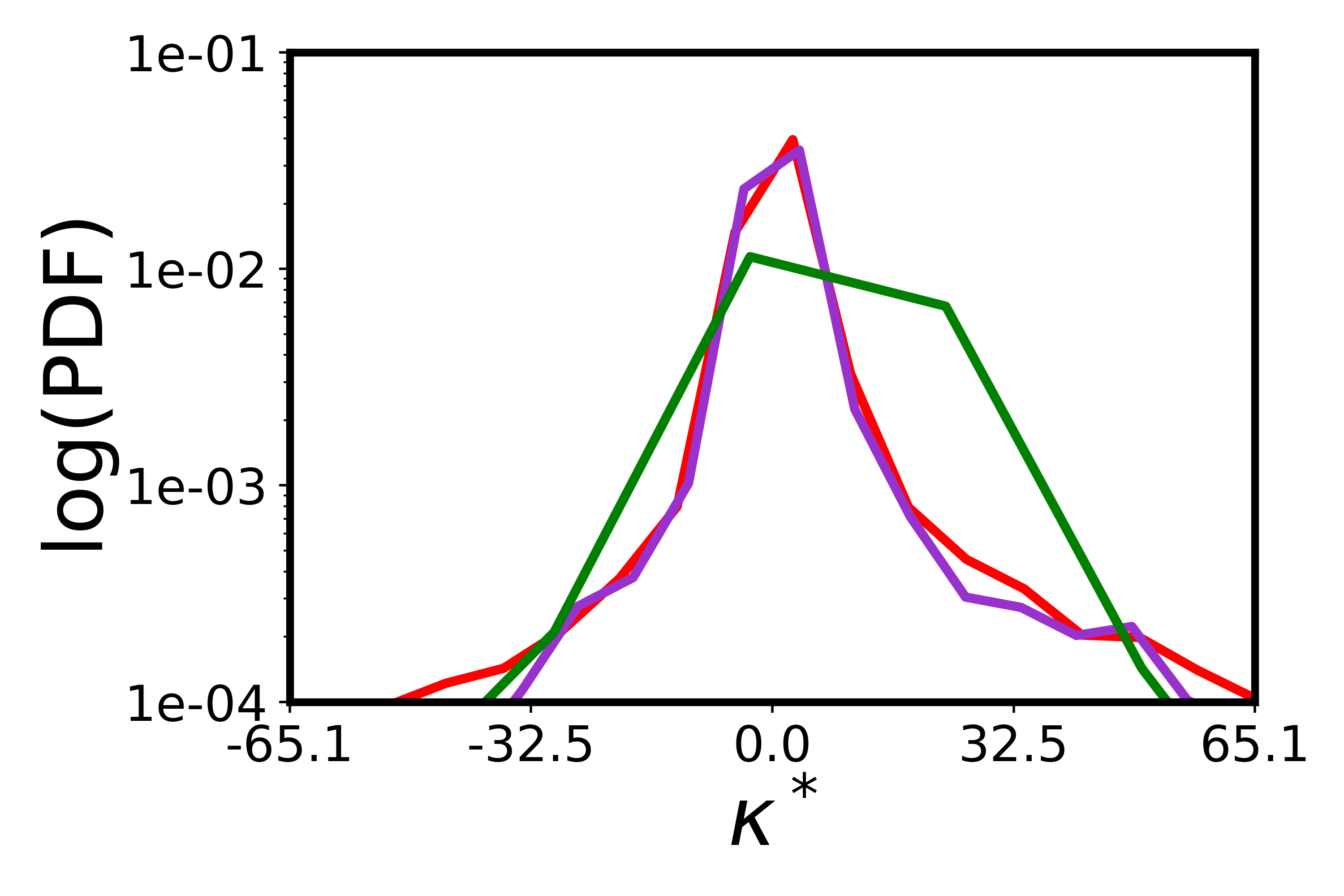}
\put(-130,115){\rotatebox{0}{\large a) $\theta=45^\circ$}}
\includegraphics[width=0.45\textwidth]{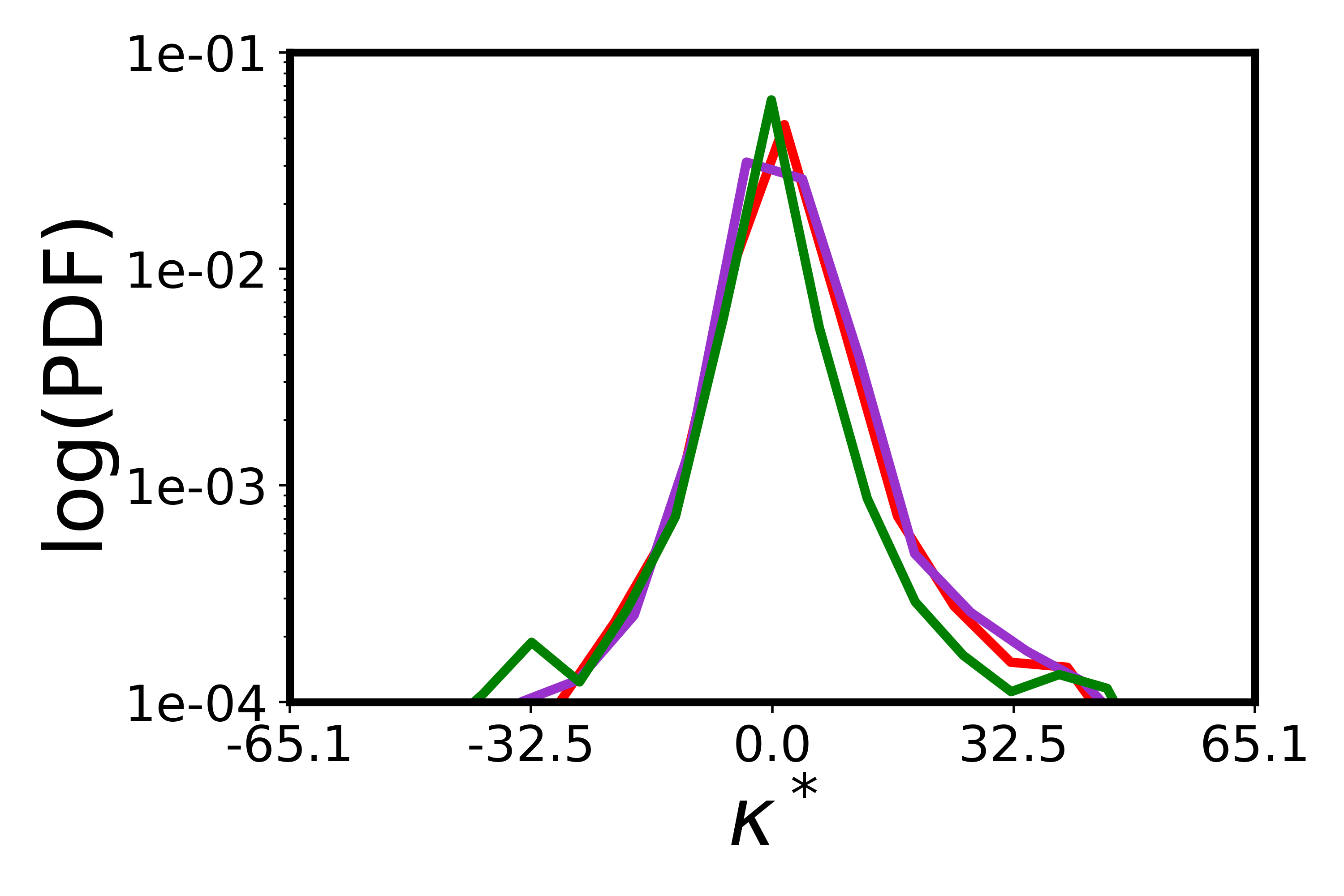}
\put(-130,115){\rotatebox{0}{\large b) $\theta=67.5^\circ$}}

\includegraphics[width=0.45\textwidth]{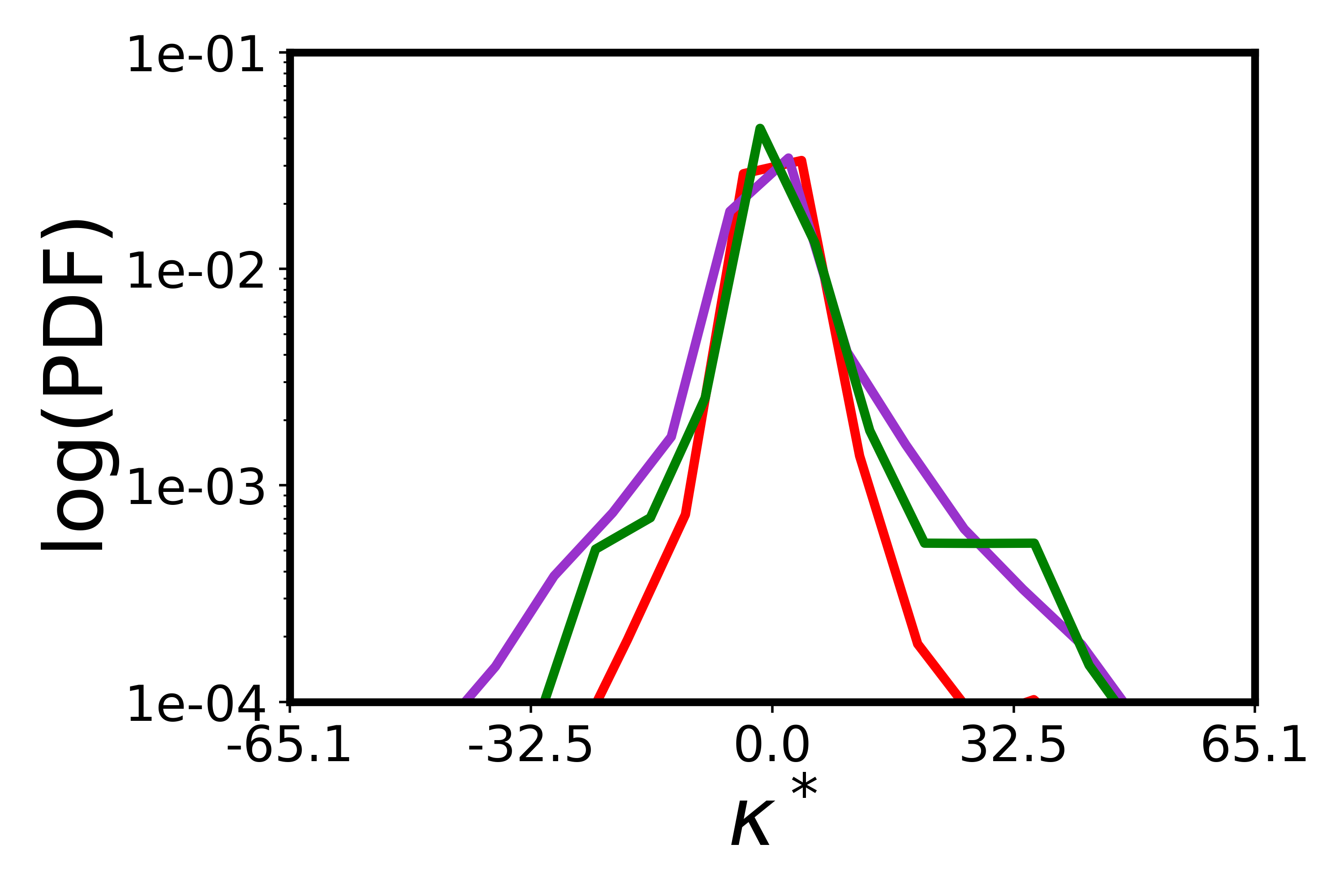}
\put(-130,115){\rotatebox{0}{\large c) $\theta=90^\circ$}}
\includegraphics[width=0.45\textwidth]{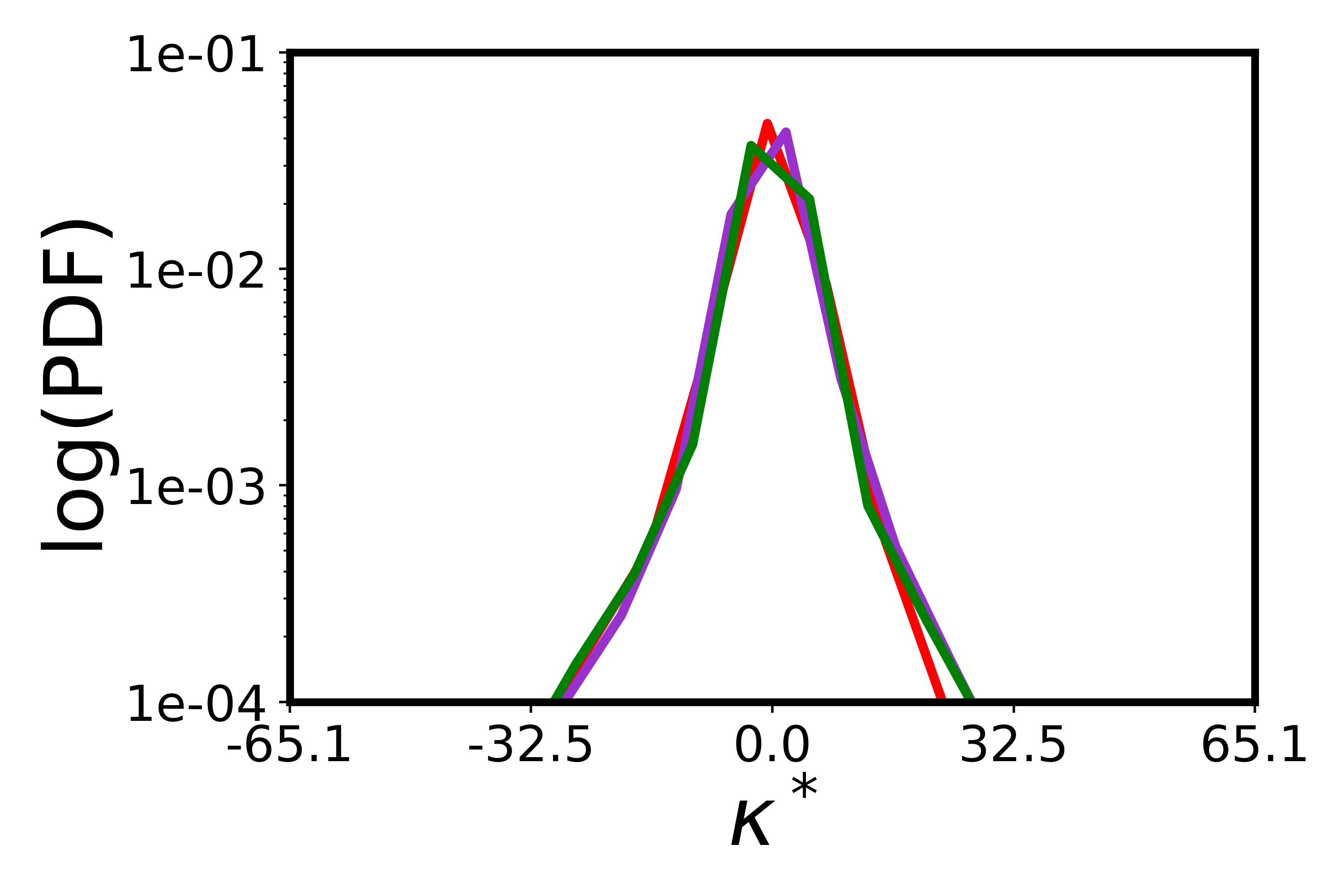}
\put(-130,115){\rotatebox{0}{\large d) $\theta=112.5^\circ$}}

\includegraphics[width=0.45\textwidth]{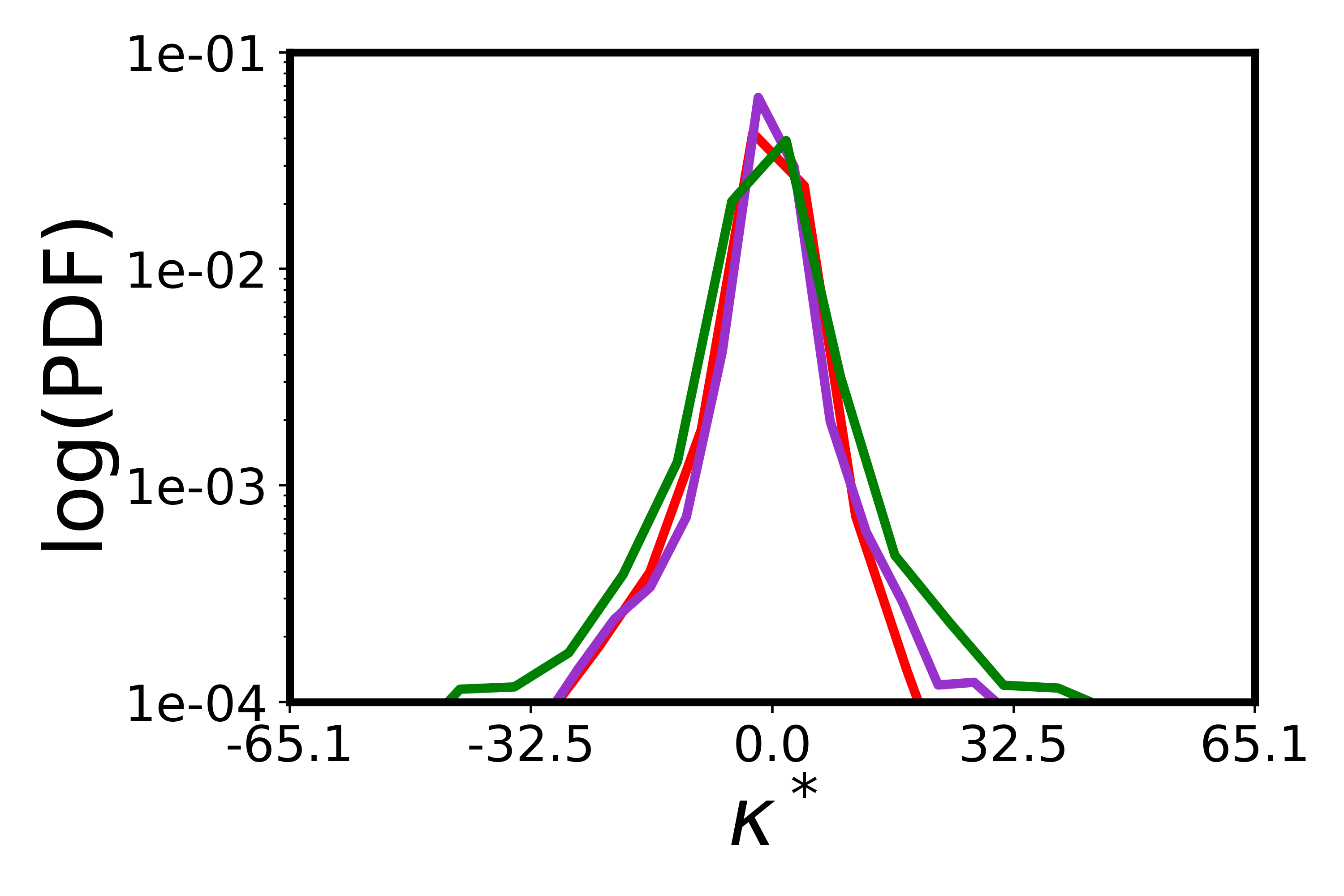}
\put(-130,115){\rotatebox{0}{\large e) $\theta=135^\circ$}}

\includegraphics[width=0.7\textwidth]{figures/Ca_legend.pdf}
\caption{\sz{Probability distribution function of the interface curvature just before saturation ($t^*=10.47$) for different values of capillary number and five values of the contact angle, panels a-e) as indicated in the legend.}}
\label{fig:Curvature_PDF}
\end{center}
\end{figure}

Figure \ref{fig:Curvature_PDF} shows the probability distribution function (PDF) of the interface curvature at saturation ($t^*=10.47$)  for different values of the capillary number and five
values of the contact angle. 
The chosen time is close to the saturation of the most stable case ($Ca=10^{-4}$ and $\theta=45^\circ$).  For this particular case, the saturation domain would be completely filled with the invading phase at any later time  and the interface curvature would be limited to that of the small pockets of defending phase trapped in the invading phase.

Comparing the different panels in figure \ref{fig:Curvature_PDF}
 (corresponding to different contact angle values) 
 several observations can be made. Firstly, as the capillary number decreases, the distribution of the interface curvature becomes narrower. This is explained by the increasing importance of the surface tension which makes the interface more stable and attenuates the strength of the fingers. 
To be more specific, let us consider figure \ref{fig:Curvature_PDF}a pertaining a contact angle $\theta=45^\circ$. 

As the capillary number decreases and the surface tension forces become more significant, the fingers becomes shorter and thicker (i.e., we observe a transition from \sz{capillary fingers} to a stable interface for lower contact angles), which decreases the range of values attained by the interface curvature. Secondly, comparing the results of the smallest capillary number in figure \ref{fig:Curvature_PDF}a with those in figure \ref{fig:Curvature_PDF}e (green lines), we observe that for the hydrophilic case (panel a), the PDF of the curvature significantly tends towards the positive values, indicating a predominantly concave shape of the stable interface, whereas  the curvature displays  a more symmetric distribution for the hydrophobic case, which indicates the presence of few thick fingers, with both positive and negative curvature (see figure \ref{fig:VisCa0.0001}).
  
\begin{figure}
\begin{center}
\includegraphics[width=0.45\textwidth]{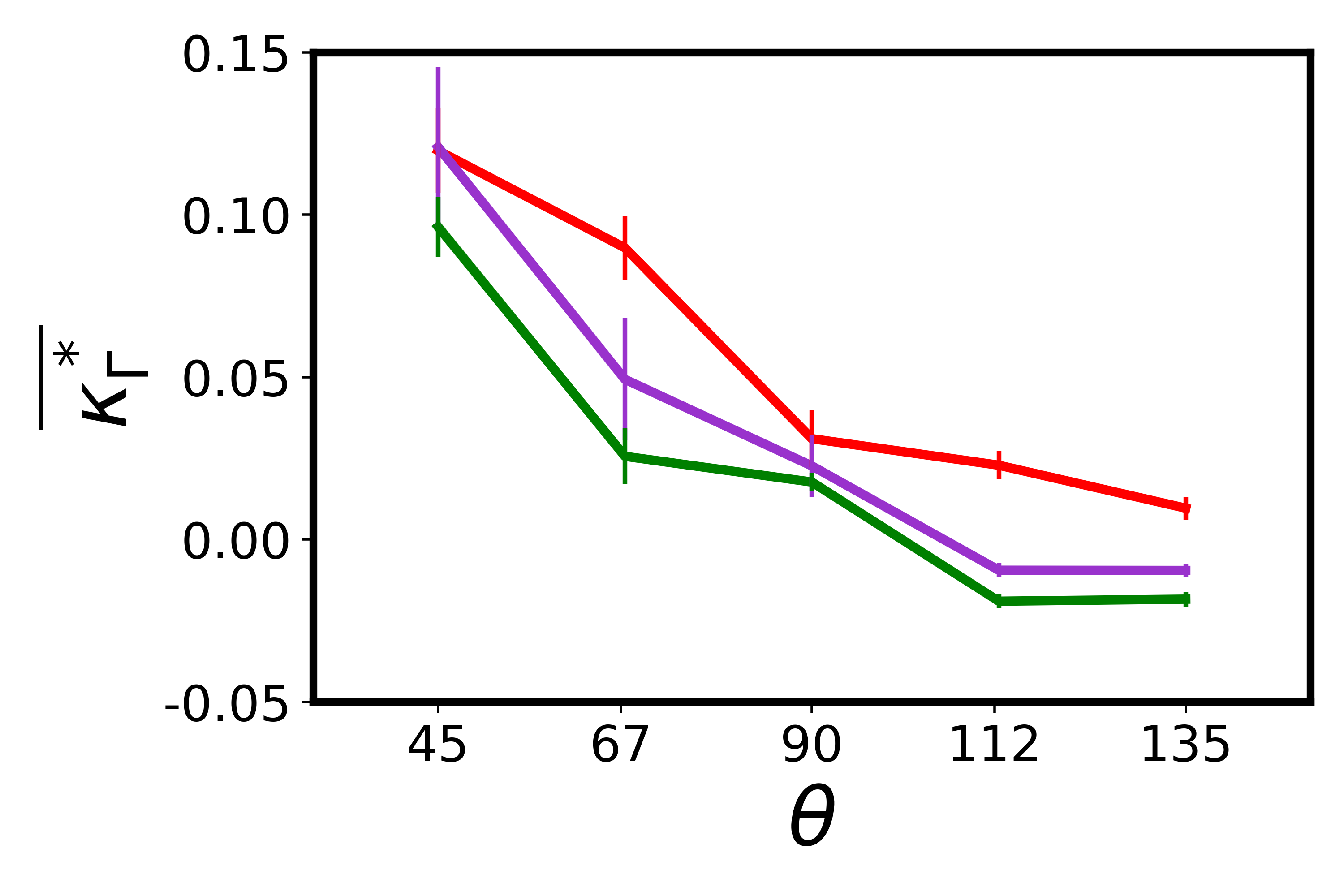}
\put(-130,115){\rotatebox{0}{\large a)}}
\includegraphics[width=0.45\textwidth]{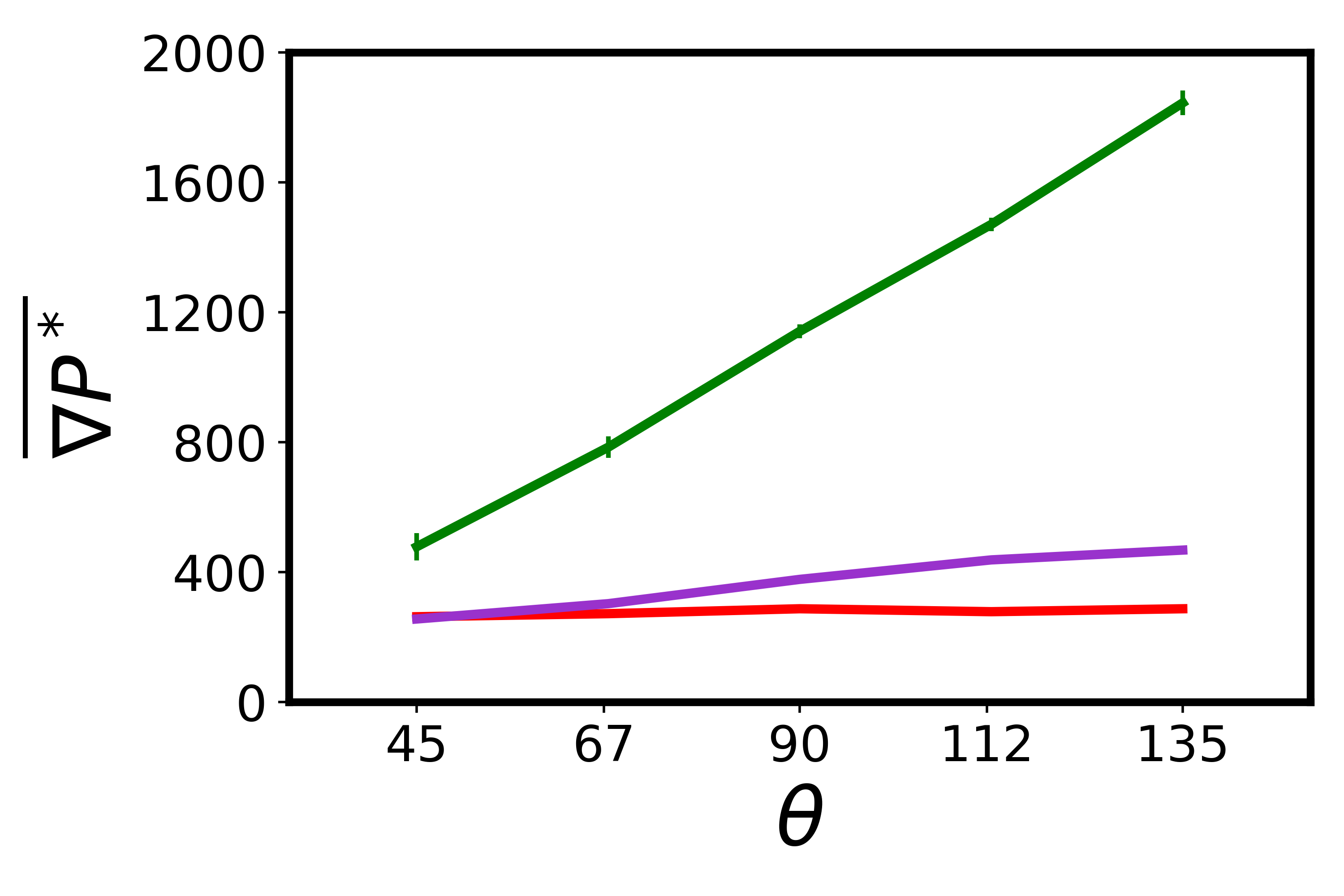}
\put(-130,115){\rotatebox{0}{\large b)}}

\includegraphics[width=0.7\textwidth]{figures/Ca_legend.pdf}
\caption{\sz{Time average of a) the mean interface curvature, and b) the pressure gradient versus the contact angle and the values of the capillary number indicated in the legend. The vertical bars on each plot represent the standard error of the samples.}}
\label{fig:MeanPressure}
\end{center}
\end{figure}

To conclude the analysis of the flow at saturation,
we display in figure \ref{fig:MeanPressure}a) the time averaged value of the interface curvature. The data in the figure reveal that, decreasing the capillary number, when the surface tension forces become relevant, the dynamics change with the solid-wall wettability.
 In detail, let us consider the smallest capillary number under investigation. As shown earlier in figure \ref{fig:Curvature_PDF}, the PDF of the mean curvature presents a narrow range of curvature values. As the interface moves within the medium and approaches the next cylinder, the interface becomes concave (positive curvature)  for the hydrophilic cases whereas  it is prevalently convex (negative curvature) for the hydrophobic cases, as it needs to adjust to the prescribed contact angle. 
 Therefore, for the cases with moderate or small values of the capillary number, the mean curvature of the interface decreases as the pore surface becomes more hydrophobic. 
 Moreover, as a consequence of the above discussion and 
%
 \sz{for the selected capillary numbers in this study}, decreasing the capillary number results in a gradual shift from unstable capillary fingering to stable penetration for the hydrophilic cases or  to \sz{thicker but fewer capillary fingers} for hydrophobic configurations, and, thus, to a reduction in the mean curvature of the interface.

Finally, figure \ref{fig:MeanPressure}b depicts the averaged pressure drop, defined as $\nabla P = (P_{in}-P_{out})/L_{sat}$, across the saturated domain for all the cases under consideration, 
recalling that in our simulations the invading phase is injected into the medium at constant flow rate. Firstly, the data reveal that, irrespective of the contact angle, the pressure gradient increases as the capillary number decreases. 
This increase can be explained considering the contribution of the capillary pressure ($P_c=\kappa\sigma$), or interfacial stresses, to the total  pressure gradient needed to drive the flow. 
At fixed capillary number, the mean curvature of the interface decreases as the contact angle increases (see figure \ref{fig:MeanPressure}b) which may suggest that 
the capillary pressure decreases with the contact angle.
This is however not the case:
the observed increase of the pressure drop with $\theta$ is attributed to the length of the interface, which increases significantly as the surface become more hydrophobic, i.e.\  increasing  $\theta$ (see figure \ref{fig:Mean_Void_Length}b).
Thus, the integral of the capillary pressure over the whole interface length is greater for the hydrophobic cases despite the curvature is locally reduced.

\begin{figure} 
\begin{center}
\includegraphics[width=0.45\textwidth]{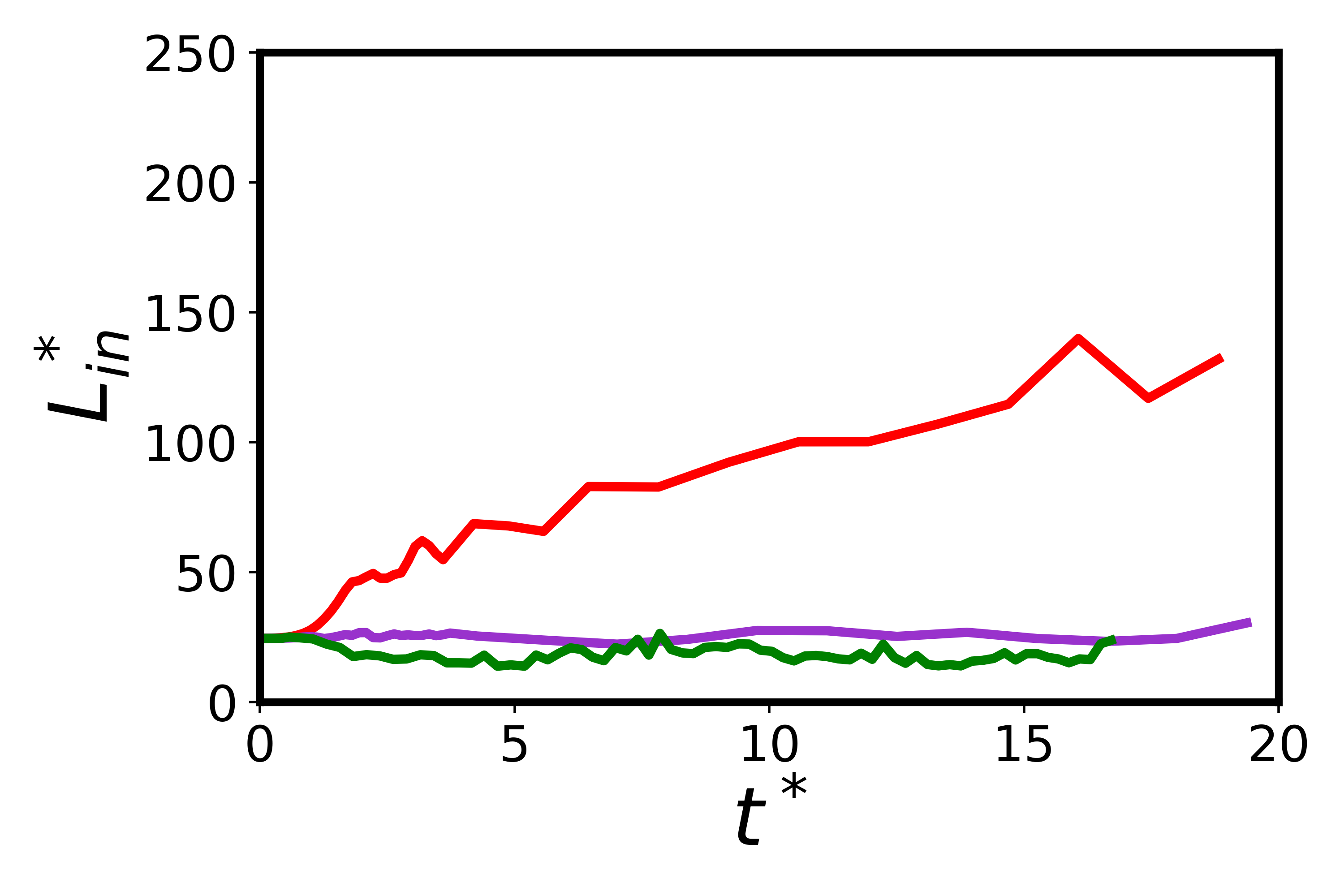}
\put(-130,115){\rotatebox{0}{\large a) $\theta=45^\circ$}}
\includegraphics[width=0.45\textwidth]{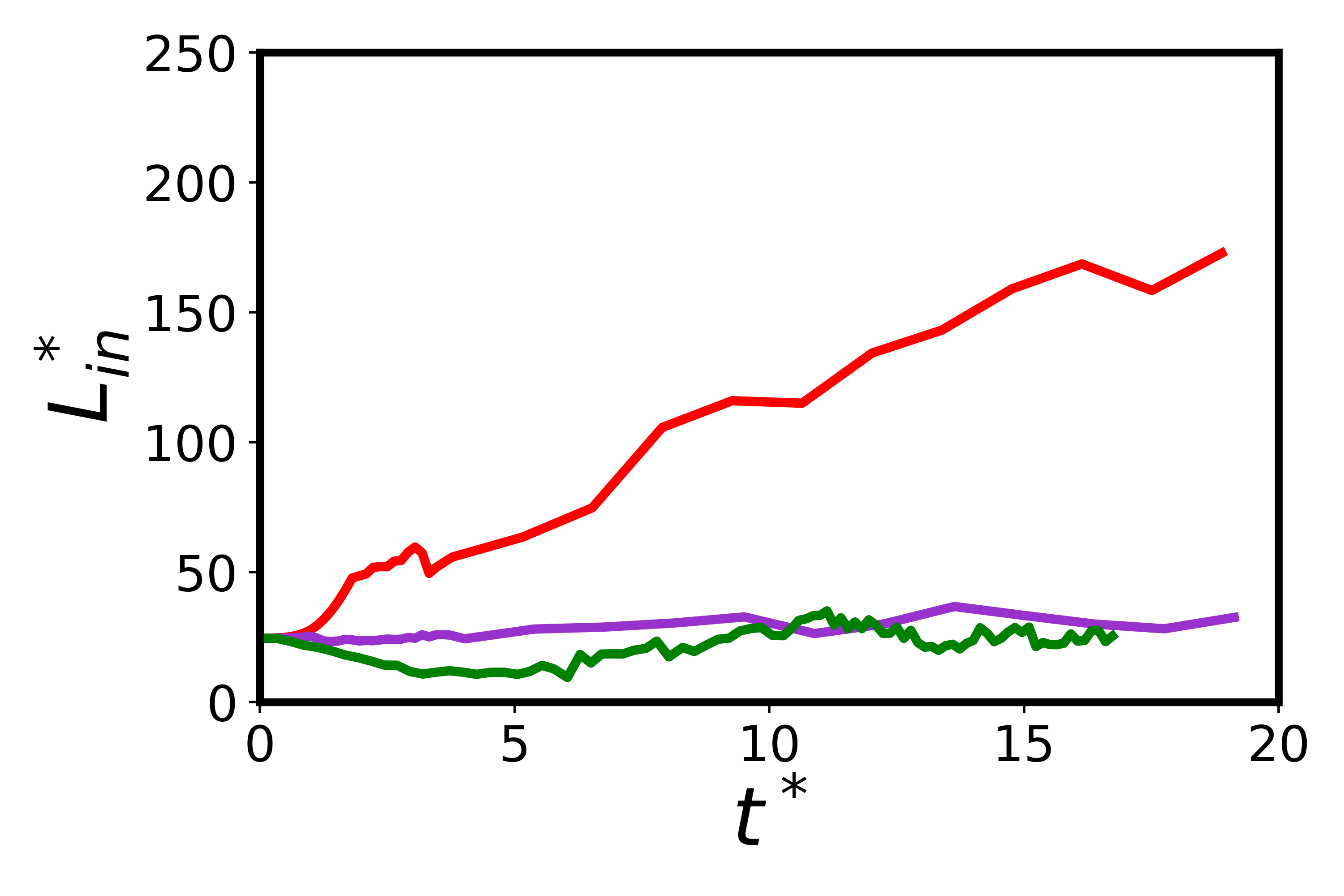}
\put(-130,115){\rotatebox{0}{\large b) $\theta=67.5^\circ$}}

\includegraphics[width=0.45\textwidth]{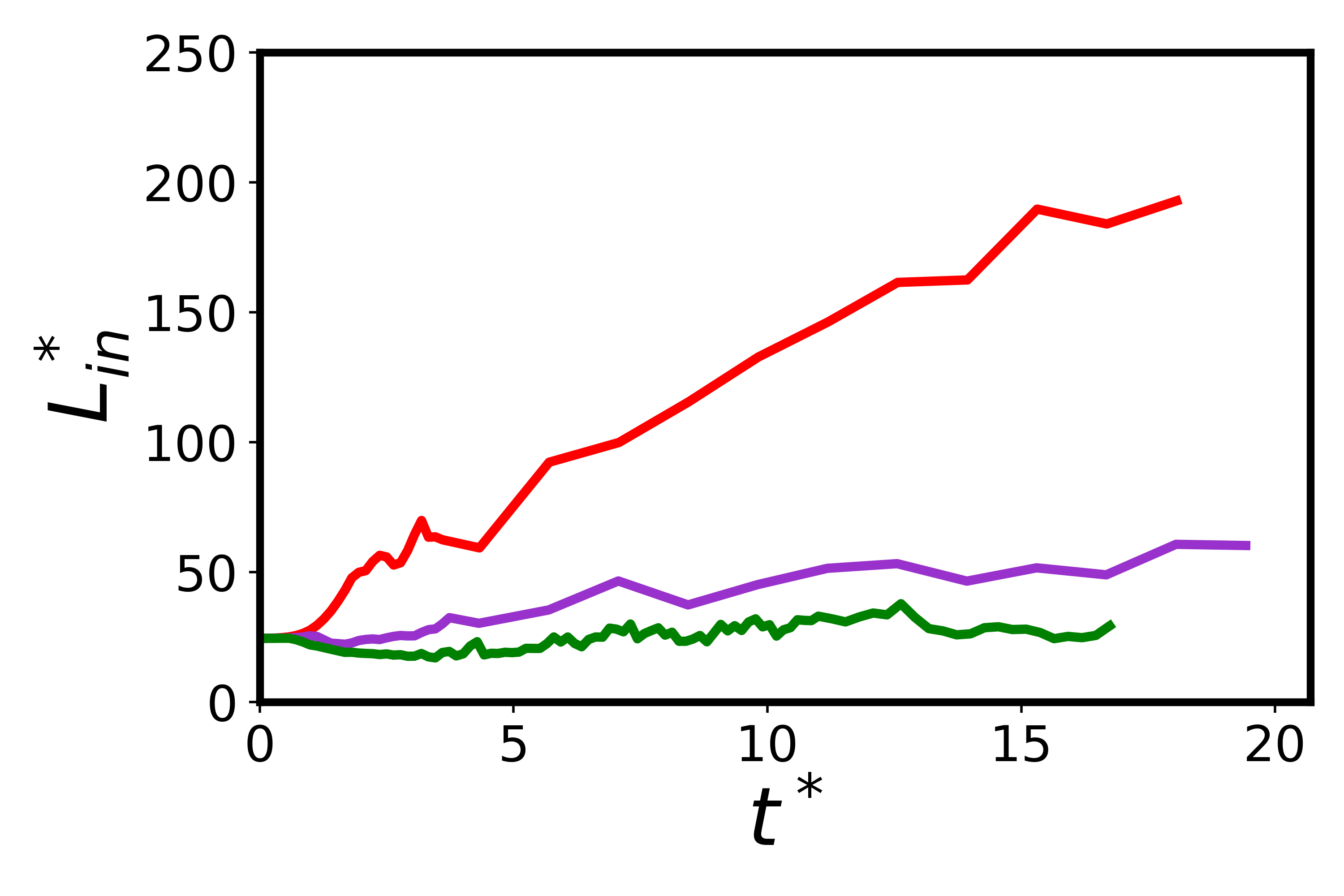}
\put(-130,115){\rotatebox{0}{\large c) $\theta=90^\circ$}}
\includegraphics[width=0.45\textwidth]{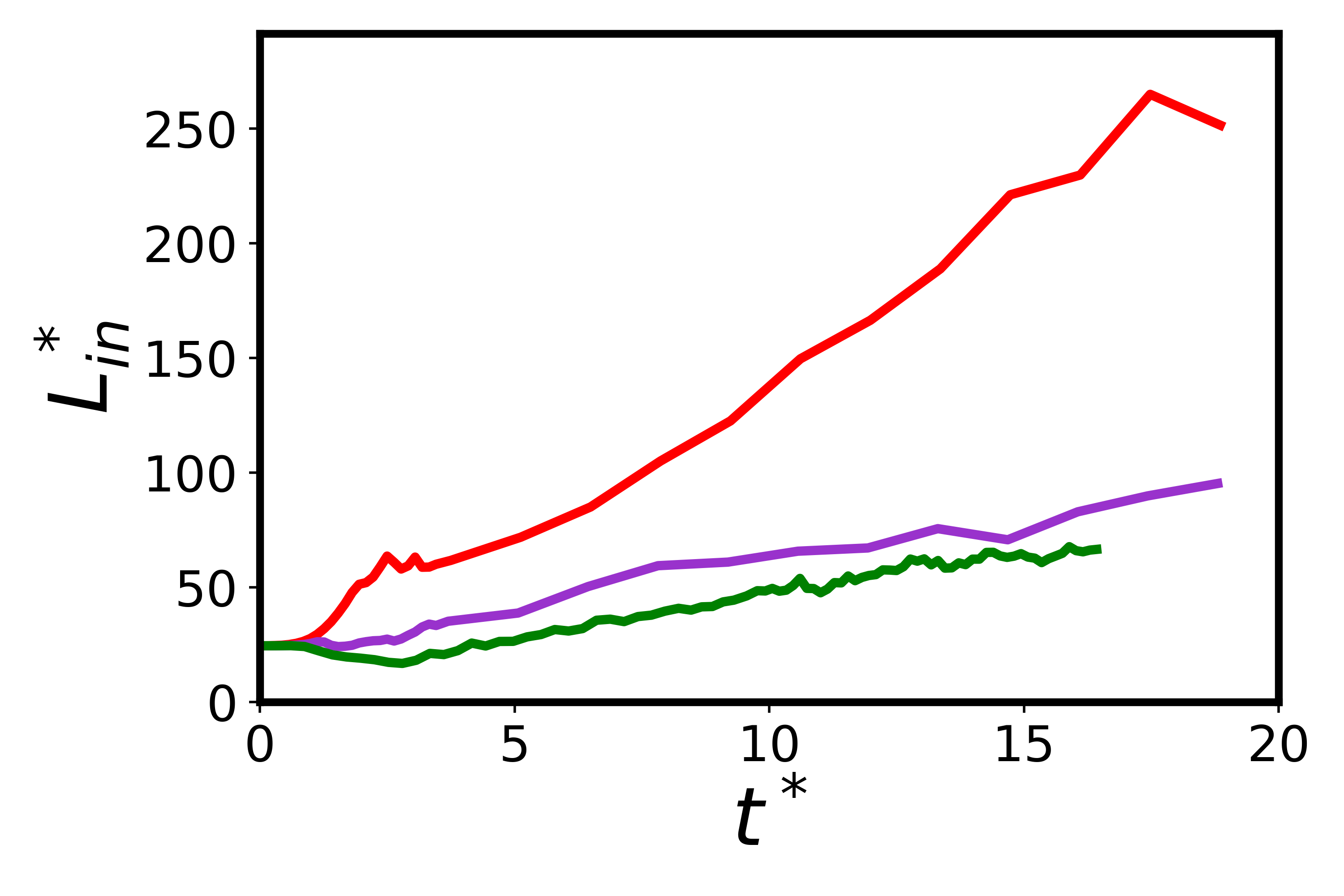}
\put(-130,115){\rotatebox{0}{\large d) $\theta=112.5^\circ$}}

\includegraphics[width=0.45\textwidth]{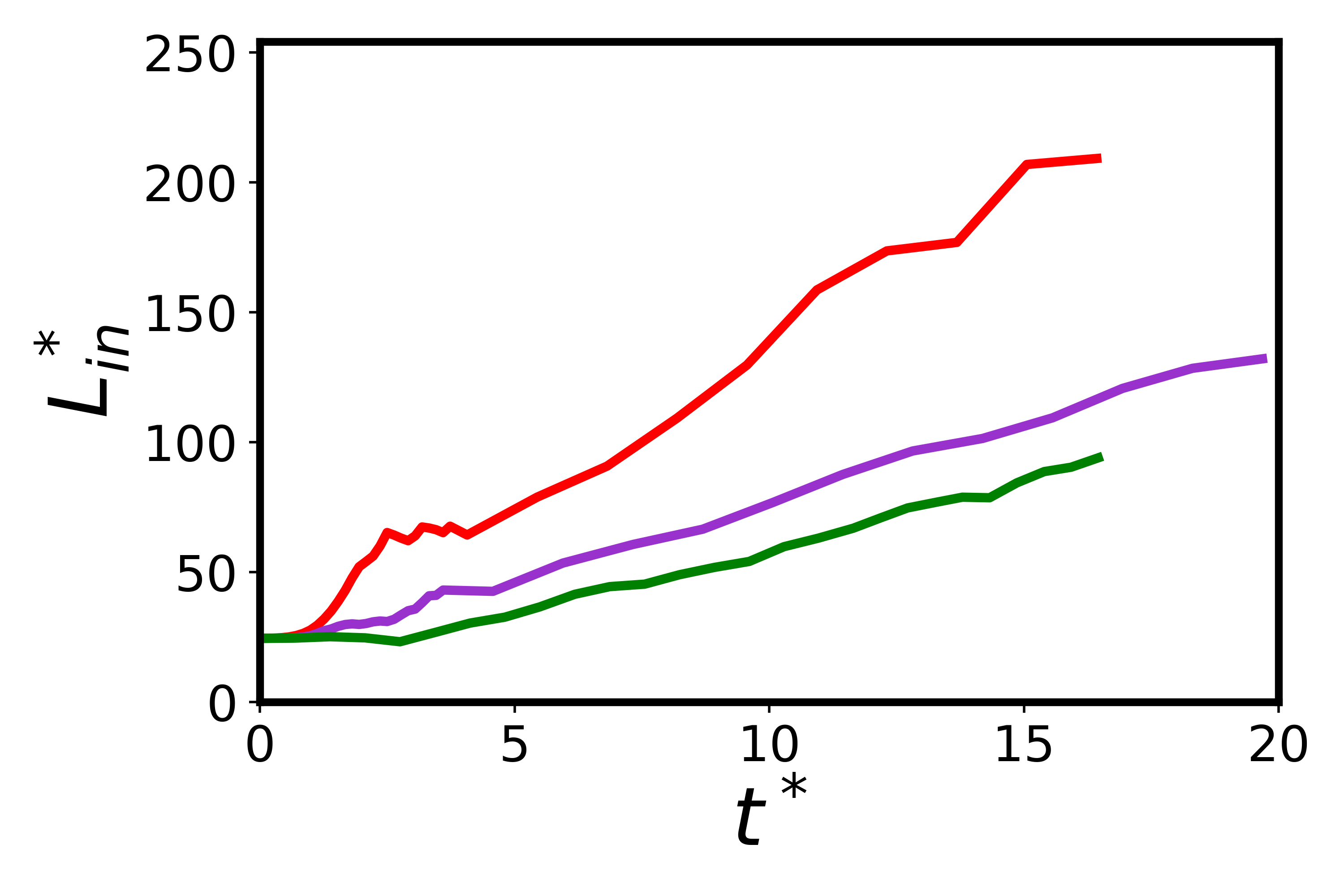}
\put(-130,115){\rotatebox{0}{\large e) $\theta=135^\circ$}}
\includegraphics[width=0.45\textwidth]{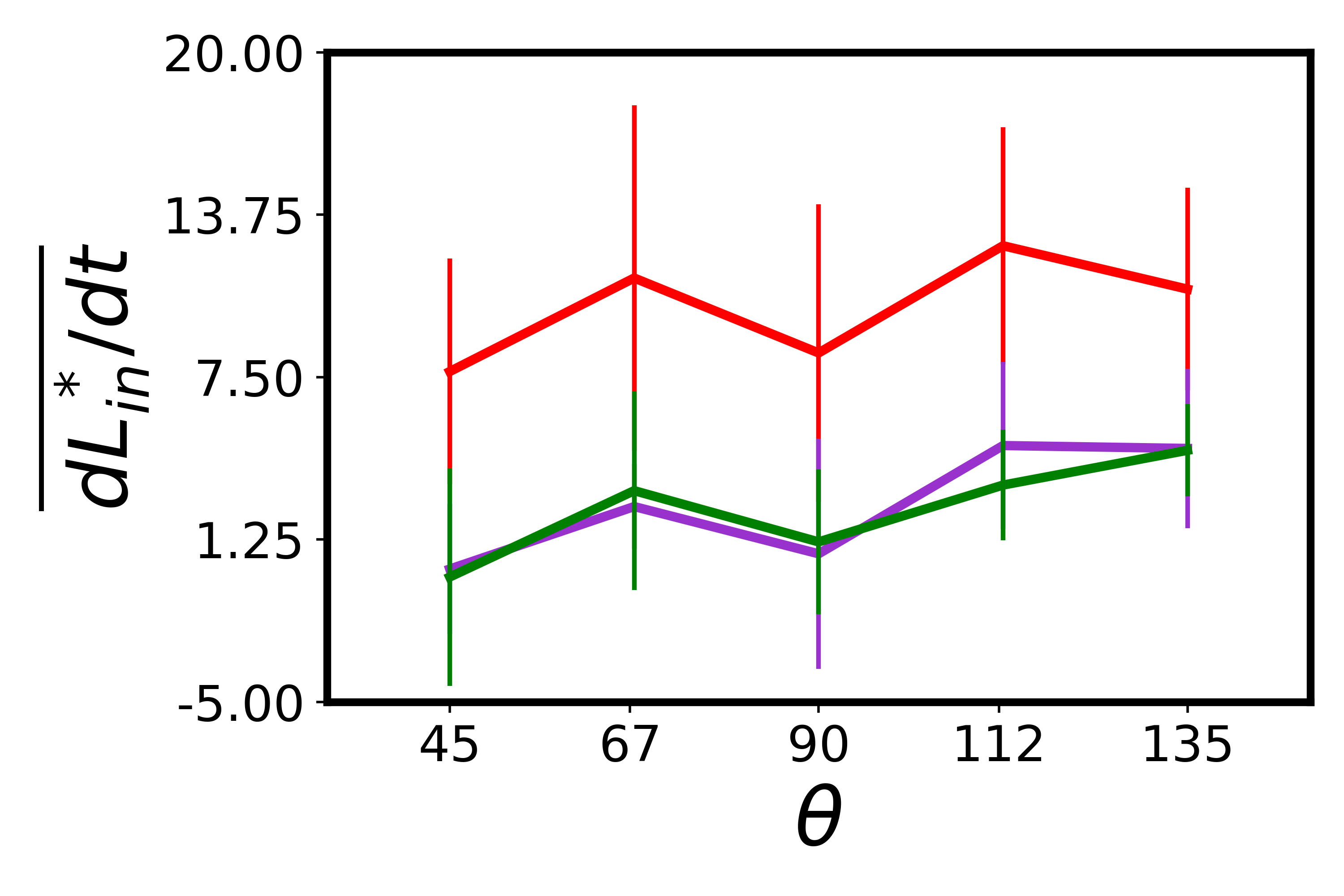}
\put(-130,115){\rotatebox{0}{\large f)}}

\includegraphics[width=0.75\textwidth]{figures/Ca_legend.pdf}
\caption{\sz{a) to e) Evolution of the interface length for different values of the capillary number (see legend) and of the contact angle (see panel title). f) time-averaged rate of change of the interface length. The vertical bars on each plot of frame f represent the standard error of the samples. }}
\label{fig:Tran_length}
\end{center}
\end{figure}
\subsection{Transient analysis} \label{Sec:Transient}

The transient dynamics of an invading fluid penetrating into a porous medium is more relevant than the quasi-steady condition in several applications.
For instance, 
in the paper industry, the largest distance travelled by the invading fluid (ink)  and the penetration velocity within the medium is important because 
the ink should evaporate and dry before reaching the other side of the paper; another example of the same problem is face masks in medical applications  \cite{yi2005numerical,maggiolo2021respiratory}.
In this section, we therefore investigate the evolution of the system properties over time and before saturation, considering the full computational domain as depicted in figure \ref{fig:Setup}. 

\begin{figure} 
\begin{center}
\includegraphics[width=0.5\textwidth]{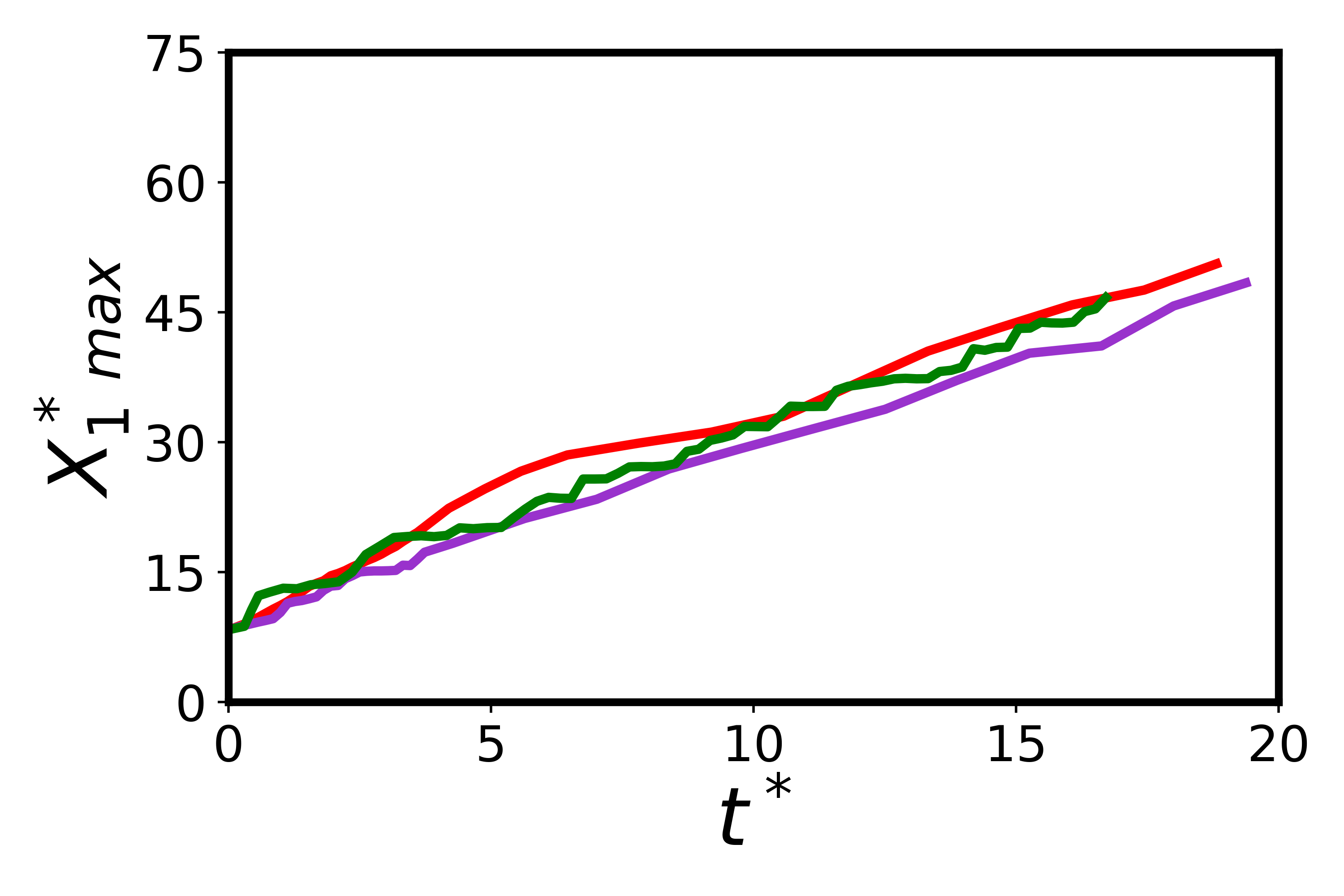}
\put(-132,130){\rotatebox{0}{\large a) $\theta=45^\circ$}}
\includegraphics[width=0.5\textwidth]{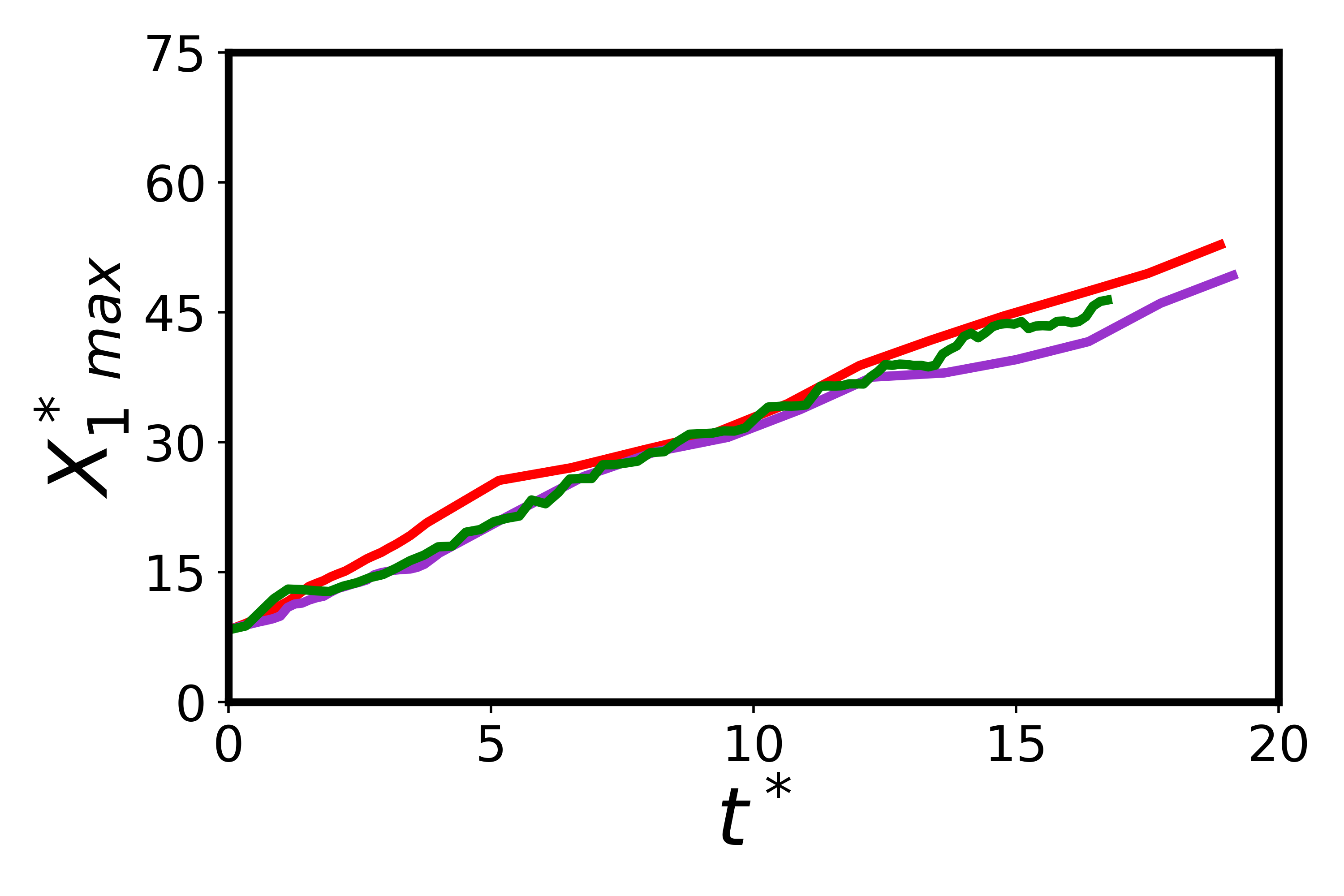}
\put(-132,130){\rotatebox{0}{\large b) $\theta=67.5^\circ$}}

\includegraphics[width=0.5\textwidth]{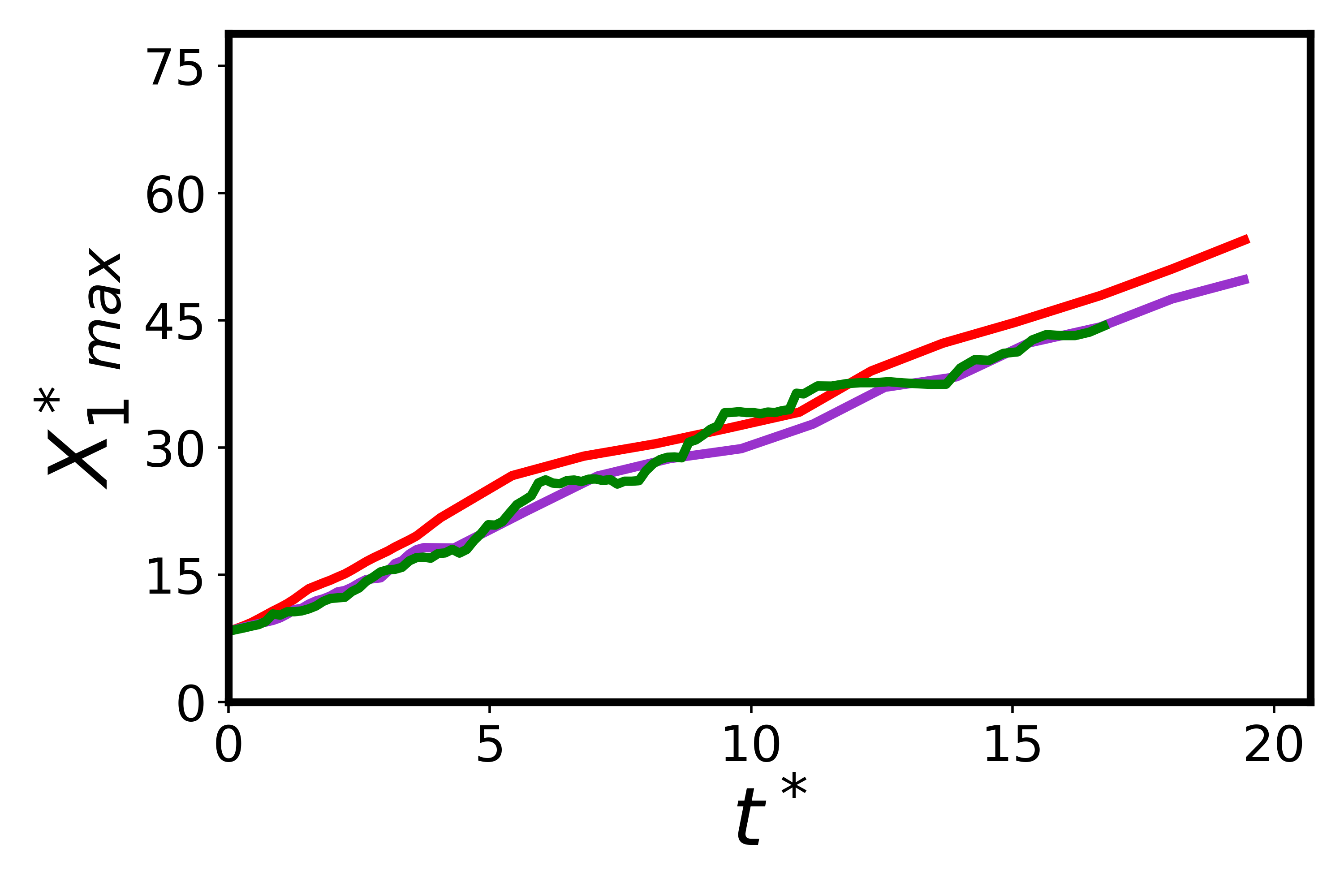}
\put(-132,130){\rotatebox{0}{\large c) $\theta=90^\circ$}}
\includegraphics[width=0.5\textwidth]{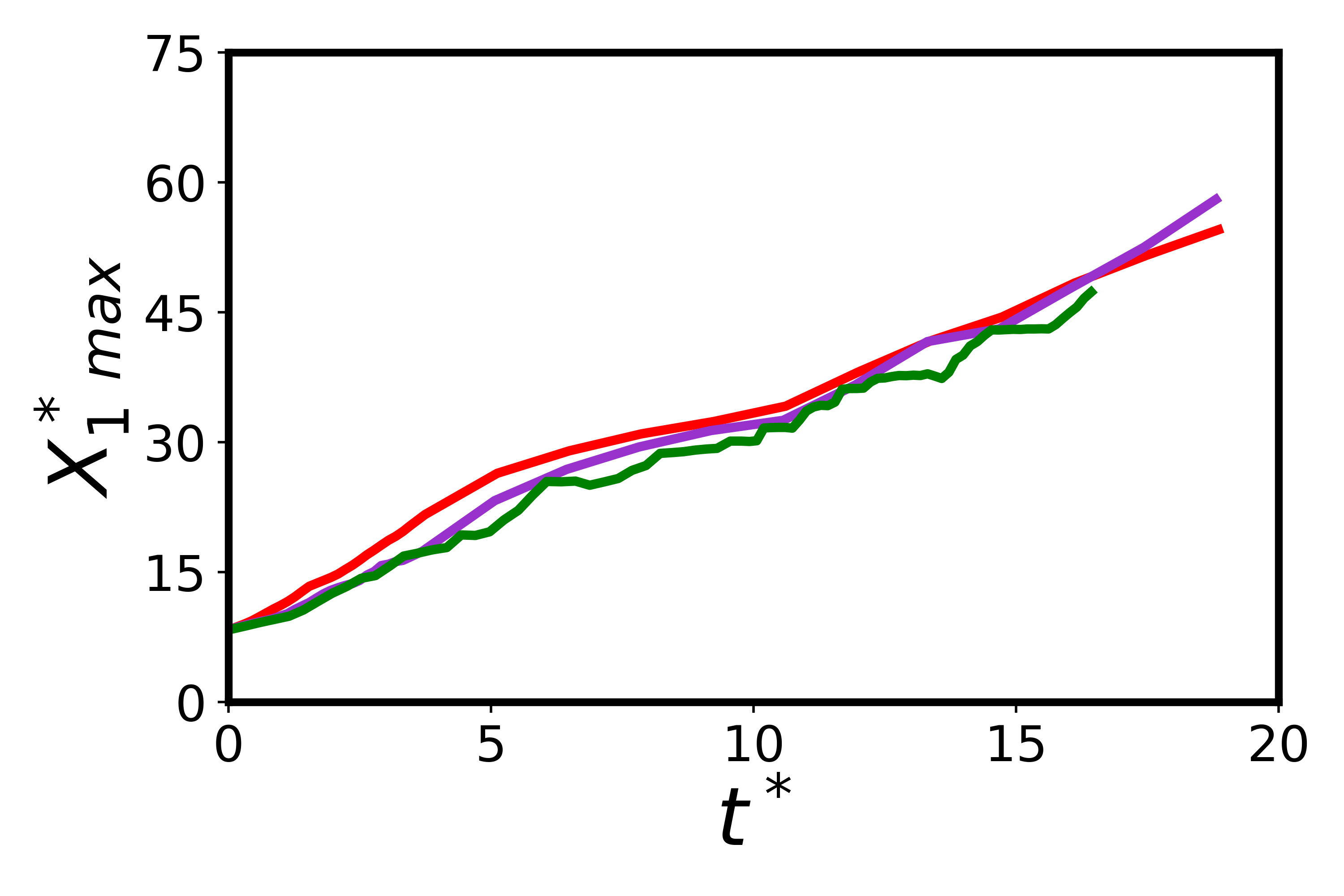}
\put(-132,130){\rotatebox{0}{\large d) $\theta=112.5^\circ$}}

\includegraphics[width=0.5\textwidth]{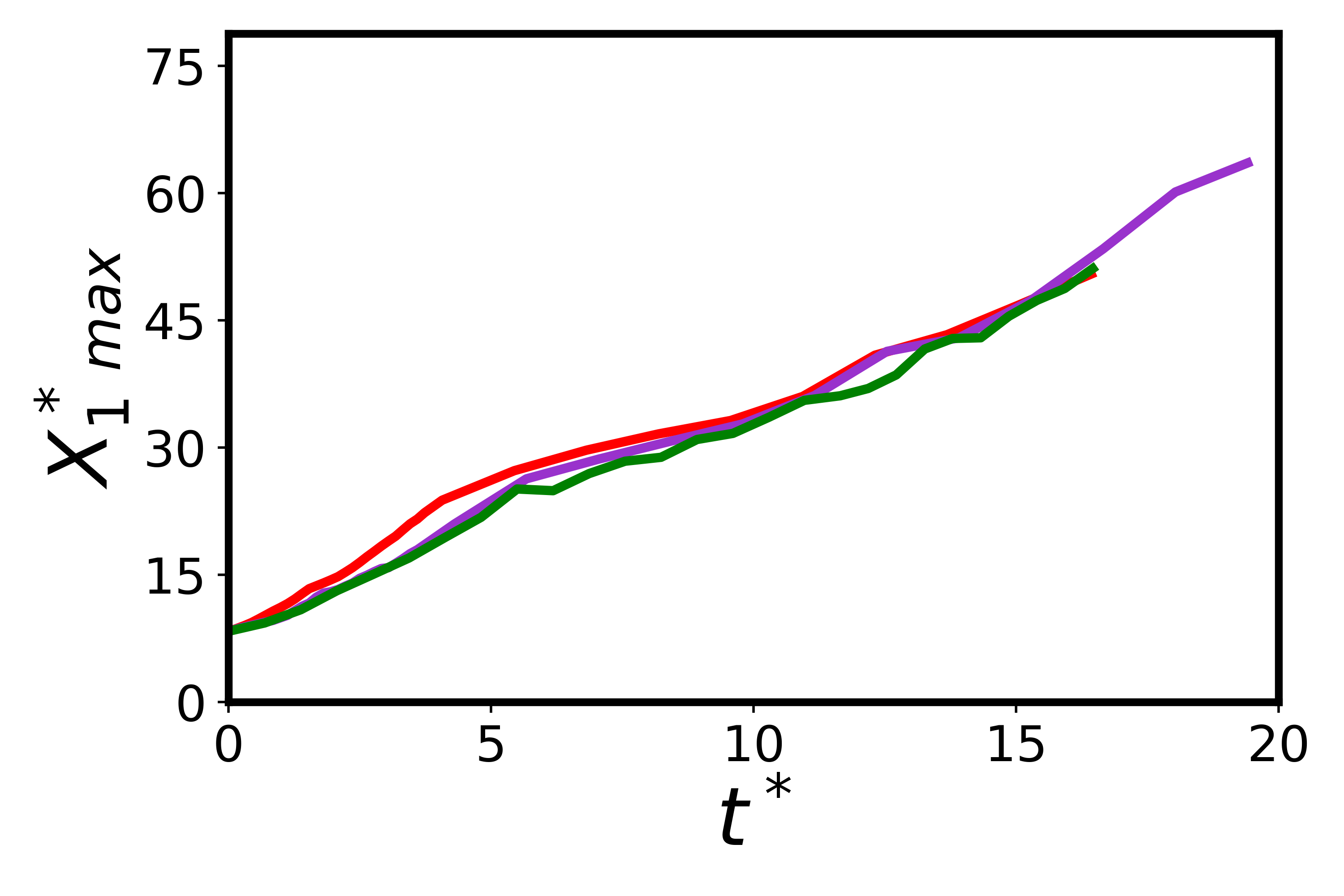}
\put(-132,130){\rotatebox{0}{\large e) $\theta=135^\circ$}}

\includegraphics[width=0.8\textwidth]{figures/Ca_legend.pdf}
\caption{\sz{Evolution of the maximum location of the interface 
for different values of the capillary number (see legend) and of the contact angle (see panel title).}
}
\label{fig:Tran_max}
\end{center}
\end{figure}

We start our analysis of the transient filling by revisiting the length of the interface for all the cases. Figure \ref{fig:Tran_length} represents the evolution of the interface length for different values of the contact angle and capillary number. As also discussed in the previous section, decreasing the capillary number reduces the length of the interface at fixed contact angle because of the shift from \sz{many thin capillary fingers to stable penetration or fewer and thicker fingers.}  
According to figure \ref{fig:Tran_length}, the length of the interface increases almost linearly with time for all the cases, but with different slopes. Figure \ref{fig:Tran_length}f)
therefore illustrates the averaged rate of change of the interface length. Considering the smallest values of the capillary number (purple and green lines in the figure) we observe that for the hydrophilic cases (figures \ref{fig:Tran_length}a and b), characterised by the continuous stable motion of the interface, the length of the interface does not change significantly over time and is almost equal to the effective width of the channel, i.e., the width of the channel without the portion occupied by the solid objects. 
As the surface becomes more hydrophobic instead, the length of the interface increases, which is the indication of the transition from stable penetration to the formation of capillary fingers.

 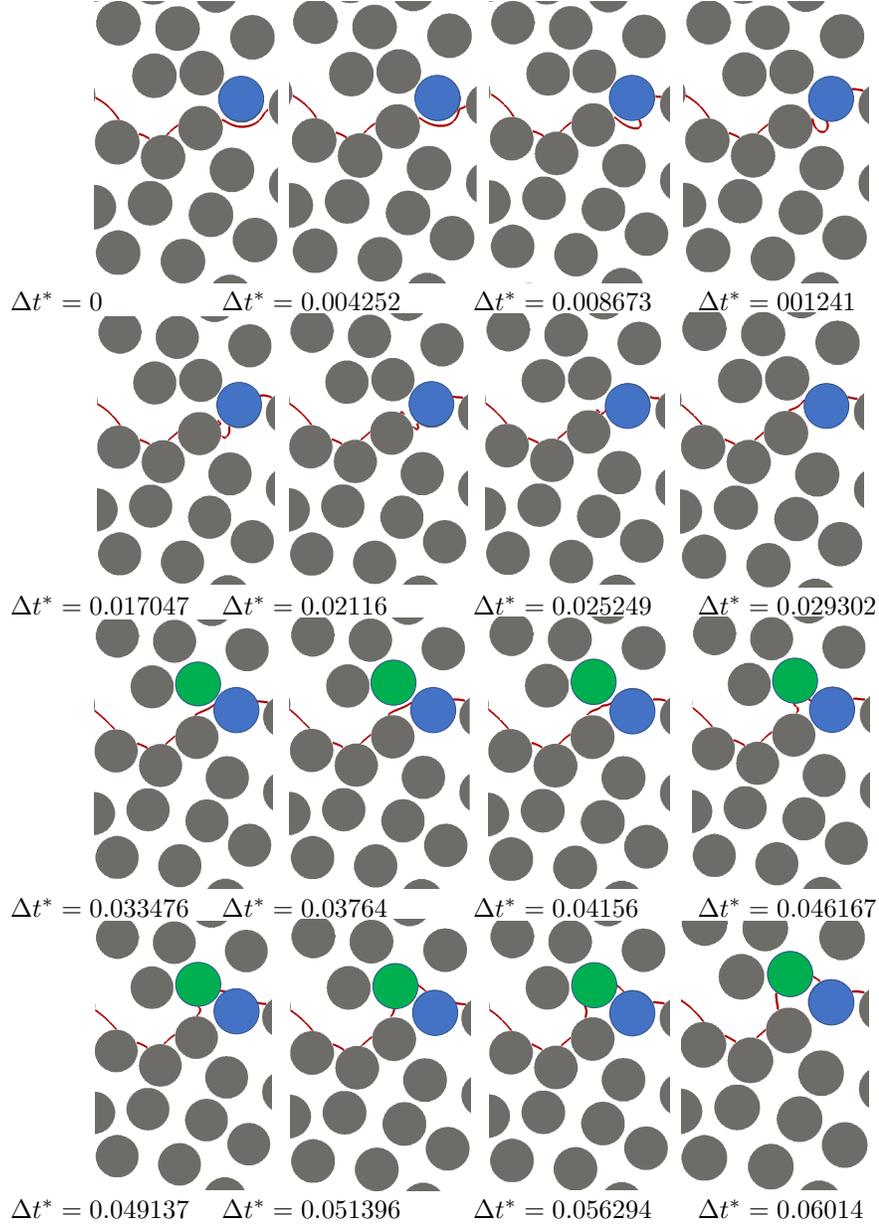
\begin{figure} 
\begin{center}
\includegraphics[width=0.85\textwidth]{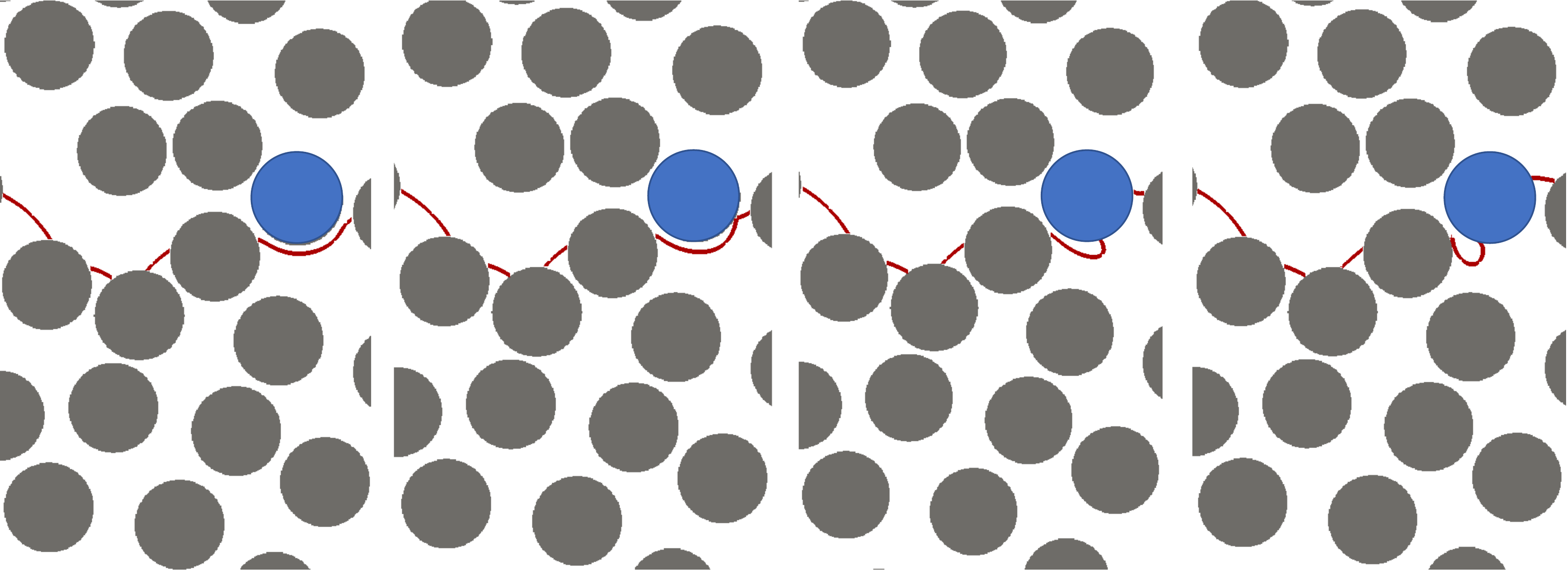}
\put(-325,-10){\rotatebox{0}{\normalsize $\Delta t^*=0$}}
\put(-245,-10){\rotatebox{0}{\normalsize $\Delta t^*=0.004252$}}
\put(-150,-10){\rotatebox{0}{\normalsize $\Delta t^*=0.008673$}}
\put(-65,-10){\rotatebox{0}{\normalsize $\Delta t^*=001241$}}

\includegraphics[width=0.85\textwidth]{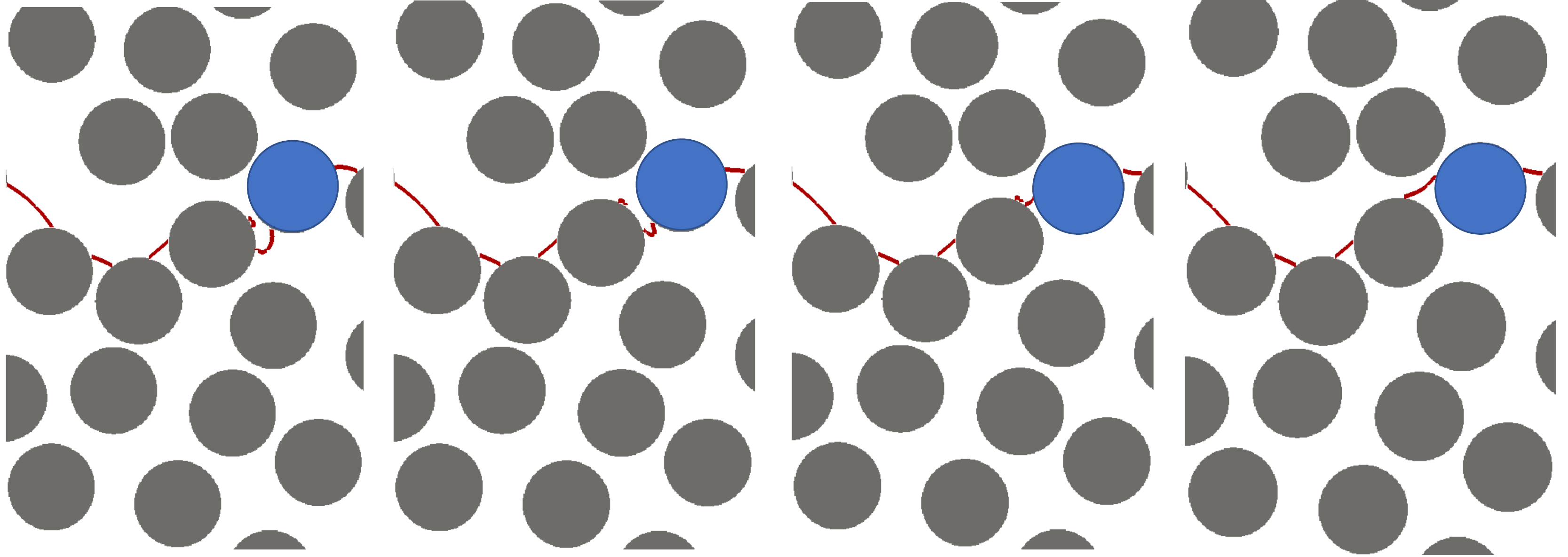}
\put(-325,-10){\rotatebox{0}{\normalsize $\Delta t^*=0.017047$}}
\put(-245,-10){\rotatebox{0}{\normalsize $\Delta t^*=0.02116$}}
\put(-150,-10){\rotatebox{0}{\normalsize $\Delta t^*=0.025249$}}
\put(-65,-10){\rotatebox{0}{\normalsize $\Delta t^*= 0.029302$}}

\includegraphics[width=0.85\textwidth]{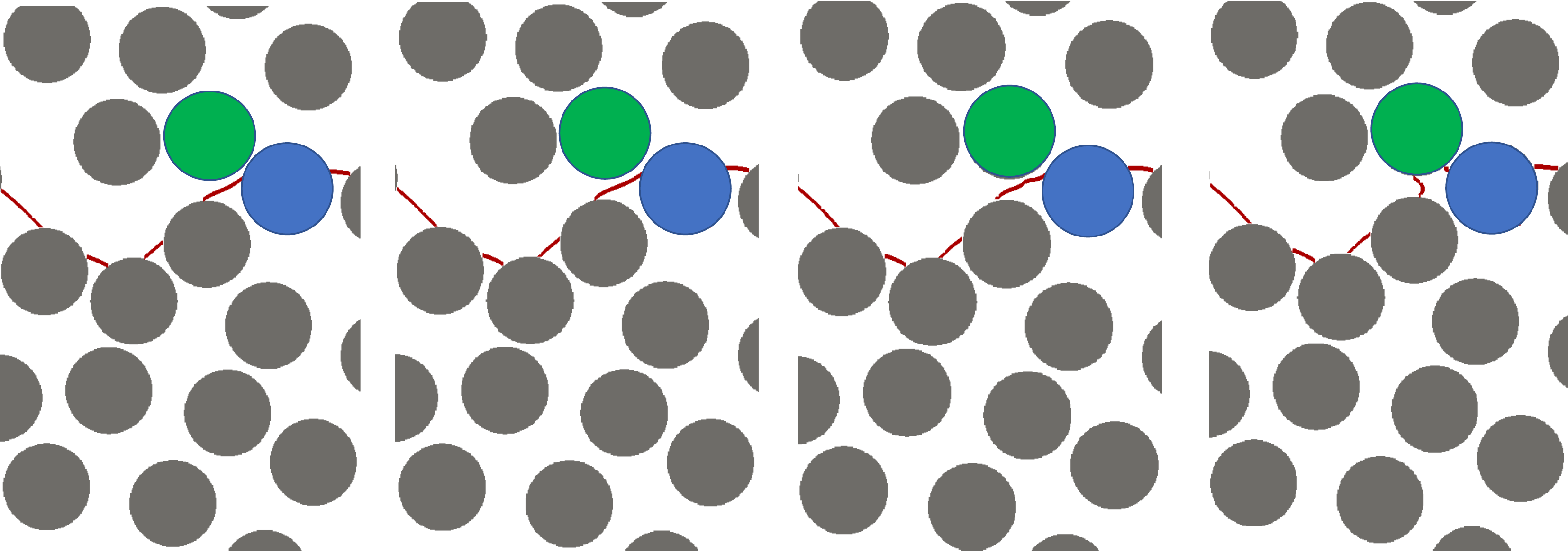}
\put(-325,-10){\rotatebox{0}{\normalsize $\Delta t^*=0.033476$}}
\put(-245,-10){\rotatebox{0}{\normalsize $\Delta t^*=0.03764$}}
\put(-150,-10){\rotatebox{0}{\normalsize $\Delta t^*=0.04156$}}
\put(-65,-10){\rotatebox{0}{\normalsize $\Delta t^*=0.046167$}}

\includegraphics[width=0.85\textwidth]{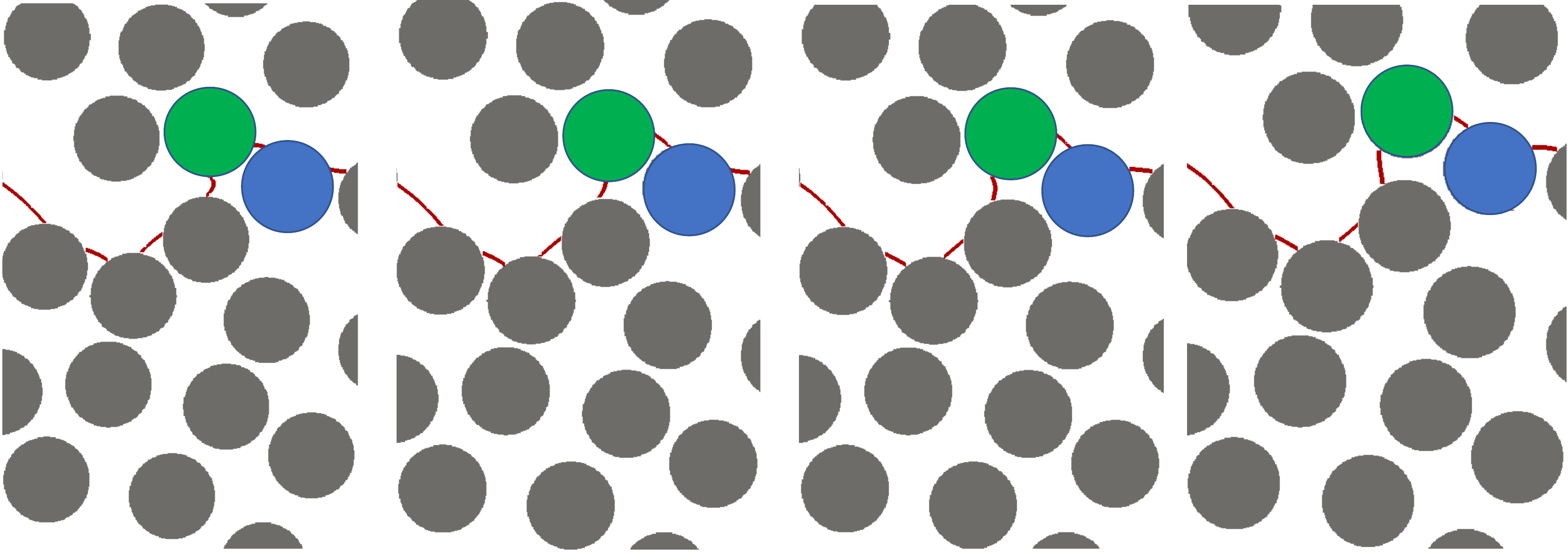}
\put(-325,-10){\rotatebox{0}{\normalsize $\Delta t^*=0.049137$}}
\put(-245,-10){\rotatebox{0}{\normalsize $\Delta t^*=0.051396$}}
\put(-150,-10){\rotatebox{0}{\normalsize $\Delta t^*=0.056294$}}
\put(-65,-10){\rotatebox{0}{\normalsize $\Delta t^*=0.06014$}}

\caption{Occurrence of consecutive jumps at the interface during the short time interval between $t^*=16.001260$ and  $t^*=16.0614$ for $Ca=10^{-4}$ and $\theta = 45^\circ$. The flow is upwards and the blue and the green colours indicate the solid object subjected to the first and the second jump within the selected subdomain.  We define $\Delta t^* = t^*- {t_0}^*$ with ${t_0}^*=16.00126$.}
\label{fig:Jumps}
\end{center}
\end{figure}

Figure \ref{fig:Tran_max} reports the evolution of the maximum distance penetrated by the interface (the point furthest away from the inlet boundary)  for all the cases under study. In general, irrespective of the contact angle at the pores, the larger the capillary number, the more the invading phase penetrates into the defending fluid, except for the cases with the smallest value of capillary number (green lines corresponding to $Ca=10^{-4}$); for this extreme capillary number, when the medium is hydrophilic (see figures \ref{fig:Tran_max}a and b) the invading phase penetrates further into the domain compared to the cases with \sz{larger} values of the capillary number  (red and purple lines in figures \ref{fig:Tran_max}). This is due to the large solid-fluid interaction forces which facilitate the motion of the interface within the medium to adjust the prescribed contact angle. 
Moreover, the oscillatory behaviour in the plots suggests the presence of successive Haines jumps of the three phase contact line as the interface moves within the medium. As the medium becomes less hydrophilic, these oscillations attenuate implying the disappearance of the Haines jumps, 
which results in a reduced penetration of the invading phase when compared to the cases with \sz{higher} capillary numbers, see panels \ref{fig:Tran_max}d) and e).

\begin{figure} 
\begin{center}
\includegraphics[width=0.45\textwidth]{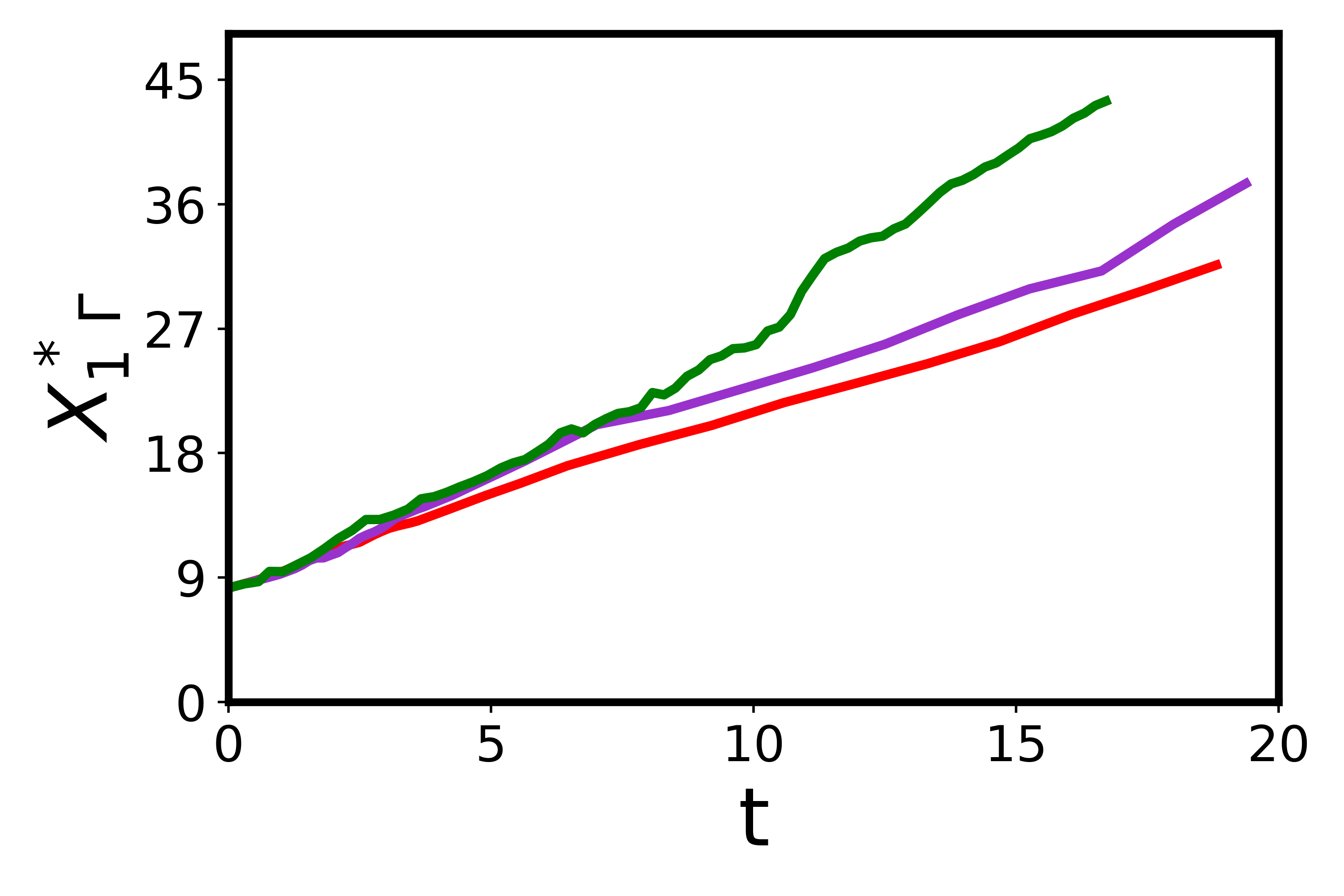}
\put(-130,115){\rotatebox{0}{\large a) $\theta=45^\circ$}}
\includegraphics[width=0.45\textwidth]{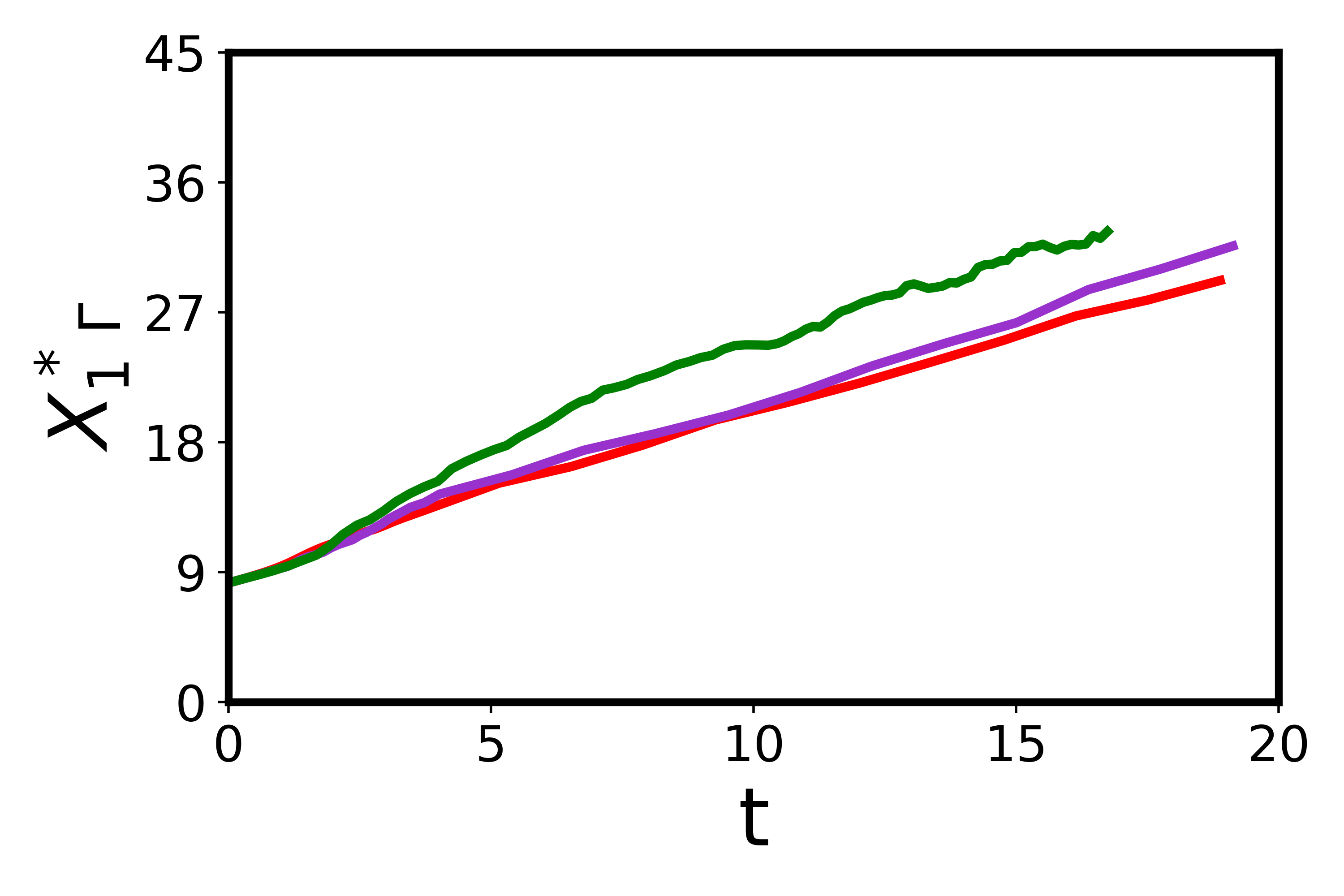}
\put(-130,115){\rotatebox{0}{\large b) $\theta=67.5^\circ$}}

\includegraphics[width=0.45\textwidth]{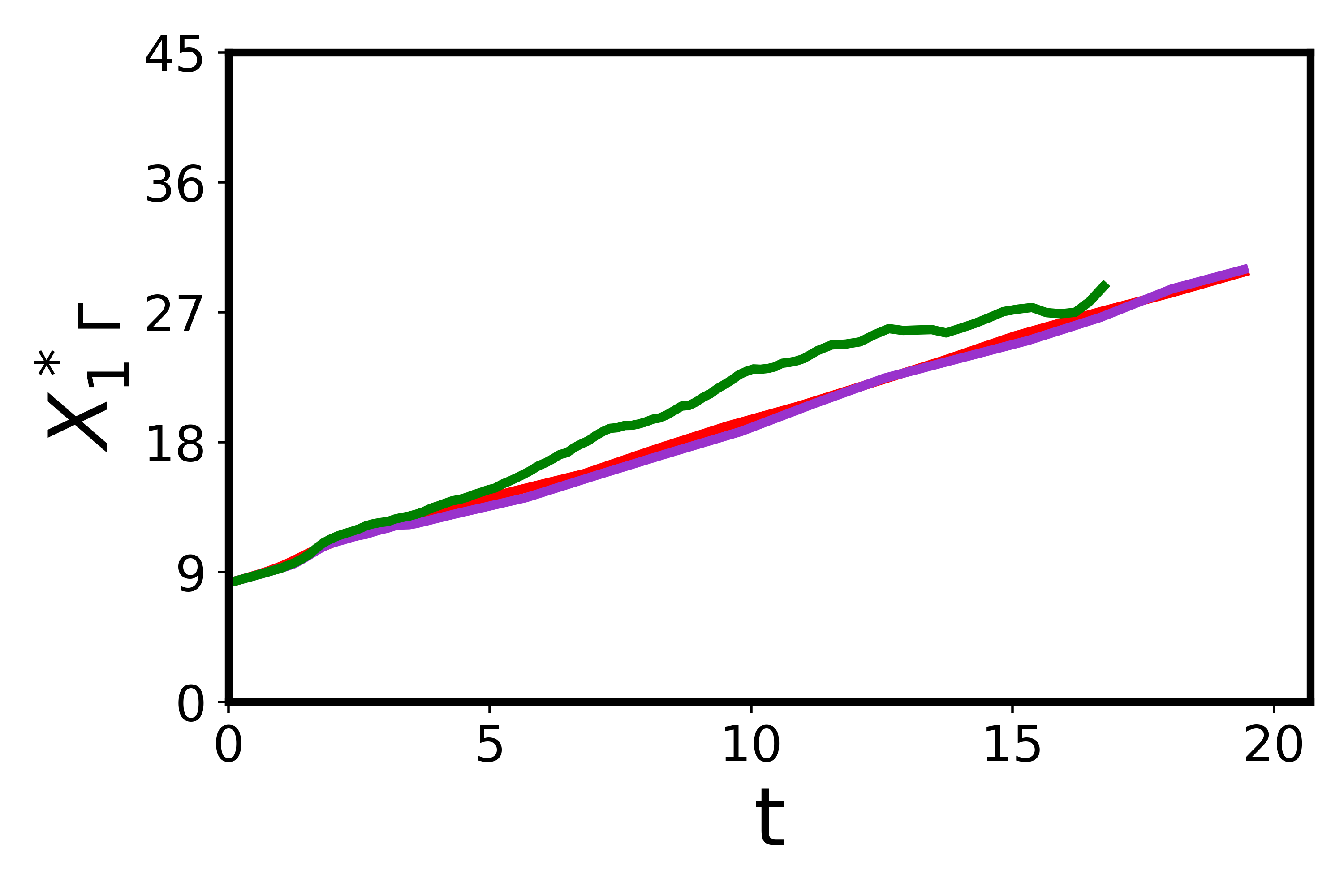}
\put(-130,115){\rotatebox{0}{\large c) $\theta=90^\circ$}}
\includegraphics[width=0.45\textwidth]{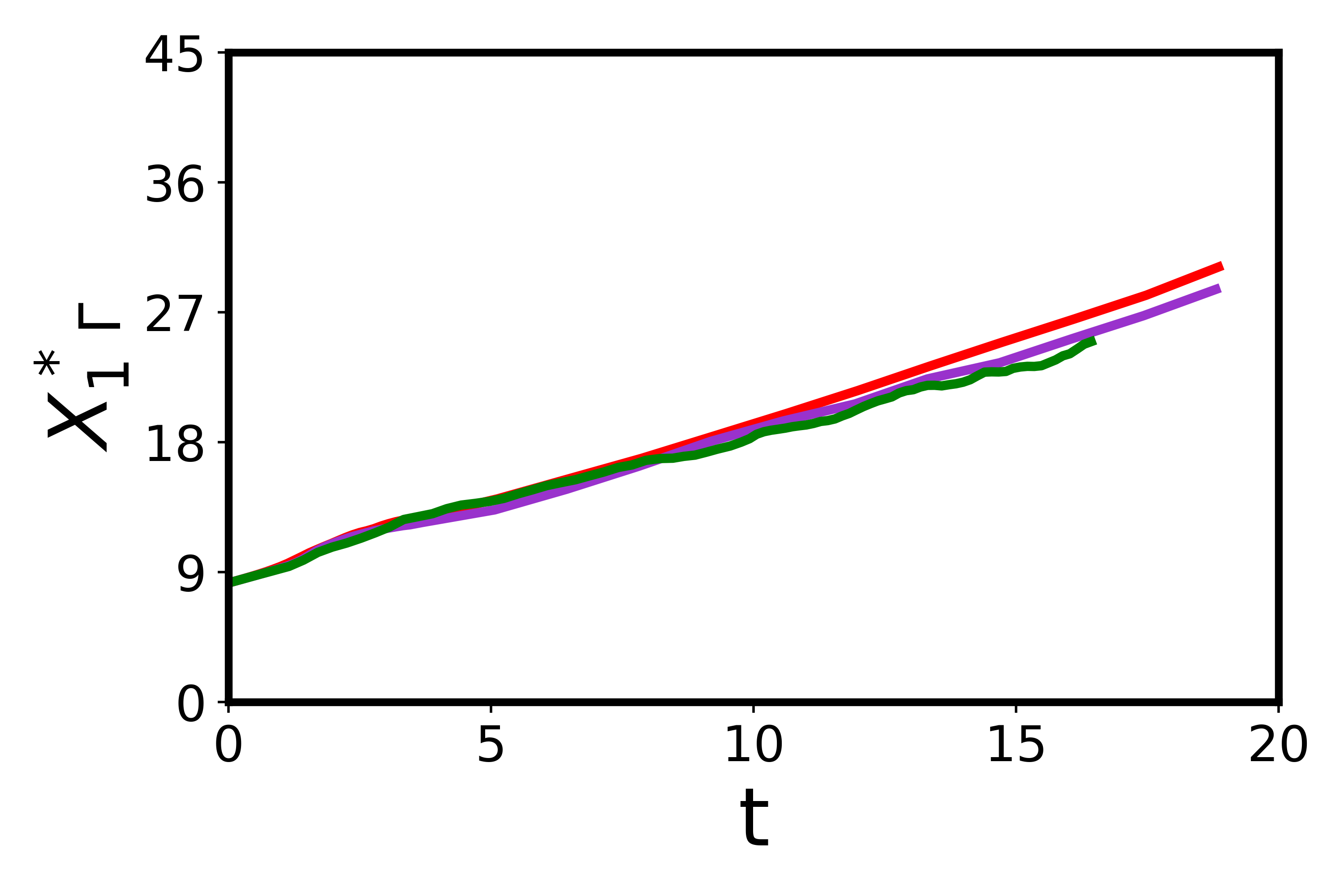}
\put(-130,115){\rotatebox{0}{\large d) $\theta=112.5^\circ$}}

\includegraphics[width=0.45\textwidth]{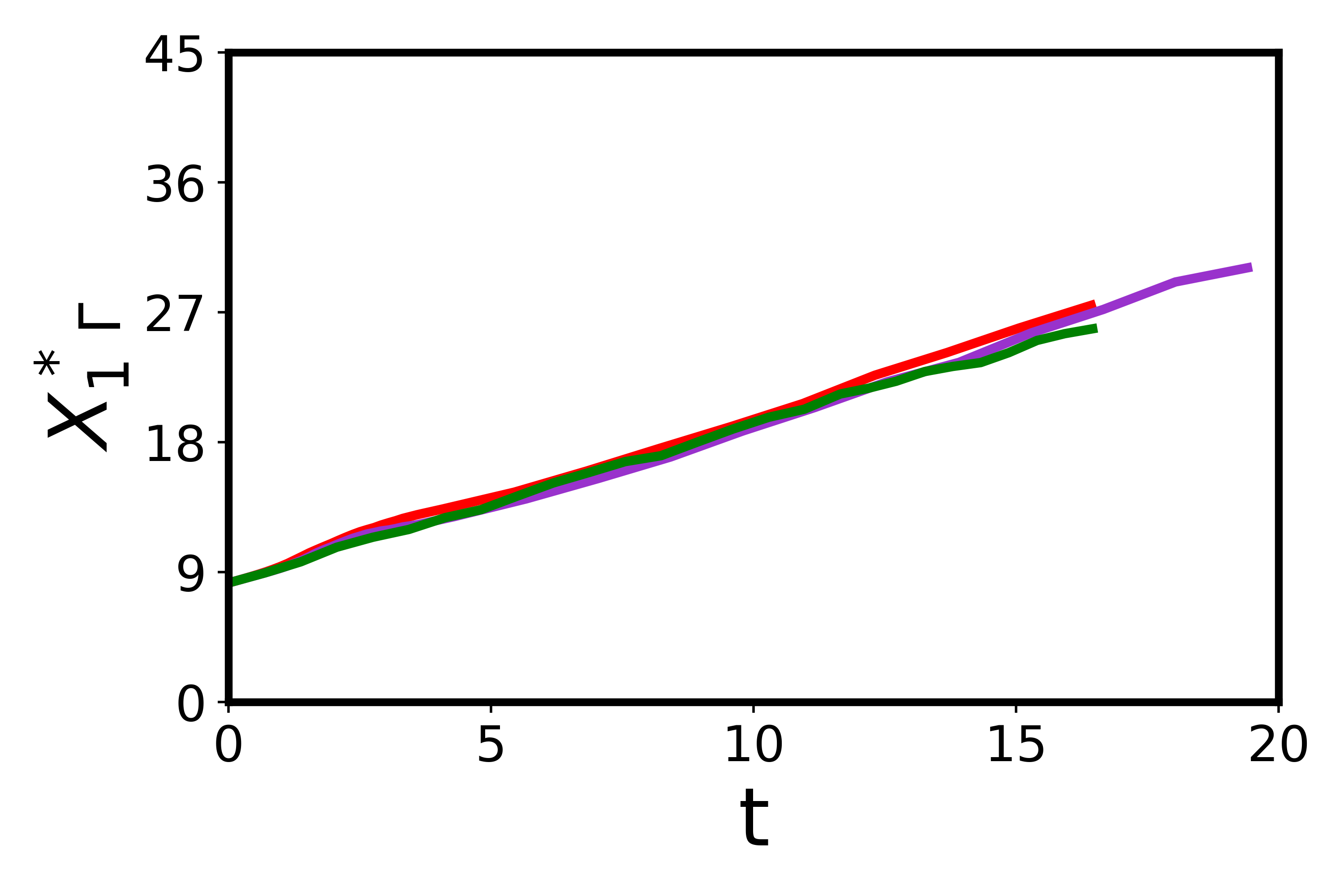}
\put(-130,115){\rotatebox{0}{\large e) $\theta=135^\circ$}}

\includegraphics[width=0.8\textwidth]{figures/Ca_legend.pdf}
\caption{\sz{Evolution of the mean location of the interface
for different values of the capillary number (see legend) and of the contact angle (see panel title).}
}
\label{fig:Tran_mean}
\end{center}
\end{figure}

\sloppy
Figure \ref{fig:Jumps} illustrates consecutive jumps of the interface in the part of the domain marked by the red box in figure \ref{fig:Setup} and
during the time interval between $t^* \in [16.001260, 16.0614]$ for the flow with $Ca=10^{-4}$ and $\theta=45^\circ$. 
 The first row of figure \ref{fig:Jumps} shows the occurrence of the first quick jump on the solid cylinder represented in blue color at $t^*=16.001260$ and  $t^*=16.01367$. To adjust to the prescribed contact angle and at the same time minimise the interfacial free energy, as evident in the second row, the interface shape changes abruptly around the blue solid object. As the interface moves along the blue solid cylinder and approaches the next solid region  
 (displayed in green),  a second jump soon occurs (see the third row of figure \ref{fig:Jumps}).  Once again, after the second jump,  
  the interface undergoes rapid deformations to accommodate the desired contact angle and the requirement of minimum free energy of the system. This scenario (successive jumps followed by rapid changes in the shape of the interface) takes place in a hydrophilic porous medium when the capillary number is small and explains the oscillations in the behaviour of the mean/maximum interface position in time.

Figure \ref{fig:Tran_mean} reports the evolution of the averaged position of the interface, calculated with $\alpha = X_1$ in equation (\ref{averaged_over_length}), for different values of the contact angle and capillary number. \sz{We first note that the data in Figure \ref{fig:Tran_mean}e suggests that the averaged positions of the interface are independent of the capillary number. This can be understood by considering the formation of capillary fingers as the advancing mechanism of the interface, where solid-fluid interaction forces are weak.}
As the contact angle decreases, \sz{ this binary classification becomes however less evident. 
In particular,} we notice that 
the configuration with the smallest capillary number (green line) does not follow the first class (capillary fingering) but introduces a new class of interface motion (stable penetration, see \ref{fig:Tran_mean}c); this is associated with a high level of fluctuations due to the contact line jumps at the interface. 
Further decreasing the contact angle (figures \ref{fig:Tran_mean}b to c), also the flow with $Ca=10^{-3}$ (purple line) moves from the fi   rst class (capillary fingering) to the \sz{second} class (stable penetration), showing an oscillatory behavior.

\section{Conclusion} \label{sec:conclusion}
We have investigated the role of the capillary number and of the surface wettability on the dynamics of the interface between an invading and a defending phase in a porous medium  by means of numerical simulations.  In particular, we consider \sz{three} values of the capillary number ($Ca =10^{-4}, 10^{-3}, 10^{-2}$) and five values of the equilibrium contact angle  ($\theta = 45^\circ,   67.5^\circ,  90^\circ,  112.5^\circ, 135^\circ$) and defined the porous medium by a random arrangement of circular cylinder. 
We identify the signature of the dominant interface motion mechanisms, namely, capillary fingering, and stable penetration, in most common quantitative measurements such as the length of the interface, the pressure drop, interface curvature, etc. The analysis is performed considering two scenarios:   first, we focus on the upstream portion of the porous medium, where the flow properties eventually saturate and hence, the fluid flow is in a quasi-steady state.  
Next, we investigate the transient dynamics and examine the evolution of the different observable over the full length of the domain. In addition, we provide flow visualisations of the  interface between the two immiscible fluids inside the porous medium for the different cases under study and a map of the dominant interface motion mechanisms as function of the capillary number and contact angle.

Our results reveal that for hydrophobic media the invading fluid penetrates the defending phase by forming few thick fingers (so-called capillary fingering).   In a hydrophilic medium, the dominant interface motion 
shifts from capillary fingering to stable penetration  while decreasing the capillary number.

We also show that  the volume of the defending phase decreases linearly as the invading phase is injected into the medium until the system reaches the quasi-steady saturated state when the volume of the defending phase is constant. 
The results suggest that the volume of the defending phase remaining in the observation window can be directly related to the dominant interface motion: for a stable penetration of the interface, the amount of trapped defending phase in the saturated domain is smaller than for the capillary fingering. Our results therefore confirm that the volume of the trapped defending phase is larger for the  capillary fingers \sz{ than for stable penetration of the interface}. 

The analyses of the saturated state reveal that the averaged length of the interface increases as the capillary number increases. We explain this observation by the change of the mode of interface motion with the capillary number: as the capillary number increases, the interface motion shifts from a stable penetration (for the hydrophilic medium) or \sz{few and thick capillary fingers} (for a hydrophobic medium) to the formation of many \sz{but thinner fingers}.

We conclude the saturated-state analyses by examining the average interface curvature and the pressure drop in the system. 
Our results show that by decreasing the capillary number, surface tension forces become more important and the average interface curvature changes with the wettability of the solid pores\sz{:} the interface is concave for a hydrophilic pore and convex for a hydrophobic pore to adjust to the prescribed contact angle, with a decrease in the mean curvature as the surface becomes more hydrophobic. 
As concerns the pressure drop through the saturated domain, it  increases as the capillary number decreases regardless of the prescribed contact angle. This is explained by examining the capillary pressure ($P_c=\kappa \sigma$), or interfacial stresses, which increase as the surface tension $\sigma$ increases. However, at low capillary numbers, we see a significant increase of the pressure drop with the contact angle. This is explained by the increase of the length of the interface, despite the curvature is on average larger and positive for the hydrophillic pores. Indeed, as we increase $\theta$ at smallest $Ca$, the dominant interface mode changes from stable penetration to capillary fingering, so increasing the interface length.

The analysis of the transient regime focuses
on the evolution of the interface length, of the maximum penetration length of the invading phase against the defending fluid, and the mean interface location. 
The evolution of the mean interface location
enables us to identify which of 
the \sz{two} dominant modes of interface motion is at work. 
In particular, 
we \sz{observe an identical behavior} for the interface motion in the hydrophobic medium. For the  smallest capillary number under study, capillary fingering disappears in favor of a stable penetration of the interface which is identified by high level of fluctuations associated to the contact line jumps at the interface.
 
 To conclude, we have presented a numerical framework for the study of immiscible fluids in a porous medium under controlled wetting conditions (static and dynamic contact angle and slip at the wall). We hope this work may trigger new experimental work while future studies may consider non-uniform porosity and permeability and include additional complexities as a non-Newtonian fluid and mass transfer between the phases.

\section*{Acknowledgements}
The computation resources were provided by SNIC (Swedish National Infrastructure for Computing), by the National Infrastructure for High Performance Computing and Data Storage in Norway (project no. NN9561K), and by the Scientific Computing Section of the Core Facilities at OIST. AS and LB are supported by the Swedish Research Council, via the multidisciplinary research environment INTERFACE (VR  2016-06119 ``Hybrid multiscale modelling of transport phenomena for energy efficient processes''), MER by the Okinawa Institute of Science and Technology Graduate University (OIST) with subsidy funding from the Cabinet Office, Government of Japan, and by the JSPS KAKENHI Grant Number JP20K22402, whereas OT gratefully acknowledges the support from the European Research Council through grant No 852529 MUCUS, and from the Swedish Research Council through grant No 2021-04820
\section*{Declaration of Interests}
 The authors report no conflict of interest.
\bibliography{mybibfile}
\end{document}